\magnification1200

%\rightline{KCL-MTH-14-05}
%\rightline{hep-th/yymmnnn}

\vskip 2cm
\centerline
{\bf The string little algebra}
\vskip 1cm
\centerline{Keith Glennon and Peter West}
\centerline{Department of Mathematics}
\centerline{King's College, London WC2R 2LS, UK}
\vskip 2cm
%\centerline{and}
%\vskip 0.5cm
%\centerline{Peter West,}
%\centerline {,}
%\centerline{}
\leftline{\sl Abstract}

\noindent

\vskip .5cm

We consider the string, like point particles and branes, to be an irreducible representation of the semi-direct product of the Cartan involution invariant subalgebra of E11 and its vector representation. We show that the subalgebra that preserves the string charges, the string little algebra, is essentially the Borel subalgebra of E9. We also show that the known string physical states carry a representation of parts of this algebra. 
\vskip 10cm
emails: keith.glennon@kcl.ac.uk, peter.west540@gmail.com
\vfill
\eject

\medskip
{{\bf 1. Introduction}}
\medskip
Quite some time ago it was conjectured  that the theory of strings and branes had a Kac-Moody symmetry that was very extended $E_8$, usually called  $E_{11}$. In particular it was proposed that the non-linear realisation of the semi-direct product of $E_{11}$ and its vector representation, denoted $E_{11}\otimes_s l_1$,  was the low energy effective action for strings and branes [1,2]. The resulting equations of motion are unique and, at low levels they are just the maximally supergravity theories. One finds the different theories by taking different decomposition of $E_{11}$ [3,4]. Indeed one can find all maximally supersymmetric supergravity theories and also all  the gauged maximal supergravities, for a review see reference  [5]. Put simply  the low energy behaviour of strings and branes is encoded in the symmetry $E_{11}\otimes_s l_1$. 
\par
The $E_{11}$ theory treats strings and branes on an equal footing and also encodes an infinite number of  duality symmetries. In particular the vector representation of $E_{11}$ contains all the brane charges [2,6,7,8]. At low levels these charges are those that we are familiar with, such as  the M2 and M5 branes, but   at higher levels one finds an infinite number of new brane charges which were not encountered in other approaches. Thus E theory contains  an infinite number of previously unknown branes and so new degrees of freedom. It  also provides a framework in which to investigate these new branes. 
\par
There is a natural correspondence between the fields in the non-linear realisation of  $E_{11}\otimes_s l_1$ and the brane charges which allows one to identify which field is the source for the brane charge in the vector represenation [6,7,8]. In the non-linear realisation a given  field arises from a positive root of $E_{11}$ and this corresponds to a weight in the vector representation that in turn corresponds to the brane charge. 
Indeed one can go further,  from this root one can construct a group element of $E_{11}$ and which encodes  a putative solution to the field equation with the corresponding brane charge [10]. This method gives the known solutions for the half BPS branes such as the M2 and M5 branes, but at higher levels it provides putative solutions for the infinite number of higher level branes encoded in the vector representation. 
\par
A general formula for the tension  of  these  branes was found in [8].  In the case of the IIA  and IIB theories this formula included  the way the tensions depended on the string coupling constants of these two theories. While this dependence for the familiar branes, such as the  D branes,  was reproduced,  the higher branes often had quite different dependencies on the string coupling. The construction was generalised to find solutions corresponding to  two positive roots in eleven [11] and ten dimensions  [12]. These solutions for low level roots reproduced all the quarter BPS branes in these dimensions. A further generalisation to include three and more numbers of $E_{11}$ roots was given in reference [11]. 
\par
While the brane charges for the familiar branes carry Lorentz indices which are contained in one anti-symmetrised block, for example $Z^{a_1a_2}$ and $Z^{a_1\ldots a_5}$ for the M2 and M5 branes, the brane charges at higher levels contain many more indices and these are usually not contained in one anti-symmetrised block.  Such branes have become known as exotic branes. The existence such charges was first indicated  when examining the action of T duality on brane charges in lower dimensions,  such as in three dimensions [9]. Although there is some literature on exotic branes it usually only utilises a part of the underlying $E_{11}$ structure. Much remains to be understood about the higher level branes and indeed all branes. 
\par
An important step in the development of quantum field theory was  the construction  by Wigner of  all of the  irreducible representations of the Poincare algebra,  which is just the semi-direct product of the Lorentz algebra and the translations [13]. Indeed one can think of a particle as being defined as an irreducible representation of the Poincare algebra.  From this approach one can deduce the on-shell states of a given particle and even the covariant field equations of motion for the free particle. 
\par
The symmetries of E theory are similar in structure to the Poincare algebra which can be  written as $SO(1,D-1)\otimes_s T^D$ where $T^D$ are the translations.  The role of the Lorentz algebra  is played by the Cartan involution subalgebra of $E_{11}$, denoted by  $I_c(E_{11})$,  and the role of the translations is played by the elements of the vector representation of $E_{11}$, denoted by $l_1$. Thus the  algebra $I_c(E_{11})\otimes_s l_1$ has the same semi-direct structure as the Poincare algebra. While the Poincare algebra only contains  the momentum generators for the point particles, the vector representation of $E_{11}$ contains all brane charges. 
\par
As such it is natural to consider irreducible representations of the $I_c(E_{11})\otimes_s l_1$ and one may hope that they define what is meant by the point particles, strings and branes. Indeed mimicking the procedure for Poincare algebra we can choose a  brane charge that we are interested in and then compute the little algebra that leaves the charge  inert. We can then  choose an irreducible representation of this little algebra and then hope that these contain  the degrees of freedom corresponding to the  brane of  interest. Finally one can then boost the representation to find the irreducible representation of $I_c(E_{11})\otimes_s l_1$ and so the on-shell conditions that define the brane and even find  the covariant equations of motion of the free brane [14]. 
\par
 The programme just outlined was carried out for the massless particle, that is, the only non-zero charge in the vector representation was chosen to be the momenta and this was  massless.  The particle little algebra was found to be  $I_c(E_{9})$ and this was shown to have an irreducible  representation that has dimension 128. It contains precisely the bosonic degrees of freedom of eleven dimensional supergravity  [14,15]. The free equations of motion must be those of the above non-linear realisation of $E_{11}\otimes_s l_1$. This implies that this non-linear realisation only has degrees of freedom of eleven dimensional supergravity.
\par
It would be interesting to further carry out the programme outlined above and in this paper we take the next step.  We will consider the the IIA theory and in particular the IIA string from this view point. We will  take the 
only non-zero charges in the vector representation to be those corresponding to the IIA string, that is, we will take  only  the momenta $P_a$ and the string charge $Z^a$ to be non-zero. We will do this  in such a way as to preserve the SO(1,1) symmetry of the string. We find the  subalgebra of $I_c(E_{11})$ that preserves this choice and so the string  little algebra. It has an algebra that has essentially the same algebra as the Borel subalgebra of $E_9$. We will then argue that the known string physical states carry a representation of at least part of this little algebra. 
\par
In section two we will give the $E_{11}\otimes_s l_1$ algebra at low levels in the decomposition which leads to the IIA string. In section three we will  derive the string little algebra at low levels and in section five we find it at all levels. In section six we 
explain how the string little algebra sits in $E_{11}$ and discuss how  the physical  states of the superstring carry a representation of parts of the string little algebra. Finally in section four we consider the same steps for a toy string model based on the algebra $SO(D,D)\otimes_s T^{2D}$ where $D$ is the dimension of spacetime.

%%%%%%%%%%%%%%%%%%%%%%%
\medskip
{{\bf 2. The  IIA algebra }}

\medskip
In this section we summarize the $E_{11}$  algebra in the form corresponding to the IIA theory. The IIA theory arises from the non-linear realisation of $E_{11} \otimes_s l_1$ when we delete node 10 and decompose the $E_{11}$ algebra in terms of the residual ${\rm SO}(10,10)$ algebra. We will refer to node 10 as the IIA node. The resulting listing of the generators in this decomposition,  which  belong to representations of ${\rm SO}(10,10)$,  are more  easily understood if also delete node 11 and further decompose with respect to the remaining ${\rm SL}(10)$ algebra. The  resulting Dynkin diagram where the  deleted nodes are indicated with a $\oplus$    is given by
$$
\matrix{
& & & & & & & & & & & & & & \oplus  & 11 & \oplus & 10 &  \cr
& & & & & & & & & & & & & & | & & | & & \cr
\bullet & - & \bullet & - & \bullet & - & \bullet & - & \bullet & - & \bullet & - & \bullet & - & \bullet & - & \bullet & & \cr
1 & & 2 & & 3 & & 4 & & 5 & & 6 & & 7 & & 8 & & 9 & &  \cr
}
$$
\par
The list of generators resulting from this decomposition can be found using the program SimpLie [16]. The generators are  organized in terms of two levels $(l_{(1)},l_{(2)})$ associated to nodes ten and eleven respectively. We will refer to the first level as the IIA level, or sometimes just the level. The number of up minus down indices on a given generator is equal to  $l_{(1)} + 2 l_{(2)}$, and this can be used to determine the eleven level $l_{(2)}$ in a generator in the list below once the IIA level is known.
\par
One can  also find the generators in the above IIA decomposition directly from the $E_{11}$ algebra in its eleven dimension, M theory, formulation which appears by deleting just the node eleven. We will denote the latter generators with a hat $\hat{}$ below. At level zero ($l_{(1)}=0$) the $E_{11}$  algebra contains the generators [17]
$$
K^{\underline{a}}{}_{\underline{b}}\   , \ \ \tilde{R} \  , \ \ R^{{\underline{a}} {\underline{b}}} \  , \ \ R_{{\underline{a}} {\underline{b}}} \  ,
\eqno(2.1)$$
where $\underline{a},\underline{b},\ldots = 0,1,\ldots,9$. The relation between the two sets of  generators is given by
$$
K^{\underline{a}}{}_{\underline{b}} = \hat{K}^{\underline{a}}{}_{\underline{b}} + {1 \over 6} \delta^{\underline{a}}{}_{\underline{b}} \tilde{R} \  , \ \  \tilde{R} = - \sum_{{\underline{a}}=0}^{9} \hat{K}^{\underline{a}}{}_{\underline{a}} + 2 \hat{K}^{11}{}_{11} \ ,
R^{{\underline{a}} {\underline{b}}} = \hat{R}^{{\underline{a}} {\underline{b}} 11} \   , \ \ R_{{\underline{a}} {\underline{b}}} = \hat{R}_{{\underline{a}} {\underline{b}} 11}.
\eqno(2.2)$$
These generators belong to the adjoint representation of ${\rm SO}(10,10)$, as indeed they must.  As illustrated in this equation we will often use the number 11, rather than 10, to denote the eleventh dimension. 
\par
At IIA  level one we have the  generators
$$
R^{\underline{a}} = \hat{K}^{\underline{a}}{}_{11}   \ , \ R^{{\underline{a}}_1 {\underline{a}}_2 {\underline{a}}_3} = \hat{R}^{{\underline{a}}_1 {\underline{a}}_2 {\underline{a}}_3} \ , \ R^{{\underline{a}}_1 \ldots {\underline{a}}_5} = \hat{R}^{{\underline{a}}_1 \ldots {\underline{a}}_5 11} \ ,
$$
$$
R^{{\underline{a}}_1 \ldots {\underline{a}}_7} = \hat{R}^{{\underline{a}}_1 \ldots {\underline{a}}_7 11,11} \ , \ R^{{\underline{a}}_1 \ldots {\underline{a}}_9} = \hat{R}^{{\underline{a}}_1 \ldots {\underline{a}}_9 11, 11 \, 11}  \ , \
\eqno(2.3)$$
At IIA level level minus one the    generators are
$$
R_{\underline{a}} = \hat{K}^{11}{}_{\underline{a}}   \ , \ R_{{\underline{a}}_1 {\underline{a}}_2 {\underline{a}}_3} = \hat{R}_{{\underline{a}}_1 {\underline{a}}_2 {\underline{a}}_3} \ , \ R_{{\underline{a}}_1 \ldots {\underline{a}}_5} = \hat{R}_{{\underline{a}}_1 \ldots {\underline{a}}_5 11} \ ,
$$
$$
R_{{\underline{a}}_1 \ldots {\underline{a}}_7} = \hat{R}_{{\underline{a}}_1 \ldots {\underline{a}}_7 11,11} \ , \ R_{{\underline{a}}_1 \ldots {\underline{a}}_9} = \hat{R}_{{\underline{a}}_1 \ldots {\underline{a}}_9 11, 11 \, 11}   \ , \
\eqno(2.4) $$
These 512  generators at level one belong to the spinor representation of ${\rm SO}(10,10)$ as do the generators at level minus one.
\par
While at level two the  IIA generators are given by
$$
R^{{\underline{a}}_1 \ldots {\underline{a}}_6} = \hat{R}^{{\underline{a}}_1 \ldots {\underline{a}}_6}  \ , \ R^{{\underline{a}}_1 \ldots {\underline{a}}_8} = \hat{R}^{{\underline{a}}_1 \ldots {\underline{a}}_8,11} \ , \ R^{{\underline{a}}_1 \ldots {\underline{a}}_7,\underline{b}} = \hat{R}^{{\underline{a}}_1 \ldots {\underline{a}}_7 11,\underline{b}} \ ,
$$
$$
\ R^{{\underline{a}}_1 \ldots {\underline{a}}_{10}} _{(1)}= \hat{R}^{{\underline{a}}_1 \ldots {\underline{a}}_{10} , (11 \, 11)}  \ , \
{R}^{{\underline{a}}_1 \ldots {\underline{a}}_{10}}  _{(2)}= \hat{R}^{{\underline{a}}_1 \ldots {\underline{a}}_{10} 11 , 11}  \ ,
$$
$$
R^{{\underline{a}}_1 \ldots {\underline{a}}_{9},\underline{b}} = \hat{R}^{{\underline{a}}_1 \ldots {\underline{a}}_{9} 11, \underline{b} 11} \ , \ R^{{\underline{a}}_1 \ldots {\underline{a}}_{8},\underline{b}_1 \underline{b}_2} = \hat{R}^{{\underline{a}}_1 \ldots {\underline{a}}_{8} 11, \underline{b}_1 \underline{b}_2 11} \ , \
$$
$$
R^{{\underline{a}}_1 \ldots {\underline{a}}_{10},\underline{b}_1 \underline{b}_2} = \hat{R}^{{\underline{a}}_1 \ldots {\underline{a}}_{10} 11 , \underline{b}_1 {\underline{b}}_2 11, 11} \ , \ R^{{\underline{a}}_1 \ldots {\underline{a}}_{9},\underline{b}_1 \underline{b}_2 \underline{b}_3} = \hat{R}^{{\underline{a}}_1 \ldots {\underline{a}}_{9} 11  , \underline{b}_1 {\underline{b}}_2 {\underline{b}}_3 11, 11} \ , $$
$$
R^{{\underline{a}}_1 \ldots {\underline{a}}_{10},\underline{b}_1 \ldots \underline{b}_4} = \hat{R}^{{\underline{a}}_1 \ldots {\underline{a}}_{10} 11, \underline{b}_1 \ldots {\underline{b}}_4 11, (11, 11)}  \ , \ \eqno(2.5) $$
and at level minus two the  IIA  generators are
$$
R_{{\underline{a}}_1 \ldots {\underline{a}}_6} = \hat{R}_{{\underline{a}}_1 \ldots {\underline{a}}_6}  \ , \ R_{{\underline{a}}_1 \ldots {\underline{a}}_8} = \hat{R}_{{\underline{a}}_1 \ldots {\underline{a}}_8,11} \ , \ R_{{\underline{a}}_1 \ldots {\underline{a}}_7,\underline{b}} = \hat{R}_{{\underline{a}}_1 \ldots {\underline{a}}_7 11,\underline{b}} \ , $$
$$
R_{(1)}{}_{{\underline{a}}_1 .. {\underline{a}}_{10}} = \hat{R}_{{\underline{a}}_1 .. {\underline{a}}_{10}, (11 \, 11)}  \ \ , \ \ R_{(2)}{}_{{\underline{a}}_1 .. {\underline{a}}_{10}} = \hat{R}_{{\underline{a}}_1 .. {\underline{a}}_{10} 11 , 11}  \ \ , $$
$$
R_{{\underline{a}}_1 .. {\underline{a}}_{9},\underline{b}} = \hat{R}_{{\underline{a}}_1 .. {\underline{a}}_{9} 11, \underline{b} 11} \  \ , \ \ R_{{\underline{a}}_1 .. {\underline{a}}_{8},\underline{b}_1 \underline{b}_2} = \hat{R}_{{\underline{a}}_1 .. {\underline{a}}_{8} 11, \underline{b}_1 \underline{b}_2 11} \ \ , $$
$$
R_{{\underline{a}}_1 \ldots {\underline{a}}_{10},\underline{b}_1 \underline{b}_2} = \hat{R}_{{\underline{a}}_1 \ldots {\underline{a}}_{10} 11, \underline{b}_1 {\underline{b}}_2 11, 11} \ \ , \ \ R_{{\underline{a}}_1 \ldots {\underline{a}}_{9},\underline{b}_1 \underline{b}_2 \underline{b}_3} = \hat{R}_{{\underline{a}}_1 \ldots {\underline{a}}_{9} , {\underline{b}}_1 {\underline{b}}_2 {\underline{b}}_3 11,11} \ \ ,  $$
$$
R_{{\underline{a}}_1 \ldots {\underline{a}}_{10},\underline{b}_1 \ldots \underline{b}_4} = \hat{R}_{{\underline{a}}_1 \ldots {\underline{a}}_{10} 11, {\underline{b}}_1 \ldots {\underline{b}}_4 \ 11 , (11 , 11)} \ \ .
\eqno(2.6) $$
Here the indices of the IIA generators in a given block are anti-symmetric. The generators at level two belong to the four rank tensor and a singlet representations  of ${\rm SO}(10,10)$ which have dimensions 4845 and 1 respectively. We note that there are two generators with ten anti-symmetric indices which are distinguished by the subscripts 1 and 2. The algebra of the IIA  generators up to   level one is given in Appendix A. We note that at IIA  level three we have a spinor and a two tensor with a spinor index representations  of ${\rm SO}(10,10)$ which have dimensions 512 and 87040 respectively. 
\par
We now consider the action of the Cartan involution in the IIA theory. We recall that the Cartan involution $I_c$ is an automorphism of the algebra which acts as $I_c(AB) = I_c(A) I_c(B)$, for any two elements of the algebra $A$ and $B$, and satisfies $I_c^2(A) = A$ for all elements $A$. Its action on the IIA generators is inherited from its action in the eleven dimensional theory where $I_c(\hat R^{\alpha})= - (-1)^{l+1} \hat R_{\alpha}$ where $l$ is the  level of $R^{\alpha}$ except at level zero where $I_c(K^{\underline{a}}{}_{\underline{b}}) = - \eta^{\underline{a} \underline{c}} \eta_{\underline{b} \underline{d}} K^{\underline{d}}{}_{\underline{c}} $ and then using the relationship between the eleven dimensional and IIA generators of equations (2.2)-(2.6).  Acting on  the  IIA  level zero   algebra generators we find that
$$
I_c(K^{\underline{a}}{}_{\underline{b}}) = - K^{\underline{b}}{}_{\underline{a}} \ \ , \ \ I_c(\tilde{R}) = - \tilde{R} \ \ , \ \  I_c(R^{\underline{a}_1 \underline{a}_2}) =  -\eta ^{\underline{a}_1\underline{b}_1}  \eta ^{\underline{a}_2\underline{b}_2}   R_{\underline{b}_1 \underline{b}_2} , 
\eqno(2.7) $$
 on the level one generators as
$$
I_c(R^{\underline{a}}) = - \eta^{\underline{a}\underline{b}}R_{\underline{b}} \ \ , \ \  I_c(R^{{\underline{a}}_1 {\underline{a}}_2 {\underline{a}}_3} ) = -
\eta^{\underline{a}_1\underline{a}_2\underline{a}_3 | \underline{b}_1\underline{b}_2\underline{b}_3}R_{{\underline{b}}_1 {\underline{b}}_2 {\underline{b}}_3} \ ,  
$$
$$
I_c(R^{{\underline{a}}_1 \ldots {\underline{a}}_5}) = + \eta^{\underline{a}_1 \ldots \underline{a}_5 | \underline{b}_1 \ldots \underline{b}_5} R_{{\underline{b}}_1 \ldots {\underline{b}}_5} \ ,
 \ I_c(R^{{\underline{a}}_1 \ldots {\underline{a}}_7} ) = - \eta^{\underline{a}_1 \ldots \underline{a}_7 | \underline{b}_1 \ldots \underline{b}_7}  R_{{\underline{b}}_1 \ldots {\underline{b}}_7}  \ , $$
$$
I_c(R^{{\underline{a}}_1 \ldots {\underline{a}}_9}) = + \eta^{\underline{a}_1 \ldots \underline{a}_9 | \underline{b}_1 \ldots \underline{b}_9} R_{{\underline{b}}_1 \ldots {\underline{b}}_9} \ , \ \eqno(2.8)$$
and  on the level two generators as
$$
I_c(R^{{\underline{a}}_1 \ldots {\underline{a}}_6}) = + \eta^{{\underline{a}}_1 \ldots \underline{a}_6 | \underline{b}_1 \ldots \underline{b}_6} R_{{\underline{b}}_1 \ldots {\underline{b}}_6} \ , \ I_c(R^{{\underline{a}}_1 \ldots {\underline{a}}_8} ) = - \eta^{\underline{a}_1 \ldots \underline{a}_8 | \underline{b}_1 \ldots \underline{b}_8}  R_{{\underline{b}}_1 \ldots {\underline{b}}_8} \ , $$
$$
I_c(R^{{\underline{a}}_1 .. {\underline{a}}_7,{\underline{b}}}) = - \eta^{{\underline{a}}_1 .. {\underline{a}}_7 , {\underline{b}} | \underline{c}_1 .. \underline{c}_7 {\underline{d}}} R_{{\underline{c}}_1 .. {\underline{c}}_7,\underline{d}} \ , \ I_c(R_{(1)}^{\underline{a}_1 .. \underline{a}_{10}}) = + \eta^{\underline{a}_1 .. \underline{a}_{10} | {\underline{b}}_1 .. {\underline{b}}_{10}}  R_{(1)}{}_{{\underline{b}}_1 .. {\underline{b}}_{10}} \ , \ $$
$$
I_c(R_{(2)}^{{\underline{a}}_1 .. {\underline{a}}_{10}}) = + \eta^{{\underline{a}}_1 .. \underline{a}_{10} | {\underline{b}}_1  .. \underline{b}_{10}} R_{(2)}{}_{\underline{b}_1 .. \underline{b}_{10}} \ , \ I_c(R^{{\underline{a}}_1 .. {\underline{a}}_{9},\underline{b}}) = + \eta^{{\underline{a}}_1 .. \underline{a}_{9} {\underline{b}} | {\underline{c}}_1  .. \underline{b}_{9} {\underline{d}}} R_{{\underline{c}}_1 \ldots {\underline{c}}_{9},\underline{d}} \ , $$
$$
I_c(R^{{\underline{a}}_1 \ldots {\underline{a}}_{8},\underline{b}_1 \underline{b}_2}) = \eta^{{\underline{a}}_1 .. \underline{b}_{2} | {\underline{c}}_1 .. \underline{d}_{2}} R_{{\underline{c}}_1 \ldots {\underline{c}}_{8},\underline{d}_1 \underline{d}_2} , I_c(R^{\underline{a}_1 \ldots \underline{a}_{10}, \underline{b}_1 \underline{b}_2}) = - \eta^{\underline{a}_1 .. {\underline{b}}_2 | {\underline{c}}_1 .. {\underline{d}}_2}  R_{{\underline{c}}_1 \ldots {\underline{c}}_{10}, {\underline{d}}_1 {\underline{d}}_2} , $$
$$
I_c(R^{{\underline{a}}_1 \ldots {\underline{a}}_{9}, {\underline{b}}_1 {\underline{b}}_2 {\underline{b}}_3}) = - \eta^{{\underline{a}}_1 \ldots \underline{b}_{3} | {\underline{c}}_1 \ldots {\underline{d}}_{3}} R_{\underline{c}_1 \ldots \underline{c}_{9}, \underline{d}_1 {\underline{d}}_2 \underline{d}_3} , $$
$$
I_c(R^{{\underline{a}}_1 \ldots {\underline{a}}_{10}, {\underline{b}}_1 \ldots {\underline{b}}_4}) = - \eta^{{\underline{a}}_1 \ldots \underline{b}_{4} | {\underline{c}}_1 \ldots {\underline{d}}_{4}} R_{\underline{c}_1 \ldots \underline{c}_{9}, \underline{d}_1 \ldots \underline{d}_4} . \eqno(2.9) $$
In these equations $\eta^{\underline{a}_1\underline{a}_2 \ldots |\underline{b}_1\underline{b}_2\ldots }\equiv \eta^{\underline{a}_1 \underline{b}_1} \eta^{\underline{a}_2 \underline{b}_2} \ldots $.
\par
We now construct a subalgebra using Cartan involution invariant combinations of the  generators. At level zero the combinations
$$
J_{\underline{a}_1 \underline{a}_2} = \eta_{\underline{a}_1 \underline{c}} K^{\underline{c}}{}_{\underline{a}_2} - \eta_{\underline{a}_2 \underline{c}} K^{\underline{c}}{}_{\underline{a}_1} \ , S_{\underline{a}_1 \underline{a}_2} = R^{\underline{c_1} \underline{c}_2} \eta_{\underline{c}_1 \underline{a}_1} \eta_{\underline{c}_2 \underline{a}_2} - R_{\underline{a}_1 \underline{a}_2} \ , \eqno(2.10) $$
are invariant under $I_c$. Similarly at level one the combinations
$$
S_{\underline{a}} = R^{\underline{c}} \eta_{\underline{c} \underline{a}} - R_{\underline{a}}  \ , \ S_{\underline{a}_1 \underline{a}_2 \underline{a}_3} = R^{\underline{c}_1 \underline{c}_2 \underline{c}_3} \eta_{\underline{c}_1 \underline{c}_2 \underline{c}_3 | \underline{a}_1  \underline{a}_2\underline{a}_3} - R_{\underline{a}_1 \underline{a}_2 \underline{a}_3} \ ,$$
$$
S_{\underline{a}_1 \ldots \underline{a}_5} = R^{\underline{c}_1 \ldots \underline{c}_5} \eta_{\underline{c}_1 \ldots \underline{c}_5 | \underline{a}_1  \ldots \underline{a}_5} + R_{\underline{a}_1 \ldots \underline{a}_5}  ,
S_{\underline{a}_1 \ldots \underline{a}_7} = R^{\underline{c}_1 \ldots \underline{c}_7} \eta_{\underline{c}_1 \ldots \underline{c}_7 | \underline{a}_1  \ldots \underline{a}_7}  - R_{\underline{a}_1 \ldots \underline{a}_7} \ ,$$
$$
S_{\underline{a}_1 \ldots \underline{a}_9} = R^{\underline{c}_1 \ldots \underline{c}_9} \eta_{\underline{c}_1 \ldots \underline{c}_9 | \underline{a}_1  \ldots \underline{a}_9}  + R_{\underline{a}_1 \ldots \underline{a}_9} \ ,
\eqno(2.11)$$
are also invariant. At level two the combinations
$$
S_{\underline{a}_1 \ldots \underline{a}_6} = R^{\underline{c}_1 \ldots \underline{c}_6} \eta_{\underline{c}_1 \ldots \underline{c}_6 | \underline{a}_1  \ldots \underline{a}_6} + R_{\underline{a}_1 \ldots \underline{a}_6}  \ , \ S_{\underline{a}_1 \ldots \underline{a}_8} = R^{\underline{c}_1 \ldots \underline{c}_8} \eta_{\underline{c}_1 \ldots \underline{c}_8 | \underline{a}_1  \ldots \underline{a}_8}  - R_{\underline{a}_1 \ldots \underline{a}_8} \ ,
$$
$$
S_{\underline{a}_1 \ldots \underline{a}_7,\underline{b}} = R^{\underline{c}_1 \ldots \underline{c}_7,\underline{d}} \eta_{\underline{c}_1 \ldots \underline{c}_7, \underline{d} | \underline{a}_1  \ldots \underline{a}_7,\underline{b}}  - R_{\underline{a}_1 \ldots \underline{a}_7,\underline{b}} \ ,
$$
$$
S_{(1)}{}_{\underline{a}_1 \ldots \underline{a}_{10}} = R_{(1)}^{\underline{c}_1 \ldots \underline{c}_{10}} \eta_{\underline{c}_1 \ldots \underline{c}_{10} | \underline{a}_1  \ldots \underline{a}_{10}}  + R_{(1)}{}_{\underline{a}_1 \ldots \underline{a}_{10}} \ , 
$$
$$
S_{(2)}{}_{\underline{a}_1 \ldots \underline{a}_{10}} = R_{(2)}^{\underline{c}_1 \ldots \underline{c}_{10}} \eta_{\underline{c}_1 \ldots \underline{c}_{10} | \underline{a}_1  \ldots \underline{a}_{10}}  + R_{(2)}{}_{\underline{a}_1 \ldots \underline{a}_{10}} \ , $$
$$
S_{\underline{a}_1 \ldots \underline{a}_{9},\underline{b}} = R^{\underline{c}_1 \ldots \underline{c}_{9},\underline{d}} \eta_{\underline{c}_1 \ldots \underline{c}_9, \underline{d} | \underline{a}_1  \ldots \underline{a}_9,\underline{b}} + R_{\underline{a}_1 \ldots \underline{a}_{9},\underline{b}} \ , $$
$$
S_{\underline{a}_1 \ldots \underline{a}_{8},\underline{b}_1 \underline{b}_2} = R^{\underline{c}_1 \ldots \underline{c}_{8},\underline{d}_1 \underline{d}_2} \eta_{\underline{c}_1 \ldots  \underline{d}_2 | \underline{a}_1  \ldots  \underline{b}_2}  + R_{{\underline{a}}_1 \ldots {\underline{a}}_{8},{\underline{b}}_1 {\underline{b}}_2} \ , $$
$$
S_{\underline{a}_1 \ldots \underline{a}_{10},\underline{b}_1 \underline{b}_2} = R^{\underline{c}_1 \ldots \underline{c}_{10},\underline{d}_1 \underline{d}_2} \eta_{\underline{c}_1 \ldots \underline{d}_2 | \underline{a}_1  \ldots \underline{b}_2}  - R_{\underline{a}_1 \ldots \underline{a}_{10},\underline{b}_1 \underline{b}_2} \ ,$$
$$
S_{\underline{a}_1 \ldots \underline{a}_{9},\underline{b}_1 \underline{b}_2 \underline{b}_3} = R^{\underline{c}_1 \ldots \underline{c}_{9},\underline{d}_1 \underline{d}_2 \underline{d}_3} \eta_{\underline{c}_1 \ldots \underline{d}_3 | \underline{a}_1  \ldots \underline{b}_3} - R_{\underline{a}_1 \ldots \underline{a}_{9},\underline{b}_1 \underline{b}_2 \underline{b}_3} \ ,  $$
$$
S_{\underline{a}_1 \ldots \underline{a}_{10},\underline{b}_1 \ldots \underline{b}_4} = R^{\underline{c}_1 \ldots \underline{c}_{10},\underline{d}_1 \ldots \underline{d}_4} \eta_{\underline{c}_1 \ldots \underline{d}_4 | \underline{a}_1  \ldots \underline{b}_4} - R_{\underline{a}_1 \ldots \underline{a}_{10},\underline{b}_1 \ldots \underline{b}_4} \ . 
 \eqno(2.12) $$
are invariant under $I_c$. 
\par
The commutators of  the above  generators, which are in  the IIA formulation,  can be found by the generators  in terms of their eleven dimensional formulation using  equations (2.2) to (2.6) and the known commutation relations for these latter generators.   At level zero we find that
$$
[J^{\underline{a}_1 \underline{a}_2},J_{\underline{b}_1 \underline{b}_2}] = - 4\delta^{[\underline{a}_1}{}_{[\underline{b}_1} J^{\underline{a}_2]}{}_{\underline{b}_2]} \ , \ [J^{\underline{a}_1 \underline{a}_2},S_{\underline{b}_1 \underline{b}_2}] = - 4 \delta^{[\underline{a}_1}{}_{[\underline{b}_1} S^{\underline{a}_2]}{}_{\underline{b}_2]} \ , 
$$
$$
[S^{\underline{a}_1 \underline{a}_2},S_{\underline{b}_1 \underline{b}_2}] = - 4\delta^{[\underline{a}_1}{}_{[\underline{b}_1} J^{\underline{a}_2]}{}_{\underline{b}_2]} \ ,
\eqno(2.13)$$
The commutators of the $I_c(E_{11})$ generators formed from !IA  level plus and minus one $E_{11}$  generators are  given by
$$
[S^{\underline{a}},S_{\underline{b}}] = - J^{\underline{a}}{}_{\underline{b}} \ , \ [S^{\underline{a}},S_{\underline{b}_1 \underline{b}_2 \underline{b}_3}] = - {3 \over 2} \delta^{\underline{a}}_{[\underline{b}_1} S_{\underline{b}_2 \underline{b}_3]} \ , \ [S^{\underline{a}},S_{\underline{b}_1 \ldots \underline{b}_5}] = - 5 \delta^{\underline{a}}_{[\underline{b}_1} S_{\underline{b}_2 \ldots \underline{b}_5]} \ ,$$
$$
[S^{\underline{a}},S_{\underline{b}_1 \ldots \underline{b}_7}] = - 2 S^{\underline{a}}{}_{\underline{b}_1 \ldots \underline{b}_7} + 7 S^{\underline{a}}{}_{[\underline{b}_1 \ldots \underline{b}_6,\underline{b}_7]}  \ , $$
$$
[S^{\underline{a}_1 \underline{a}_2 \underline{a}_3},S_{\underline{b}_1 \underline{b}_2 \underline{b}_3}] = - 18 J^{[\underline{a}_1}{}_{[\underline{b}_1} \delta^{\underline{a}_2 \underline{a}_3}_{\underline{b}_2 \underline{b}_3]} + 2 S^{\underline{a}_1 \underline{a}_2 \underline{a}_3}{}_{\underline{b}_1 \underline{b}_2 \underline{b}_3} \ , $$
$$
[S^{\underline{a}_1 \underline{a}_2 \underline{a}_3},S_{\underline{b}_1 \ldots \underline{b}_5}] = - 30 S_{[\underline{b}_1 \underline{b}_2} \delta_{\underline{b}_3 \underline{b}_4 \underline{b}_5]}^{\underline{a}_1 \underline{a}_2 \underline{a}_3} + S^{\underline{a}_1 \underline{a}_2 \underline{a}_3}{}_{\underline{b}_1 \ldots \underline{b}_5} - 5 S^{\underline{a}_1 \underline{a}_2 \underline{a}_3}{}_{[\underline{b}_1 \ldots \underline{b}_4,\underline{b}_5]} \ , $$
$$
[S^{\underline{a}_1 \underline{a}_2 \underline{a}_3},S_{\underline{b}_1 \ldots \underline{b}_7}] = 0 \ , \ [S^{\underline{a}_1 \ldots \underline{a}_5},S_{\underline{b}_1 \ldots \underline{b}_5}] = - 5 \cdot 30 J^{[\underline{a}_1}{}_{[\underline{b}_1} \delta^{\underline{a}_2 \ldots \underline{a}_5]}_{\underline{b}_2 \ldots \underline{b}_5]} \ ,$$
$$
[S^{\underline{a}_1 \ldots \underline{a}_5},S_{\underline{b}_1 \ldots \underline{b}_7}] = 9 \cdot 70 S_{[\underline{b}_1 \underline{b}_2} \delta_{\underline{b}_3 .. \underline{b}_7]}^{\underline{a}_1 .. \underline{a}_5} \ , \ [S^{\underline{a}_1 \ldots \underline{a}_7},S_{\underline{b}_1 \ldots \underline{b}_7}] = - 7 \cdot 7 \cdot 180 J^{[\underline{a}_1}{}_{[\underline{b}_1} \delta^{\underline{a}_2 \ldots \underline{a}_7]}_{\underline{b}_2 \ldots \underline{b}_7]} , \dots 
\eqno(2.14) $$
and at level zero $I_c(E_{11})$ with the $I_c(E_{11})$ generators formed from !IA  level plus and minus one $E_{11}$  generators  as
$$
[J_{\underline{a}_1 \underline{a}_2},S_{\underline{b}}] = - 2 \eta_{\underline{b} [\underline{a}_1} S_{\underline{a}_2]}  , \ [J^{\underline{a}_1 \underline{a}_2},S_{\underline{b}_1 \underline{b}_2 \underline{b}_3}] = - 2 \cdot 3 \delta^{[\underline{a}_1}{}_{[\underline{b}_1} S^{\underline{a}_2]}{}_{\underline{b}_2 \underline{b}_3]} \ ,
$$
$$
[J^{\underline{a}_1 \underline{a}_2},S_{\underline{b}_1 \ldots \underline{b}_5}] = - 2 \cdot 5 \delta^{[\underline{a}_1}_{[\underline{b}_1} S^{\underline{a}_2]}{}_{\underline{b}_2 \ldots \underline{b}_5]} \ , \ [J^{\underline{a}_1 \underline{a}_2},S_{\underline{b}_1 .. \underline{b}_7}] = 16 S_{[\underline{b}_1 .. \underline{b}_6}{}^{[\underline{a}_1} \delta^{\underline{a}_2]}{}_{\underline{b}_7]} \ , $$
$$
[S_{\underline{a}_1 \underline{a}_2},S_{\underline{b}}] = S_{\underline{a}_1 \underline{a}_2 \underline{b}} \ , \ [S^{\underline{a}_1 \underline{a}_2},S_{\underline{b}_1 \underline{b}_2 \underline{b}_3}] = + 6 \delta^{\underline{a}_1 \underline{a}_2}_{[\underline{b}_1 \underline{b}_2} S_{\underline{b}_3]} - 2 S^{\underline{a}_1 \underline{a}_2}{}_{\underline{b}_1 \underline{b}_2 \underline{b}_3} \ , $$
$$
[S^{\underline{a}_1 \underline{a}_2},S_{\underline{b}_1 \ldots \underline{b}_5}] = + 10 \delta^{\underline{a}_1 \underline{a}_2}_{[\underline{b}_1 \underline{b}_2} S_{\underline{b}_3 \underline{b}_4 \underline{b}_5]} - S^{\underline{a}_1 \underline{a}_2}{}_{\underline{b}_1 \ldots \underline{b}_5} \ , \ $$
$$
[S^{\underline{a}_1 \underline{a}_2},S_{\underline{b}_1 \ldots \underline{b}_7}] = 42 S_{[\underline{b}_1 \ldots \underline{b}_5} \delta_{\underline{b}_6 \underline{b}_7]}^{\underline{a}_1 \underline{a}_2} \ , \ldots 
 \eqno(2.15) $$
\par
We now consider how  the vector ($l_1$) representation of $E_{11}$ appears in the   IIA theory. These generators  can be related to the way the $l_1$ generators appear in the eleven dimensional theory  which we will again denoted by a hat.i.e.  $\hat{}{}$. The generators in the $l_1$ representation in its IIA presentation at  IIA level zero  are [17]
$$
P_{\underline{a}} = \hat{P}_{\underline{a}} \ \ , \ \ Q^{\underline{a}} = - \hat{Z}^{\underline{a} 11} \ \ .
 \eqno(2.16) $$
We note that the coefficient in $Q^{\underline{a}} = - Z^{\underline{a} 11}$ is different to that in the  conventions in reference [17], see  below (A.15) of that paper
\par
The generators in the $l_1$ representation at IIA  level one are given by
$$
Z = \hat{P}_{11} \ , \ Z^{\underline{a}_1 \underline{a}_2} = \hat{Z}^{\underline{a}_1 \underline{a}_2} \ , \ Z^{\underline{a}_1 \ldots \underline{a}_4} = \hat{Z}^{\underline{a}_1 \ldots \underline{a}_4 11} \ , \ Z^{\underline{a}_1 \ldots \underline{a}_6}  = \hat{Z}^{\underline{a}_1 \ldots \underline{a}_6 11,11}  \ ,    $$
$$
Z^{\underline{a}_1 \ldots \underline{a}_8}  = \hat{Z}^{\underline{a}_1 \ldots \underline{a}_8}  \ , \ Z^{\underline{a}_1 \ldots \underline{a}_{10}}  = \hat{Z}^{\underline{a}_1 \ldots \underline{a}_{10} 11}  \ ,
\eqno(2.17)$$
and at  IIA level two  as
$$
Z^{\underline{a}_1 \ldots \underline{a}_5} = \hat{Z}^{\underline{a}_1 \ldots \underline{a}_5} \ , \ Z_{(1)}^{\underline{a}_1 \ldots \underline{a}_7}  = \hat{Z}^{\underline{a}_1 \ldots \underline{a}_7,11} \ , \ Z_{(2)}^{\underline{a}_1 \ldots \underline{a}_7}  = \hat{Z}^{\underline{a}_1 \ldots \underline{a}_7 11} \ , \
Z^{\underline{a}_1 \ldots \underline{a}_6,\underline{b}}  = \hat{Z}^{\underline{a}_1 \ldots \underline{a}_6 11,\underline{b}} \ ,
$$
$$
\ Z^{\underline{a}_1 \ldots \underline{a}_7,\underline{b}_1 \underline{b}_2}  = \hat{Z}^{\underline{a}_1 \ldots \underline{a}_8 11,\underline{b}_1 \underline{b}_2 11} \ , \ Z_{(1)}^{\underline{a}_1 \ldots \underline{a}_9}  = \hat{Z}_{(1)}^{\underline{a}_1 \ldots \underline{a}_9 11,11} \ , \ Z_{(2)}^{\underline{a}_1 \ldots \underline{a}_9}  = \hat{Z}_{(2)}^{\underline{a}_1 \ldots \underline{a}_9 11,11} \ ,
$$
$$
Z_{(3)}^{\underline{a}_1 \ldots \underline{a}_9} = \hat{Z}^{\underline{a}_1 \ldots \underline{a}_9, (11 \, 11)} \ , \ Z_{(1)}^{\underline{a}_1 \ldots \underline{a}_8,\underline{b}}  = \hat{Z}{}^{\underline{a}_1 \ldots \underline{a}_8 11,\underline{b} 11} \ , \ Z_{(2)}^{\underline{a}_1 \ldots \underline{a}_8,\underline{b}}  = \hat{Z}{}^{\underline{a}_1 \ldots \underline{a}_8 11,(\underline{b} 11)} \ ,
$$
$$
Z^{\underline{a}_1 \ldots \underline{a}_7, \underline{b}_1 \underline{b}_2} = \hat{Z}^{\underline{a}_1 \ldots \underline{a}_7 11, \underline{b}_1 \underline{b}_2 11} \ \ , \ \ Z_{(1)}^{\underline{a}_1 \ldots \underline{a}_{10}, \underline{b}} = \hat{Z}_{(1)}^{\underline{a}_1 \ldots \underline{a}_{10} 11,\underline{b} \, 11,11}  \ \ , 
$$
$$
Z_{(2)}^{\underline{a}_1 \ldots \underline{a}_{10}, \underline{b}} = \hat{Z}_{(2)}^{\underline{a}_1 \ldots \underline{a}_{10}11,\underline{b} \, 11,11} \ \ , \ \ Z_{(3)}^{\underline{a}_1 \ldots \underline{a}_{10}, \underline{b}} = \hat{Z}^{\underline{a}_1 \ldots \underline{a}_{10} 11, (\underline{b} \, 11 , 11)} \ \ ,$$
$$
Z_{(1)}^{\underline{a}_1 \ldots \underline{a}_9, \underline{b}_1 \underline{b}_2} = \hat{Z}_{(1)}^{\underline{a}_1 \ldots \underline{a}_9 11, \underline{b}_1 \underline{b}_2 \, 11,11} \ \ , \ \ Z_{(2)}^{\underline{a}_1 \ldots \underline{a}_9, \underline{b}_1 \underline{b}_2} = \hat{Z}_{(2)}^{\underline{a}_1 \ldots \underline{a}_9 11, \underline{b}_1 \underline{b}_2 \, 11,11} \ \ ,
$$
$$
Z^{\underline{a}_1 \ldots \underline{a}_8, \underline{b}_1 \underline{b}_2 \underline{b}_3} = \hat{Z}^{\underline{a}_1 \ldots \underline{a}_8 11, \underline{b}_1 \underline{b}_2 \underline{b}_3 \, 11, 11} \ , \ Z^{\underline{a}_1 \ldots \underline{a}_9, \underline{b}_1 \ldots \underline{b}_4} = \hat{Z}^{\underline{a}_1 \ldots \underline{a}_9 11, \underline{b}_1 \ldots \underline{b}_4 \, 11 , (11,11)} \ , $$
$$
Z_{(1)}^{\underline{a}_1 \ldots \underline{a}_{10}, \underline{b}_1 \underline{b}_2 \underline{b}_3} = \hat{Z}_{(1)}^{\underline{a}_1 \ldots \underline{a}_{10} 11, \underline{b}_1 \underline{b}_2 \underline{b}_3 \, 11, (11,11)} \ , \ Z_{(2)}^{\underline{a}_1 \ldots \underline{a}_{10}, \underline{b}_1 \underline{b}_2 \underline{b}_3} = \hat{Z}_{(2)}^{\underline{a}_1 \ldots \underline{a}_{10} 11, \underline{b}_1 \underline{b}_2 \underline{b}_3 \, 11,(11,11)} \ .
\eqno(2.18)$$
All of the indices in a given block are anti-symmetric. The lowered subscripts 1, 2  and  3 indicate  generators that  occur with multiplicity more than one. In some cases this is because there are two ways to find a generators with the same indices from the eleven dimensional generators and in the other cases the generator in eleven dimensions has multiplicity two.
\par
The commutators of the $l_1$ generators with the $I_c(E_{11})$ are most easily computed from their eleven dimensional origin given  in the above equations. The commutators  of the  IIA  level zero $E_{11}$  with IIA level zero $l_1$ generators are given as 
$$
[J_{\underline{a}_1 \underline{a}_2},P_{\underline{c}}] = - 2 \eta_{\underline{c}[\underline{a}_1} P_{\underline{a}_2]} \ , \ [J^{\underline{a}_1 \underline{a}_2},Q^{\underline{c}}] = - 2 \eta^{\underline{c} [\underline{a}_1} Q^{\underline{a}_2]} \ ,
$$
$$
[S^{\underline{a}_1 \underline{a}_2},P_{\underline{c}}] = - 2 \delta_c^{[\underline{a}_1} Q^{\underline{a}_2]} \ , \ [S_{\underline{a}_1 \underline{a}_2},Q^{\underline{c}}] = - 2 \delta^{\underline{c}}_{[\underline{a}_1} P_{\underline{a}_2]} \ ,
\eqno(2.19)$$
The commutators of the level zero $I_c(E_{11})$  generators with level one $l_1$ generators are
$$
[J_{{\underline{a}}_1 {\underline{a}}_2},Z] = 0 \ , \ [J_{{\underline{a}}_1 \underline{a}_2},Z^{\underline{b}_1 \underline{b}_2}] = -2 \cdot 2 \delta^{[\underline{b}_1}_{[\underline{a}_1} Z_{\underline{a}_2]}{}^{\underline{b}_2]} \ ,
$$
$$
[J_{\underline{a}_1 \underline{a}_2},Z^{\underline{b}_1 \ldots \underline{b}_4}] = -4 \cdot 2 \delta^{[\underline{b}_1}_{[\underline{a}_1} Z_{\underline{a}_2]}{} ^{\underline{b}_2 \underline{b}_3 \underline{b}_4]}{}  \ , [J^{\underline{a}_1 \underline{a}_2},Z^{\underline{b}_1 \ldots \underline{b}_6}] =- 6 \cdot 2  \delta^{[\underline{b}_1}_{[\underline{a}_1}  Z_{\underline{a}_2]}{} ^{\underline{b}_2 \ldots \underline{b}_6]},
$$
$$
[S^{\underline{a}_1 \underline{a}_2},Z] = Z^{\underline{a}_1 \underline{a}_2} \ , \ [S^{\underline{a}_1 \underline{a}_2},Z_{\underline{b}_1 \underline{b}_2}] = Z^{\underline{a}_1 \underline{a}_2}{}_{\underline{b}_1 \underline{b}_2} - 2 \delta^{a_1 a_2}_{\underline{b}_1 \underline{b}_2} Z \ ,
$$
$$
[S_{\underline{a}_1 \underline{a}_2},Z^{\underline{b}_1 \ldots \underline{b}_4}] = - 12 Z^{[\underline{b}_1 \underline{b}_2} \delta^{\underline{b}_3 \underline{b}_4]}_{\underline{a}_1 \underline{a}_2}  + {1 \over 3} Z^{\underline{b}_1 \ldots \underline{b}_4 \underline{a}_1 \underline{a}_2} \ ,
$$
$$
[S_{\underline{a}_1 \underline{a}_2},Z^{\underline{b}_1 \ldots \underline{b}_6}] = - 2 \cdot 45 Z^{[\underline{b}_1 \ldots \underline{b}_4} \delta^{\underline{b}_5 \underline{b}_6]}_{\underline{a}_1 \underline{a}_2} \ , \ldots
\eqno(2.20)$$
The level plus and minus one $E_{11}$ generators  with the level zero $l_1$ generators have the commutators
$$
\ [S_{\underline{a}},P_{\underline{c}}] = - \eta_{\underline{a} \underline{c}} Z , \ [S^{\underline{a}_1 \underline{a}_2 \underline{a}_3},P_{\underline{b}}] = 3 \delta^{[\underline{a}_1}_{\underline{b}} Z^{\underline{a}_2 \underline{a}_3]} , \ [S^{\underline{a}_1 \ldots \underline{a}_5},P_{\underline{b}}] = - {5 \over 2} \delta^{[\underline{a}_1}_{\underline{b}} Z^{\underline{a}_2 .. \underline{a}_5]} \ ,
$$
$$
[S^{\underline{a}_1 \ldots \underline{a}_7},P_{\underline{b}}] = {7 \over 6} \delta^{[\underline{a}_1}_{\underline{b}} Z^{\underline{a}_2 \ldots \underline{a}_7]}  \ , \ [S^{\underline{a}},Q^{\underline{b}}] = Z^{\underline{a} \underline{b}} \ , \ [S^{\underline{a}_1 \underline{a}_2 \underline{a}_3},Q^{\underline{b}}] = - Z^{\underline{a}_1 \underline{a}_2 \underline{a}_3 \underline{b}} \ ,
$$
$$
[S^{\underline{a}_1 \ldots \underline{a}_5},Q^{\underline{b}}] = {1 \over 6} Z^{\underline{a}_1 \ldots \underline{a}_5 \underline{b}} \ , \ [S^{\underline{a}_1 \ldots \underline{a}_7},Q^{\underline{b}}] = 0 \  , \ldots
\eqno(2.21)$$
The commutators of the $I_c(E_{11})$ generators formed from !IA  level plus and minus one $E_{11}$  generators 
with the  level one $l_1$ generators  are given by
$$
[S_{\underline{a}},Z] = P_{\underline{a}} \ , \  [S_{\underline{a}},Z^{\underline{b}_1 \underline{b}_2}] = - 2 \delta_{\underline{c}}^{[\underline{b}_1} Q^{\underline{b}_2]} \ , \  [S^{\underline{a}},Z^{\underline{b}_1 \ldots \underline{b}_4}] = Z^{\underline{a} \underline{b}_1 \ldots \underline{b}_4} \ , $$
$$
[S^{\underline{a}},Z^{\underline{b}_1 \ldots \underline{b}_6}] = Z_{(1)}^{\underline{a} \underline{b}_1 \ldots \underline{b}_6} + Z^{\underline{b}_1 \ldots \underline{b}_6,\underline{a}} \ , \  [S_{\underline{a}_1 \underline{a}_2 \underline{a}_3},Z] = 0 \ ,  $$
$$
[S_{\underline{a}_1 \underline{a}_2 \underline{a}_3},Z^{\underline{b}_1 \underline{b}_2}] = - 6 \delta_{[\underline{a}_1 \underline{a}_2}^{\underline{b}_1 \underline{b}_2} P_{\underline{a}_3]} + Z_{\underline{a}_1 \underline{a}_2 \underline{a}_3}{}^{\underline{b}_1 \underline{b}_2} \ ,
$$
$$
[S_{\underline{a}_1 \underline{a}_2 \underline{a}_3},Z^{\underline{b}_1 \ldots \underline{b}_4}] = 4! \delta^{[\underline{b}_1 \underline{b}_2 \underline{b}_3}_{\underline{a}_1 \underline{a}_2 \underline{a}_3} Q^{\underline{b}_4]} - Z_{(2)}{}_{\underline{a}_1 \underline{a}_2 \underline{a}_3}{}^{\underline{b}_1 \ldots \underline{b}_4} + Z^{\underline{b}_1 \ldots \underline{b}_4}{}_{[\underline{a}_1 \underline{a}_2 ,\underline{a}_3]} \ ,
$$
$$
[S_{\underline{a}_1 \underline{a}_2 \underline{a}_3},Z^{\underline{b}_1 \ldots \underline{b}_6}] = 0 \ , \ [S^{\underline{a}_1 \ldots \underline{a}_5},Z] = {1 \over 2} Z^{\underline{a}_1 \ldots \underline{a}_5} \ , 
$$
$$
[S^{\underline{a}_1 \ldots \underline{a}_5},Z^{\underline{b}_1 \underline{b}_2}] = - Z_{(2)}^{\underline{a}_1 \ldots \underline{a}_5 \underline{b}_1 \underline{b}_2} - {1 \over 3} Z^{\underline{a}_1 \ldots \underline{a}_5[\underline{b}_1 , \underline{b}_2]} \ ,
$$
$$
[S_{\underline{a}_1 \ldots \underline{a}_5},Z^{\underline{b}_1 \ldots \underline{b}_4}] = 60 P_{[\underline{a}_1} \delta_{\underline{a}_2 \ldots \underline{a}_5]}^{\underline{b}_1 \ldots \underline{b}_4} \ , \ [S_{\underline{a}_1 \ldots \underline{a}_5},Z^{\underline{b}_1 \ldots \underline{b}_6}] = {6! \over 2} Q^{[\underline{b}_1} \delta^{\underline{b}_2 \ldots \underline{b}_6]}_{\underline{a}_1 \ldots \underline{a}_5} \ , 
$$
$$
[S^{\underline{a}_1 \ldots \underline{a}_7},Z] = - {1 \over 6} Z_{(1)}^{\underline{a}_1 \ldots \underline{a}_7} - {3 \over 2} Z_{(2)}^{\underline{a}_1 \ldots \underline{a}_7}   \ , \ [S^{\underline{a}_1 \ldots \underline{a}_7},Z^{\underline{b}_1 \underline{b}_2}] = 0 \ ,
$$
$$
[S^{\underline{a}_1 \ldots \underline{a}_7},Z^{\underline{b}_1 \ldots \underline{b}_4}] = 0 \ , \ [S^{\underline{a}_1 \ldots \underline{a}_7},Z^{\underline{b}_1 \ldots \underline{b}_6}] = {3 \cdot 7! \over 2} P_{[\underline{a}_1} \delta_{\underline{a}_2 \ldots \underline{a}_7]}^{\underline{b}_1 \ldots \underline{b}_6} \ , \ldots
 check\eqno(2.22)$$
\par
The general  element of $I_c(E_{11})$, can be written as
$$
\Lambda^{\underline{\alpha}} S_{\underline{\alpha}} = \Lambda^{\underline{a}_1 \underline{a}_2} J_{\underline{a}_1 \underline{a}_2} + \tilde{\Lambda}^{\underline{a}_1 \underline{a}_2} S_{\underline{a}_1 \underline{a}_2} + \Lambda^{\underline{a}} S_{\underline{a}} + \Lambda^{\underline{a}_1 \underline{a}_2 \underline{a}_3} S_{\underline{a}_1 \underline{a}_2 \underline{a}_3} + \Lambda^{\underline{a}_1 \ldots \underline{a}_5} S_{\underline{a}_1 \ldots \underline{a}_5}  $$
$$
+ \Lambda^{\underline{a}_1 \ldots \underline{a}_6} S_{\underline{a}_1 \ldots \underline{a}_6}  + \Lambda^{\underline{a}_1 \ldots \underline{a}_7} S_{\underline{a}_1 \ldots \underline{a}_7} + \Lambda^{\underline{a}_1 \ldots \underline{a}_7,\underline{b}} S_{\underline{a}_1 \ldots \underline{a}_7,\underline{b}} ,\
+ \Lambda^{\underline{a}_1 \ldots \underline{a}_8} S_{\underline{a}_1 \ldots \underline{a}_8}  + \ldots
\eqno(2.23) $$
The transformation of the vector representation under $I_c(E_{11})$ transformations  is given by
$$
\delta l_A = [\Lambda^{\underline{\alpha}} S_{\underline{\alpha}},l_A] .
\eqno(2.24)$$
Evaluating equation (2.24) on the $l_1$ generators we find, at IIA level zero:
$$
\delta P_{\underline{b}} = - 2 \Lambda_{\underline{b}}{}^{\underline{e}} P_{\underline{e}} - 2 \tilde{\Lambda}_{\underline{b} \underline{e}} Q^{\underline{e}} - \Lambda_{\underline{b}} Z + 3 \Lambda_{\underline{b} \underline{e}_1 \underline{e}_2} Z^{\underline{e}_1 \underline{e}_2} - {5 \over 2} \Lambda_{\underline{b} \underline{e}_1 \ldots \underline{e}_4} Z^{\underline{e}_1 \ldots \underline{e}_4} + \ldots
$$
$$
\delta Q^{\underline{b}} = - 2 \Lambda^{\underline{b}}{}_{\underline{e}} Q^{\underline{e}} - 2 \tilde{\Lambda}^{\underline{b} \underline{e}} P_{\underline{e}} + \Lambda_{\underline{e}} Z^{\underline{e} \underline{b}} - \Lambda_{\underline{e}_1 \underline{e}_2 \underline{e}_3} Z^{\underline{e}_1 \underline{e}_2 \underline{e}_3 \underline{b}} + {1 \over 6} \Lambda_{\underline{e}_1 \ldots \underline{e}_5} Z^{\underline{e}_1 \ldots \underline{e}_5 \underline{b}} + \ldots \ ;  check\eqno(2.25)$$
and at IIA  level one
$$
\delta Z = \Lambda^{\underline{e}} P_{\underline{e}} + \tilde{\Lambda}_{\underline{e}_1 \underline{e}_2} Z^{\underline{e}_1 \underline{e}_2} + {1 \over 2} \Lambda_{\underline{e}_1 \ldots \underline{e}_5} Z^{\underline{e}_1 \ldots \underline{e}_5} + \ldots
$$
$$
\delta Z^{\underline{b}_1 \underline{b}_2} = - 6 \Lambda^{\underline{e} \underline{b}_1 \ldots \underline{b}_4} P_{\underline{e}} - 2 \Lambda^{[\underline{b}_1} Q^{\underline{b}_2]} - 2 \Lambda^{\underline{b}_1 \underline{b}_2} Z + 4 \Lambda^{[\underline{b}_1}{}_{\underline{e}} Z^{\underline{b}_2] \underline{e}} + \tilde{\Lambda}_{\underline{e}_1 \underline{e}_2} Z^{\underline{e}_1 \underline{e}_2 \underline{b}_1 \underline{b}_2}
$$
$$
+ \Lambda_{\underline{e}_1 \underline{e}_2 \underline{e}_3} Z^{\underline{e}_1 \underline{e}_2 \underline{e}_3 \underline{b}_1 \underline{b}_2} - \Lambda_{\underline{e}_1 \ldots \underline{e}_5} Z_{(2)}^{\underline{e}_1 \ldots \underline{e}_5 \underline{b}_1 \underline{b}_2}
- {1 \over 3} \Lambda_{\underline{e}_1 \ldots \underline{e}_5} Z^{\underline{e}_1 \ldots \underline{e}_5 [\underline{b}_1 , \underline{b}_2]} + \ldots
$$
$$
\delta Z^{\underline{b}_1 \ldots \underline{b}_4} = 60 \Lambda^{\underline{e} \underline{b}_1 \ldots \underline{b}_4} P_{\underline{e}} + 24 \Lambda^{[\underline{b}_1 \underline{b}_2 \underline{b}_3} Q^{\underline{b}_4]} - 8 \Lambda_{\underline{e}}{}^{[\underline{b}_1} Z^{\underline{b}_2 \underline{b}_3 \underline{b}_4] \underline{e}} - 12 \tilde{\Lambda}^{[\underline{b}_1 \underline{b}_2} Z^{\underline{b}_3 \underline{b}_4]} $$
$$
+ {1 \over 3} \Lambda_{\underline{e}_1 \underline{e}_2} Z^{\underline{e}_1 \underline{e}_2 \underline{b}_1 \ldots \underline{b}_4} + \Lambda_{\underline{e}} Z^{\underline{e} \underline{b}_1 \ldots \underline{b}_4} - \Lambda_{\underline{e}_1 \underline{e}_2 \underline{e}_3} Z_{(2)}^{\underline{e}_1 \underline{e}_2 \underline{e}_3 \underline{b}_1 \ldots \underline{b}_4}
+ \Lambda_{\underline{e}_1 \underline{e}_2 \underline{e}_3} Z^{\underline{b}_1 \ldots \underline{b}_4 \underline{e}_1 \underline{e}_2 , \underline{e}_3} + \ldots
$$
$$
\delta Z^{\underline{b}_1 \ldots \underline{b}_6} = 8 \cdot 135 Q^{[\underline{b}_1} \Lambda^{\underline{b}_2 \ldots \underline{b}_6]} - 8 \cdot 7 \cdot 135 \Lambda^{\underline{e} \underline{b}_1 \ldots \underline{b}_6} P_{\underline{e}} - 6 \cdot 15 Z^{[\underline{b}_1 \ldots \underline{b}_4} \tilde{\Lambda}^{\underline{b}_5 \underline{b}_6]} $$
$$
+ \Lambda_{\underline{e}} Z^{\underline{e} \underline{b}_1 \ldots \underline{b}_6} + \Lambda_{\underline{e}} Z^{\underline{b}_1 \ldots \underline{b}_6,\underline{e}}  - 6 \cdot 2 \Lambda_{\underline{e}}{}^{[\underline{b}_1} Z^{\underline{b}_2 \ldots \underline{b}_6] \underline{e}} + \ldots \ \ ;
 \eqno(2.26)$$
where we have only give results up the generators  $Z^{\underline{b}_1 \ldots \underline{b}_6}$.
At level two we find that
$$
\delta Z^{\underline{b}_1 \ldots \underline{b}_5} = {6! \over 2} \Lambda^{\underline{e} \underline{b}_1 \ldots \underline{b}_5} P_{\underline{e}} - 60 \Lambda^{\underline{b}_1 \ldots \underline{b}_5} Z - 60 Z^{[\underline{b}_1 \underline{b}_2} \Lambda^{\underline{b}_3 \underline{b}_4 \underline{b}_5]} - 5 Z^{[\underline{b}_1 \ldots \underline{b}_4} \Lambda^{\underline{b}_5]} $$
$$
+ 10 \Lambda_{\underline{e}}{}^{[\underline{b}_1} Z^{\underline{b}_2 \ldots \underline{b}_5] \underline{e}} + \tilde{\Lambda}_{\underline{e}_1 \underline{e}_2} Z_{(2)}^{\underline{e}_1 \underline{e}_2 \underline{b}_1 \ldots \underline{b}_5} + {1 \over 3} \tilde{\Lambda}_{\underline{e}_1 \underline{e}_2} Z_{(1)}^{\underline{b}_1 \ldots \underline{b}_5 \underline{e}_1 \underline{e}_2}
- {2 \over 3} \tilde{\Lambda}_{\underline{e}_1 \underline{e}_2} Z^{\underline{b}_1 \ldots \underline{b}_5 [\underline{e}_1,\underline{e}_2]} + \ldots
$$
$$
\delta Z_{(1)}^{\underline{b}_1 \ldots \underline{b}_7} = 6 \cdot 7 \cdot 135 Q^{[\underline{b}_1} \Lambda^{\underline{b}_2 \ldots \underline{b}_7]} - 8 \cdot 7 \cdot 7 \cdot 135 \Lambda^{\underline{e} \underline{b}_1 \ldots \underline{b}_7} P_{\underline{e}} - 7 \cdot 7 \cdot 135 \Lambda^{\underline{e} [ \underline{b}_1 \ldots \underline{b}_6,\underline{b}_7]} P_{\underline{e}} 
$$
$$
+ 7 \cdot 135 Z^{[\underline{b}_1 \underline{b}_2} \Lambda^{\underline{b}_3 \ldots \underline{b}_7]} + 7 \cdot 7 \cdot 135 Z^{\underline{b}_1 \ldots \underline{b}_7} Z + \ldots
$$
$$
\delta Z_{(2)}^{\underline{b}_1 \ldots \underline{b}_7} = - 2 \cdot 3 \cdot 7 \cdot 15 Q^{[\underline{b}_1} \Lambda^{\underline{b}_2 \ldots \underline{b}_7]} - 8 \cdot 7 \cdot 15 \Lambda^{\underline{e} \underline{b}_1 \ldots \underline{b}_7} P_{\underline{e}} + 7 \cdot 7 \cdot 15 \Lambda^{\underline{e} [\underline{b}_1 \ldots \underline{b}_6,\underline{b}_7]} P_{\underline{e}}
$$
$$
+ 8 \cdot 7 \cdot 15 \Lambda^{\underline{b}_1 \ldots \underline{b}_7,\underline{e}} P_{\underline{e}} + 3 \cdot 7 \cdot 15 Z^{[\underline{b}_1 \underline{b}_2} \Lambda^{\underline{b}_3 \ldots \underline{b}_7]}
+ 9 \cdot 7 \cdot 15 \Lambda^{\underline{b}_1 \ldots \underline{b}_7} Z  + \ldots
$$
$$
\delta Z^{\underline{b}_1 \ldots \underline{b}_6,\underline{c}} = 135 \cdot 12 Q^{[\underline{b}_1} \Lambda^{\underline{b}_2 \ldots \underline{b}_6] \underline{c}} - 135 \cdot 6 Q^{\underline{c}} \Lambda^{\underline{b}_1 \ldots \underline{b}_6} - 135 \cdot 8 \cdot 7 \Lambda^{\underline{e} \underline{b}_1 \ldots \underline{b}_6 \underline{c}} P_{\underline{e}}
$$
$$
+ 135 \cdot 6 \cdot 7 \Lambda^{\underline{e} \underline{c} [\underline{b}_1 \ldots \underline{b}_5,\underline{b}_6]} P_{\underline{e}} - 135 \cdot 7 \cdot 7 \Lambda^{\underline{e}  \underline{b}_1 \ldots \underline{b}_6 , \underline{c}} P_{\underline{e}}
$$
$$
+ 135 \cdot 7 \Lambda^{\underline{b}_1 \ldots \underline{b}_6 \underline{c}} Z - 135 \cdot 5 Z^{[\underline{b}_1 \underline{b}_2} \Lambda^{\underline{b}_3 \ldots \underline{b}_6] \underline{c}}
- 135 \cdot 6 Z^{[\underline{b}_1 |\underline{c}|} \Lambda^{\underline{b}_2 \ldots \underline{b}_6] \underline{c}} + \ldots \eqno(2.27)$$
where we have only computed up to and including transformations of  the generator $Z^{\underline{b}_1 \ldots \underline{b}_6,\underline{c}}$.
\par
The tangent space algebra is  $I_c(E_{11})$. In eleven dimensions a bilinear  $I_c(E_{11})$ invariant was found to be  [18]
$$
L_{E_{11}}^2 = L_A L_B K^{AB} = P_{\tilde{a}} P^{\tilde{a}} + {1 \over 2} Z_{\tilde{a}_1 \tilde{a}_2} Z^{\tilde{a}_1 \tilde{a}_2} + {1 \over 5!} Z_{\tilde{a}_1 \ldots \tilde{a}_5} Z^{\tilde{a}_1 \ldots \tilde{a}_5} + \ldots
\eqno(2.28)$$
where $\tilde{a},\tilde{b},\ldots = 0,\ldots,10$ and $K^{AB}$ is the tangent group metric. The corresponding IIA invariant can be found by using equations (2.22) which relates  the IIA vector generators to those in the eleven dimensional theory. The result is given by
$$
L_{IIA}^2 = P_{\underline{a}} P^{\underline{a}} + Q_{\underline{a}} Q^{\underline{a}} + Z Z + {1 \over 2} Z_{\underline{a}_1 \underline{a}_2} Z^{\underline{a}_1 \underline{a}_2} + {1 \over 4!} Z_{\underline{a}_1 \ldots \underline{a}_4} Z^{\underline{a}_1 \ldots \underline{a}_4}
+ {1 \over 5!} Z_{\underline{a}_1 \ldots \underline{a}_5} Z^{\underline{a}_1 \ldots \underline{a}_5} + \ldots \eqno(2.29) $$

\medskip
{{\bf 3. The string little algebra at low levels}}
\medskip
\par
In 1939  Wigner found the  irreducible representations  of the Poincare group ${\rm SO}(1,3) \otimes_s \{{P}_{\mu} \}$, $\mu = 0,\ldots,3$. In this paper we will study the irreducible representations of the semi-direct product of $I_c(E_{11})$ with it's $l_1$ vector representation,  denote by $I_c(E_{11}) \otimes_s l_1$. Given the similarities of Poincare algebra and the algebra $I_c(E_{11}) \otimes_s l_1$, the generators of the vector representation commute,  we can follow a similar path to that taken  by Wigner. In this section we will carry out the first step and find the string little algebra at low levels. We will first  review the method for constructing irreducible representations in E theory [14] adapted to the IIA string. 
\par
In the case of the Poincare algebra, the Wigner method, at least  as practiced by physicists,  begins by choosing specific values for the momenta  $P_{\mu}$,  which are the conserved charges of the translation. We  then find  the subalgebra of 
${\rm SO}(1,3)$ which preserves this choice, that is, we find the  little algebra,  For a  massive particle we can choose the momenta  to take values  $P_{\mu} = (m,0,0,0)$ in which case the little algebra  is SO(3).  We then choose an irreducible representation of ${\rm SO}(1,3)$ which encodes the physical degrees of freedom of the particle. Finally one carries out a boost out of the rest frame to find the full representation of the Poincare group. 
\par
In E theory, the Poincare algebra is replaced by  semi-direct products such as $I_c(E_{11}) \otimes_s l_1$ and the conserved charges $P_a$ by the charges in the vector representation. The lowest level elements of $I_c(E_{11})$ and the vector representation are the Lorentz algebra and the translations respectively and so the Poincare algebra is the lowest level part of 
$I_c(E_{11}) \otimes_s l_1$. However, now the situation is much more complicated, $I_c(E_{11}) $ is an infinite dimensional algebra,  whose properties are largely only understood at low levels, while the vector representation contains an infinite number of brane charges.  Increasing in level the next elements in eleven dimensions,  are the M2, M5 brane charges and the Taub-Nut charge which are followed by an infinite number of new charges. 
\par
In E theory we begin by choosing  values for the  brane charges in the vector representation that are those of interest,   to be  non-zero. Next we find the subalgebra of  $I_c(E_{11}) $,  denoted by ${\cal H}$,   which preserves this choice of brane charges. In keeping with the Poincare algebra case we refer to  the  subalgebra ${\cal H}$  as the little algebra. The next step is to choose an irreducible representation of ${\cal H}$ and this should encode the degrees of freedom of the brane we are studying. 
\par
The irreducible representation of  $I_c(E_{11}) \otimes_s l_1$ corresponding to  the point particle was studied in reference [14]. The non-zero brane charges were choosen to be that of the usual momenta which was taken to be massless. In particular this paper took $P_a= (m, 0\ldots ,0,m$  with all the other brane charges being zero. The resulting little algebra was found to be ${\cal H} = I_c(E_9)$ which is just  affine $E_8$. Despite the infinite nature of this algebra it has a representation that only has a finite number of states and these are precisely the bosonic degrees of freedom of eleven dimensional supergravity [14]. 
\par
In this paper we will apply the same strategy to find the irreducible representation of $I_c(E_{11}) \otimes_s l_1$ that corresponds to the   string. In particular we will study the IIA string.  As explained in section two the IIA theory emerges when we consider the decomposition  that results from deleting node ten in  the $E_{11}$  Dynkin diagram to find the algebra  $D_{10}$. The generators are then arranged according to their level with respect to this node. We  often referred to this level as the IIA level. 
\par
The lowest members of the vector representation, that is the ones that have IIA level zero,   are the usual translations $P_{\underline{a}}$ and the string charge  $Q^{\underline{a}}$,  and they belong to the vector representation of $D_8$. It might be tempting to take these charges to have the values $P_{\underline{a}}= (m,0,\ldots, 0)$ and $Q^{\underline{a}}=(0,m,0,\ldots , 0)$  where $\underline a, \underline b, \ldots =0,1,\ldots , 9$. 
However, unlike for the point particle,  a string has a world volume which has a corresponding symmetry, namely SO(1,1). We will take this symmetry to  be preserved by our choice and so we  take the brane charges  in $l_A$ to obey 
$$
P_a = - \varepsilon_{ab} Q^b \ \ {\rm or \  equivalently} \ \ Q^a = - \varepsilon^{ab} P_b \ \ ,  \ \  l_A = 0 \ \ {\rm otherwise} 
\eqno(3.1)$$
In this equations $a,b,\ldots =0,1$. The choice first mentioned does satisfy these conditions but it is obviously not ${\rm SO}(1,1)$ covariant.
\par
 The adoption of a relation among the brane charges, rather than particular values, is a new and important difference required when we find irreducible representations of $I_c(E_{11}) \otimes_s l_1$ that correspond to branes rather than those of the Poincare algebra which concern the point particle. The  choice of brane charges  of equation (3.1) obeys  $L_{IIA}^2 = 0$ for the $I_c(E_{11})$  invariant  of equation (2.29). For the known brane charges this condition is just the same as the well known half BPS condition that results from the supersymmetry algebra. It also obeys the condition $P_a Q^a = 0$ which means that it obeys the $I_c(E_{11})$ constraints discussed in reference [18] where the relation of these two conditions  to the Casimirs is explained. 
 \par
 The next step is to find the little algebra  ${\cal H}$ that preserves the choice of equation (3.1). To do this we examine the 
  $I_c(E_{11}) $ variation of the brane charges given in equations (2.26) and (2.27)  and look for which of them are non-zero. Keeping only such terms, we find at IIA level zero that 
$$
\delta P_{\underline{b}} = - 2 (\Lambda_{\underline{b}}{}^{e} - \tilde{\Lambda}_{\underline{b} c} \varepsilon^{ce})P_e \ , \  \delta Q^{\underline{b}} = 2 (\Lambda^{\underline{b}}{}_c \varepsilon^{ce} - \tilde{\Lambda}^{\underline{b} e}) P_e \ ; 
\eqno(3.2)$$
at level one:
$$
\delta Z = \Lambda^e P_e \ , \ \delta Z^{\underline{b}_1 \underline{b}_2} = - 6 (\Lambda^{\underline{b}_1 \underline{b}_2 e} - {1 \over 3} \Lambda^{[\underline{b}_1} \delta^{\underline{b}_2]}{}_c \varepsilon^{ce}) P_e \ , 
$$
$$
\delta Z^{\underline{b}_1 \ldots \underline{b}_4} = 12 (5 \Lambda^{e \underline{b}_1 \ldots \underline{b}_4} - 2 \Lambda^{[\underline{b}_1 \underline{b}_2 \underline{b}_3} \delta^{\underline{b}_4]}{}_c \varepsilon^{ce}) P_e \ , $$
$$
\delta Z^{\underline{b}_1 \ldots \underline{b}_6} = - 8 \cdot 135 (\delta^{[\underline{b}_1}{}_c \Lambda^{\underline{b}_2 \ldots \underline{b}_6]} \varepsilon^{ce} + 7 \Lambda^{e \underline{b}_1 \ldots \underline{b}_6}) P_e 
  \eqno(3.3)$$
and at level two:
$$
\delta Z^{\underline{b}_1 \ldots \underline{b}_5} = {6! \over 2} \Lambda^{e \underline{b}_1 \ldots \underline{b}_5} P_e , \delta \underline{Z}^{\underline{b}_1 \ldots \underline{b}_7} = 7 \cdot 15 ( 6 \delta^{[\underline{b}_1}{}_p \Lambda^{\underline{b}_2 \ldots \underline{b}_7]} \varepsilon^{pe} -  8 \Lambda^{e \underline{b}_1 \ldots \underline{b}_7}  + 7 \Lambda^{e [\underline{b}_1 \ldots \underline{b}_6,\underline{b}_7]})P_e  , 
 $$
$$
\delta Z^{\underline{b}_1 \ldots \underline{b}_6,\underline{c}} = 135 (- 12 \delta^{[\underline{b}_1}{}_{p} \Lambda^{\underline{b}_2 \ldots \underline{b}_6] \underline{c}} \varepsilon^{p e} + 6 \delta^{\underline{c}}{}_{p} \Lambda^{\underline{b}_1 \ldots \underline{b}_6} \varepsilon^{pe} - 8 \cdot 7 \Lambda^{e \underline{b}_1 \ldots \underline{b}_6 \underline{c}} $$
$$
+ 6 \cdot 7 \Lambda^{e \underline{c} [\underline{b}_1 \ldots \underline{b}_5,\underline{b}_6]} - 7 \cdot 7 \Lambda^{e \underline{b}_1 \ldots \underline{b}_6,\underline{d}}) P_e 
$$
$$
\delta Z^{\underline{b}_1 \ldots \underline{b}_7} = 7 \cdot 135 ( - 6 \delta^{[\underline{b}_1}{}_p \Lambda^{\underline{b}_2 \ldots \underline{b}_7]} \varepsilon^{pe} - 8 \cdot 7 \Lambda^{e \underline{b}_1 \ldots \underline{b}_7} - 7 \Lambda^{e [\underline{b}_1 \ldots \underline{b}_6,\underline{b}_7]}) P_e 
 \eqno(3.4)$$
In these equations we have omitted the variation of some of the more complicated charges and we have used equation (3.1) to eliminate $Q^a$ in terms of $P_a$.
\par
Requiring that the variations of the charges  to vanish results in  restrictions on the $\Lambda^{\underline{\alpha}}$'s  and so the transformation. We first consider the variations of the level IIA zero charges $ P_{\underline{b}}$ and $ Q^{\underline{b}}$. These do not separately have to vanish but they must preserve the relation of equation (3.1), namely 
$$
\delta (P_a + \varepsilon_{ab} Q^b) = 0
 \eqno(3.5)$$
Inserting (3.2) for the case of $\underline{b} = b$ into (3.5) we see $\Lambda^{ab}$ and $\tilde{\Lambda}^{ab}$ must   satisfy
$$
2 ( - \Lambda_a{}^b + \varepsilon_{ac} \Lambda^{c}{}_d \varepsilon^{db}) P_b + 2 (\tilde{\Lambda}_{ac} \varepsilon^{cb} - \varepsilon_{ac} \tilde{\Lambda}^{cb}) P_b = 0. 
\eqno(3.6)$$
However (3.6) vanishes identically since we may write  $\Lambda^{ab} = \varepsilon^{ab} \Lambda$ and $\tilde{\Lambda}^{ab} = \varepsilon^{ab} \tilde{\Lambda}$. Hence there are no restrictions on $\Lambda^{ab}$ and $\tilde{\Lambda}^{ab}$. Part of this symmetry is the obvious SO(1,1) Lorentz world sheet symmetry that we already  insisted the charges conditions should preserve but they also preserve the transformations corresponding to $S_{ab}$. 
\par 
Setting the variations of all the other brane charges  to zero gives further restrictions on the $I_c(E_{11})$ transformation.  Examining equations  (3.2)  we find the following restrictions on the $\Lambda^{\underline{\alpha}}$'s. At level zero we find the conditions 
$$
\Lambda^{ab} \ \neq 0  \ , \ \Lambda^{ij} \ \neq \ 0 \  , \ \ 
\tilde{\Lambda}^{ab} \ \neq \ 0  \ , \ \ \tilde{\Lambda}^{ai} \ = \- \varepsilon^{ac} \Lambda_c{}^i \  , \ \ \tilde{\Lambda}^{ij} \ \neq 0  \ \ . \eqno(3.7)$$
While varying only the charges  $Z, Z^{\underline{b}_1 \underline{b}_2}, Z^{\underline{b}_1 \ldots \underline{b}_4} , Z^{\underline{b}_1 \ldots \underline{b}_6} ,  $  at level one leads to the conditions 
$$
\Lambda^a = 0 , \ \  \Lambda^{a i_1 i_2} = 0 , \ \ \Lambda^{a b i} = {1 \over 6} \varepsilon^{a b} \Lambda^{i}  , \ \ 
\Lambda^{a i_1 \ldots i_4} = 0  , 
$$
$$
 \Lambda^{a b i_1 i_2 i_3} = {2 \over 5 \cdot 4} \varepsilon^{a b} \Lambda^{i_1 i_2 i_3}, \  \ \Lambda^{i_1 \ldots i_5} \neq 0 ,\ 
\Lambda^{e i_1 \ldots i_6} = 0, 
$$
$$
 \Lambda^{ab i_1 \ldots i_5} = {1 \over 7 \cdot 6} \varepsilon^{ab} \Lambda^{i_1 \ldots i_5} , \ \ \Lambda^{i_1 \ldots i_7} \neq 0 \ \ , 
 \eqno(3.8)$$
At level two the computations become more difficult but it is straightforward to find the following restrictions 
$$
\Lambda^{a i_1 \ldots i_5} \ = \ 0, \ \ \Lambda^{a b i_1 \ldots i_4} \ = \ 0  , \ \ \Lambda^{i_1 \ldots i_6} \ \neq \ 0, 
$$
$$
\Lambda^{a i_1 \ldots i_7} = 0 , \ \ \Lambda^{a i_1 \ldots i_6,j} = 0 , \ \ \Lambda^{i_1 \ldots i_7,j} \neq 0 , \ 
\ \ \Lambda^{i_1 \ldots i_8} \neq 0
\eqno(3.9)$$
\par
The generators of the little algebra  are found by inserting these  relations for the parameters  into the generator of equation (2.23). Equations  (3.7) to (3.9) imply that
$$
S_a \ , \ S_{a i_1 i_2} \ , \ S_{a i_1 \ldots i_4} \ , \ S_{a i_1 \ldots i_6} \ , \ S_{a_1 a_2 i_1 \ldots i_4} \ , \ S_{a i_1 \ldots i_7} \notin {\cal H} 
\eqno(3.10)$$
but that   ${\cal H}$ contains the following generators 
$$
J_{ij}\ , S_{ij}\ , J_{ab}\ , S_{ab}\ , \ L_{ai}^{(0)} = J_{a i} + \varepsilon_a{}^e S_{ei} 
\eqno(3.11) $$
at level zero and $$ 
 \ L_i^{(1)} = S_i + {1 \over 2} \varepsilon^{e_1 e_2} S_{e_1 e_2 i} \ , \ \ 
L_{i_1 i_2 i_3}^{(1)} = S_{i_1 i_2 i_3} + \varepsilon^{e_1 e_2} S_{e_1 e_2 i_1 i_2 i_3} \ , \ \ 
$$
$$
L_{i_1 \ldots i_5}^{(1)} = S_{i_1 \ldots i_5} + {1 \over 2} \varepsilon^{e_1 e_2} S_{e_1 e_2 i_1 \ldots i_5} \ , 
\eqno(3.12)$$
at level one. The superscript refers to the IIA level to which the generators belong. 
\par
Finding the rest of the relations between the $\Lambda^{\underline{\alpha}}$'s resulting   from the variations of the higher level charges becomes more and more difficult as to find the variations  requires the commutation relations for higher and higher level generators which are generally not known. Instead one can exploit the fact that the little group algebra ${\cal H}$ must close and as the generators of equation (3.11) and (3.12) belong to ${\cal H}$ we can find in this way new elements by calculating their commutators. This algebra will be computed below but for completeness we now list  some of the elements of the little algebra that we find in this way: 
$$
L_{i_1 \ldots i_7}^{(1)} = S_{i_1 \ldots i_7} - {1 \over 2} \varepsilon^{a_1 a_2} S_{a_1 a_2 i_1 \ldots i_7} \ \ ,
 \eqno(3.13)$$ 
 which is also a IIA level one generator and also the level two generators 
$$
L_{i_1 \ldots i_6}^{(2)} = S_{i_1 \ldots i_6} + {1 \over 2} \varepsilon^{a_1 a_2} S_{i_1 \ldots i_6 a_1 a_2} - \varepsilon^{a_1 a_2} S_{i_1 \ldots i_6 a_1 , a_2}
- {1 \over 8} \varepsilon^{a_1 a_2} \varepsilon^{b_1 b_2} S_{i_1 \ldots i_6 a_1 a_2 , b_1 b_2} \ \ ,  $$
$$
L_{i_1 .. i_8}^{(2)} = S_{i_1 .. i_8}  + \varepsilon^{a_1 a_2} S_{i_1 .. i_8 a_1,a_2} - {1 \over 6}  \varepsilon^{a_1 a_2} S_{i_1 .. i_8,a_1 a_2} - {1 \over 2} \varepsilon^{a_1 a_2} S_{(1)}{}_{a_1 a_2 i_1 .. i_8}
$$
$$
-   {4 \over 9} \varepsilon^{a_1 a_2} S_{(2)}{}_{a_1 a_2 i_1 .. i_8}
$$
$$
L_{i_1 .. i_7,j}^{(2)} = S_{i_1 .. i_7,j} - {1 \over 2} \varepsilon^{a_1 a_2} S_{a_1 a_2 i_1 .. i_7,j} - {1 \over 2} \varepsilon^{a_1 a_2} S_{i_1 .. i_7 a_1,a_2 j} - {1 \over 12} \varepsilon^{a_1 a_2} S_{i_1 .. i_7 j,a_1 a_2} $$
$$
- {1 \over 18} \varepsilon^{a_1 a_2} S_{(2)}{}_{a_1 a_2  i_1 .. i_7 j} 
\eqno(3.14) $$
\par
The above results imply that the string  little algebra  is given by  
$$
{\cal H} = \{ J_{ab}  , \ L_{ai}^{(0)}   , \ J_{ij} \ ,  \ S_{ab}  , \ S_{ij} \ ; \ L_i^{(1)}  , \  L_{i_1 i_2 i_3}^{(1)}  \, \ L_{i_1 \ldots i_5}^{(1)}  , \ L_{i_1 \ldots i_7}^{(1)} \ , ; 
L_{i_1 \ldots i_6}^{(2)}   , L_{i_1 \ldots i_8}^{(2)}  , \ L_{i_1 \ldots i_7,j}^{(2)}  , \  \ldots \}  
\eqno(3.15)$$
Here the levels are separated by a semi-colon and the $\ldots$ denotes higher level generators that we have not computed
\par
We will now compute the algebra of the generators of ${\cal H}$ . At IIA level zero we find that 
$$
[J^{ab},J_{cd}] = - 4 \delta^{[a}_{[c} J^{b]}{}_{d]}  , \ \ [J^{ab},L_{ci}^{(0)}] = - 2 \delta^{[a}_c L^{b]}{}_i^{(0)} , \ \ [J^{ab},J_{ij}] = 0 , 
$$
$$
[J^{ab},S_{cd}] = - 4 \delta^{[a}_{[c} S^{b]}{}_{d]} , 
\ \ [J^{ab},S_{ij}] = 0, \ \ [L^{ai}{}^{(0)},L_{bj}^{(0)}] = 0 , \ \  [L^{ai}{}^{(0)},J_{jk}] = 2 \delta^i_{[j} L^a{}_{k]}^{(0)} ,
$$
$$
 \ \ [L^{ai}{}^{(0)},S_{bc}] = 2 \varepsilon^a{}_{[b} L_{c]i}^{(0)} \ , \ 
[L^{ai}{}^{(0)},S_{jk}] = 2 \varepsilon^a{}_e \delta^i_{[j} L^e{}_{k]}^{(0)} \ \ , 
\ \ [J^{ij},J_{kl}] = - 4 \delta^{[i}_{[k} J^{j]}{}_{l]}  ,  
$$
$$
[J^{ij},S_{cd}] = 0 \ \ , \ \ [J^{ij},S_{kl}] = 0 \ \ , \ \ [S^{ab},S_{cd}] = - 4 \delta^{[a}{}_{[c} S^{b]}{}_{d]} \ \ , \ \ 
$$
$$
\ [S^{ab},S_{ij}] = 0 \ \ , \ \ [S^{ij},S_{kl}] = - 4 \delta^{[i}_{[k} J^{j]}{}_{l]} \ \ . 
\eqno(3.16)$$
\par
The commutators  of the generators in  ${\cal H}$  arising from  IIA levels $\pm 1$ are given by
$$
[L_i^{(1)},L_j^{(1)}] = 0 \ \ , \ \ [L_i^{(1)},L_{j_1 j_2 j_3}^{(1)}] = 0 \ \ , \ \  [L_i^{(1)},L_{j_1 \ldots j_5}^{(1)}] = - L_{i j_1 \ldots j_5}^{(2)} \ \ ,  $$
$$
[L_i^{(1)},L_{j_1 \ldots j_7}^{(1)}] = - L_{i j_1 \ldots j_7}^{(2)} + L_{j_1 \ldots j_7,i}^{(2)} \ \ , \ \
[L_{i_1 i_2 i_3}^{(1)},L_{j_1 j_2 j_3}^{(1)}] = 2 L_{i_1 i_2 i_3 j_1 j_2 j_3}^{(2)}  \ \ , $$
$$
[L_{i_1 i_2 i_3}^{(1)},L_{j_1 \ldots j_5}^{(1)}] = L_{i_1 i_2 i_3 j_1 \ldots j_5}^{(2)} - 5 L_{i_1 i_2 i_3 [j_1 \ldots j_4,j_5]}^{(2)} \ . \eqno(3.17)$$
as well as higher level relations. 
\par 
The algebra of the level zero generators with the  level $\pm 1$ generators of ${\cal H}$ is given by
$$
[J_{ab},L_i^{(1)}] = 0 \ \ , \ \ [J_{ab},L_{i_1 i_2 i_3}^{(1)}] = 0 \ \ , \ \ [J_{ab},L_{i_1 \ldots i_5}^{(1)}] = 0 \ \ , \ \ [J_{ab},L_{i_1 \ldots i_7}^{(1)}] = 0 \ \ , $$
$$
[L_{ai}^{(0)},L_j^{(1)}] = 0 \ , \  [L_{ai}^{(0)},L_{j_1 j_2 j_3}^{(1)}] = 0 \ , \ [L_{ai}^{(0)},L_{j_1 \ldots j_5}^{(1)}] = 0 \ , \ [L_{ai}^{(0)},L_{j_1 \ldots j_7}^{(1)}] = 0 \ \ , $$
$$
[J^{ij},L_k^{(1)}] = - 2 \delta^{[i}_k L^{j]}{}^{(1)} \ , \ [J^{ij},L_{k_1 k_2 k_3}^{(1)}] = - 6 \delta^{[i}_{[k_1} L^{j]}{}_{k_2 k_3]}^{(1)} \ \ , $$
$$
[J^{ij},L_{k_1 \ldots k_5}^{(1)}] = - 10 \delta^{[i}_{[k_1} L^{j]}{}_{k_2 \ldots k_5]}^{(1)} \ \ , \ \ [J^{ij},L_{k_1 \ldots k_7}^{(1)}] = - 14 \delta^{[i}_{[k_1} L^{j]}{}_{k_2 \ldots k_7]}^{(1)} \ \ , $$
$$
[S^{ab},L_i^{(1)}] = \varepsilon^{ab} L_i^{(1)} \ \ , \ \ [S^{ab},L_{i_1 i_2 i_3}^{(1)}] = \varepsilon^{ab} L_{i_1 i_2 i_3}^{(1)} \ \ , \ \ [S^{ab},L_{i_1 \ldots i_5}^{(1)}] = \varepsilon^{ab} L_{i_1 \ldots i_5}^{(1)} \ \ , $$
$$
[S^{ab},L_{i_1 \ldots i_7}^{(1)}] = \varepsilon^{ab} L_{i_1 \ldots i_7}^{(1)} \ \ , \ \ [S_{ij},L_k^{(1)}] = - L_{ijk}^{(1)} \ \ , $$
$$
[S^{ij},L_{k_1 k_2 k_3}^{(1)}] = - 2 L^{ij}{}_{k_1 k_2 k_3}^{(1)} + 6 L_{[k_1}^{(1)} \delta_{k_2 k_3]}^{ij}  \ \ , $$
$$
[S^{ij},L_{k_1 \ldots k_5}^{(1)}] = 10 L^{(1)}_{[k_1 k_2 k_3} \delta_{k_4 k_5]}^{i j} - L^{(1)}{}^{ij}{}_{k_1 \ldots k_5} \ \ , $$
$$
[S^{ij},L_{k_1 \ldots k_7}^{(1)}] = 6 \cdot 7 L_{[k_1 \ldots k_5}^{(1)} \delta_{k_6 k_7]}^{ij} \ \ . \eqno(3.18)$$

%%%%%%%%%%%%%%%%%%%%

\medskip
{{\bf 4. A toy string as an irreducible representation }}
\medskip
Rather than study the irreducible representations of $I_c(E_{11})\otimes_s l_1$ we will, in this section,  consider the simpler case of $I_c (D_{D}^+)\otimes_s l_1$. In this context $l_1$ is the vector representation which contains the generators 
 $P_{a \, n}$, $Q^a_n$ with $n \geq 0$ and we take the real form of $D_{D}$ to be SO(D,D). The  algebra, $I_c (D_{10}^+)$ occurs  as  subalgebra of  $I_c(E_{11})$ in its IIA formulation and so our results will shed light on this latter case. In appendix B we formulated the algebra $SO(D,D)\otimes_s (P_a, Q^a)$ in detail and using these results   it is straightforward to formulate the algebra $I_c (D_D^+)\otimes_s l_1$. 
 \medskip 
 {\bf 4.1 The $D_D^+ \otimes_s l_1$ algebra}
 \medskip
In this section we will give the explicit form of the affine Lie algebra  ${{\rm SO}(D,D)}^+$ as well as  its Cartan invariant involution subalgebra.  We will also discuss its vector representation. 
 Given any finite dimensional semi-simple Lie algebra $G$ we can divide the generators into those that are even and those that are odd under the   Cartan involution operator $I_c$ that the algebra possess. Let us denote these generators by $S^a$ and $T^i$ respectively in which case $I_c(S^a)=S^a$ and $I_c(T^i)=-T^i$. In terms of these generators  the algebra takes the form 
 $$ 
 [S^a , S^b] = f^{ab}{}_c S^c ,\ [S^a , T^i] = f^{ai}{}_j T^j ,\ [T^i , T^j] = f^{ij}{}_c S^c
 \eqno(4.1.1)$$
The affine algebra $G^+$ has the generators $S_n^a$ and $T_n^i$  which obey the algebra 
$$ 
 [S_n^a , S_m^b] = f^{ab}{}_c S_{n+m}^c ,\ [S^a_n , T_m^i] = f^{ai}{}_j T_{n+m}^j ,\ [T_n^i , T_m^j] = f^{ij}{}_c S_{n+m}^c
 \eqno(4.1.2)$$
 where we omit to write  the central term. 
 \par
 As explained in reference [20] the Cartan involution acts on $G^+$ as $I_c (S_n)= S_{-n} $ and $I_c (T_n)= -T_{-n} $ and so the Cartan involution  invariant generators in $I_c(G^+)$ are given by 
 $$
 {\cal S}_n^a = S_n^a+S_{-n}^a ,\ \ \ {\rm and } \ \ {\cal T}_n^i =  T_n^i-T_{-n}^i
 \eqno(4.1.3)$$
 and obey the commutators  
 $$ 
 [{\cal S}_n^a , {\cal S}_m^b] = f^{ab}{}_c {\cal S}_{n+m}^c + f^{ab}{}_c {\cal S}_{n-m}^c ,\ [{\cal S}^a_n , {\cal T}_m^i] = f^{ai}{}_j {\cal T}_{n+m}^j 
 -f^{ai}{}_j {\cal T}_{n-m}^j ,\ 
 $$
 $$
 [{\cal T}_n^i , {\cal T}_m^j] = f^{ij}{}_c {\cal S}_{n+m}^c -f^{ij}{}_c {\cal S}_{n-m}^c
 \eqno(4.1.4)$$
 This is not an affine algebra.
 \par
  We note that  the generators ${\cal S}_n^a$ formed a closed subalgebra. The generators ${ S}^a$ form the algebra $I_c(G)$ and so the generators $S_n^a$ have the algebra $(I_c(G))^+$. However,  the generators ${\cal S}_n^a$ have the algebra   $I(I_c(G)^+)$ which is the subalgebra of $(I_c(G))^+$ that is  invariant under  the involution $I$ which acts as $I(S_n^a)= S_{-n}^a$. Obviously this is a  subalgebra of  $I_c(G^+)$.   For example, if $G=A_{D}= SL(D+1)$ with generators $K^i{}_j$,  then $I_c (SL(D+1))=SO(D+1)$ contains the generators $K^i{}_j- K^j{}_i$. In this case $(I(I_c(G^+))= I(SO(D+1)^+)$ with the generators  $K_n{}^i{}_j- K_n{}^j{}_i+K_{-n}{}^i{}_j- K_{-n}{}^j{}_i$. The algebra  $I_c(SO(D+1)^+)$ contains in addition the generators $K_n{}^i{}_j +K_n{}^j{}_i-K_{-n}{}^i{}_j- K_{-n}{}^j{}_i$
 \par
 As is well known one can formulate an affine algebra by taking $S^a_n = S^a e^{in\sigma }$  and $T^i_n = T^i e^{in\sigma }$.   One then recovers the commutations of the affine algebra of equation (4.1.2). This allows us to  interpret the affine algebra as a loop algebra, that is, as arising from a  map of the closed loop, or  closed string,  into the finite dimensional semi-simple Lie group $G$. For  the Cartan involution invariant algebra $(I_c(G^+)$ we have instead,
 $$
 {\cal S}^a_n= 2S ^a \cos n\sigma ,\ \ {\rm and }\ \ \  {\cal T}^i_n= 2i \sin n\sigma T^i  
  \eqno(4.1.5)$$
  One readily finds that these generators do obey the commutators of equation (4.1.4). We can interpret $I_c(G^+)$ as a mapping of an open string into the finite dimensional semi-simple Lie group $G$. The open string associated with the ${\cal S}^a_n$  is the usual string which obeys Neuman  boundary conditions at both ends while the  open string associated with ${\cal T}^i_n$ obeys Dirichlet at each end. 
 \par
 We will now apply the above discussion to formulate the Cartan involution algebra of $D_D^+= {SO(D, D)}^+$. In appendix B we listed the Cartan even generators of ${{\rm SO}(D,D)}$ as $J_{\underline a, \underline b}$ and $S_{\underline a, \underline b}$ and the odd generators as $T_{\underline a, \underline b}$ and $U_{\underline a, \underline b}$. We recall that  $T_{\underline a, \underline b}$ and $U_{\underline a, \underline b}$ are symmetric and anti-symmetric in their indices respectively. 
Adding the level indices 
$n,m,\ldots$ to the generators of ${{\rm SO}(D,D)}$  we find the generators of the affine algebra ${{\rm SO}(D,D)}^+$. Their commutators can easily be read off from equations (B.1),  (B.2) and (B.3)  in appendix B to be given by 
 $$
[J_n{}^{\underline{a}_1 \underline{a}_2} , J_m{}_{\underline{b}_1 \underline{b}_2 }] = - 4 \delta^{[\underline{a}_1}_{[\underline{b}_1} J_{n+m} {}^{\underline{a}_2]}{}_{\underline{b}_2}] ,\ 
[S^{\underline{a}_1 \underline{a}_2}_n,S_{\underline{b}_1 \underline{b}_2 \, m}] = - 4 \delta^{[\underline{a}_1}_{[\underline{b}_1} J_{ n+m}{}^{\underline{a}_2]}{}_{\underline{b}_2] } 
$$
$$
[J_n{}^{\underline{a}_1 \underline{a}_2},S_m{}_{\underline{b}_1 \underline{b}_2 }] = - 4 \delta^{[\underline{a}_1}_{[\underline{b}_1} S_{n+m}{}^{\underline{a}_2]}{}_{\underline{b}_2] } , 
\eqno(4.1.6) $$
 and 
$$
[J_{n}{}^{\underline{a}_1 \underline{a}_2} , T_{m}{}_{\underline{b}_1 \underline{b}_2} ]= -4 \delta ^{[\underline{a}_1 }_{(\underline{b}_1 |}T_{n+m}{}^{\underline{a}_2]}{}_{| \underline{b}_2)} , \ 
[J_{n}{}^{\underline{a}_1 \underline{a}_2} , U_{m}{}_{\underline{b}_1 \underline{b}_2} ]= -4 \delta ^{[\underline{a}_1 }_{[\underline{b}_1|}U_{n+m}{}^{\underline{a}_2]}{}_{|\underline{b}_2]} , \ 
$$
$$
[S_{n}{}^{\underline{a}_1 \underline{a}_2} , T_{m}{}_{\underline{b}_1 \underline{b}_2} ]= 4 \delta ^{[\underline{a}_1 }_{(\underline{b}_1|}U_{n+m}{}^{\underline{a}_2]}{}_{|\underline{b}_2)} , \ 
[S_{n}{}^{\underline{a}_1 \underline{a}_2} , U_{m}{}_{\underline{b}_1 \underline{b}_2} ]= 4 \delta ^{[\underline{a}_1 }_{[\underline{b}_1 |}T_{n+m}{}^{\underline{a}_2]}{}_{ |\underline{b}_2]} 
\eqno(4.1.7)$$
as well as  
$$
[T_{n}{}^{\underline{a}_1 \underline{a}_2} , T_{m}{}_{\underline{b}_1 \underline{b}_2} ]= 4 \delta ^{(\underline{a}_1 }_{(\underline{b}_1|}J_{n+m}{}^{\underline{a}_2)}{}_{| \underline{b}_2)} , \ 
[U_{n}{}^{\underline{a}_1 \underline{a}_2} , U_{m}{}_{\underline{b}_1 \underline{b}_2} ]= 4 \delta ^{[\underline{a}_1 }_{[\underline{b}_1|}J_{n+m}{}^{\underline{a}_2]}{}_{|\underline{b}_2]} , \ 
$$
$$
[T_{n}{}^{\underline{a}_1 \underline{a}_2} , U_{m}{}_{\underline{b}_1 \underline{b}_2} ]= 4 \delta ^{(\underline{a}_1 }_{[\underline{b}_1|}S_{n+m}{}^{\underline{a}_2)}{}_{|\underline{b}_2]} 
\eqno(4.1.8)$$
We have omitted the usual central terms. 
\par
The algebra can also be given in terms of the generators 
$$
M_{n}{}^{\underline{a}_1 \underline{a}_2}= {1\over 2} (J_n{}^{\underline{a}_1 \underline{a}_2}+S_n{}^{\underline{a}_1 \underline{a}_2}) , \  \bar M_{n}{}^{\underline{a}_1 \underline{a}_2}= {1\over 2} (J_n{}^{\underline{a}_1 \underline{a}_2}-S_n{}^{\underline{a}_1 \underline{a}_2}) , \ 
$$
$$
V_{n}{}^{\underline{a}_1 \underline{a}_2}= T_n{}^{\underline{a}_1 \underline{a}_2}-U_n{}^{\underline{a}_1 \underline{a}_2},\ \bar V_{n}{}^{\underline{a}_1 \underline{a}_2}= T_n{}^{\underline{a}_1 \underline{a}_2}+U_n{}^{\underline{a}_1 \underline{a}_2}
\eqno(4.1.9)$$
Their  commutators are given by 
$$
[M_{n}{}^{\underline{a}_1 \underline{a}_2} , M_{m}{}_{\underline{b}_1 \underline{b}_2} ]= -4 \delta ^{[\underline{a}_1 }_{(\underline{b}_1 |}M_{n+m}{}^{\underline{a}_2]}{}_{| \underline{b}_2)} ,\  \ [\bar M_n{}^{\underline{a}_1 \underline{a}_2} , \bar M_m{} _{\underline{b}_1 \underline{b}_2} ]= -4 \delta ^{[\underline{a}_1 }_{[\underline{b}_1}\bar M_{n+m} {}^{\underline{a}_2]}{}_{\underline{b}_2]} ,
$$
$$
[ M_n{}^{\underline{a}_1 \underline{a}_2} , \bar M_m{} _{\underline{b}_1 \underline{b}_2} ]=0
\eqno(4.1.10)$$
and 
$$
[M_n{}^{\underline{a}_1 \underline{a}_2} , V_{m}{}_{\underline{b}_1 \underline{b}_2} ]= -4 \delta ^{[\underline{a}_1 }_{\underline{b}_1}V_{n+m}{}^{\underline{a}_2]}{}_{\underline{b}_2} , \ 
[M_n{}^{\underline{a}_1 \underline{a}_2} , \bar  V_{m}{}_{\underline{b}_1 \underline{b}_2} ]= -4 \delta ^{[\underline{a}_1 }_{\underline{b}_2} V_{n+m}{}^{\underline{a}_2]}{}_{\underline{b}_1}  , \ 
$$
$$
[\bar M_n{}^{\underline{a}_1 \underline{a}_2} , \bar V_{m}{}_{\underline{b}_1 \underline{b}_2} ]= -4 \delta ^{[\underline{a}_1 }_{\underline{b}_1}\bar V_{n+m}{}^{\underline{a}_2]}{}_{\underline{b}_2} , \ 
[\bar M_n{}^{\underline{a}_1 \underline{a}_2} , V_{m}{}_{\underline{b}_1 \underline{b}_2} ]= -4 \delta ^{[\underline{a}_1}_{\underline{b}_2} \bar V_{n+m}{}^{\underline{a}_2]}{}_{\underline{b}_1}  
\eqno(4.1.11)$$
as well as  
$$
[ V_n{}^{\underline{a}_1 \underline{a}_2} ,   V_{m}{}_{\underline{b}_1 \underline{b}_2} ]= 4 \delta ^{\underline{a}_1 }_{\underline{b}_1}\bar M_{n+m}{}^{\underline{a}_2}{}_{\underline{b}_2} +4 \delta ^{\underline{a}_2 }_{\underline{b}_2}M_{n+m}{}^{\underline{a}_1}{}_{\underline{b}_1} , \ 
$$
$$
[ \bar V_n{}^{\underline{a}_1 \underline{a}_2} ,  \bar  V_{m}{}_{\underline{b}_1 \underline{b}_2} ]= 4 \delta ^{\underline{a}_1 }_{\underline{b}_1}M_{n+m}{}^{\underline{a}_2}{}_{\underline{b}_2} +4 \delta ^{\underline{a}_2 }_{\underline{b}_2} \bar M_{n+m}{}^{\underline{a}_1}{}_{\underline{b}_1} , \ 
$$
$$
[ V_n{}^{\underline{a}_1 \underline{a}_2} ,  \bar V_{m}{}_{\underline{b}_1 \underline{b}_2} ]= 4 \delta ^{\underline{a}_2 }_{\underline{b}_1}M_{n+m}{}^{\underline{a}_1}{}_{\underline{b}_2} +4 \delta ^{\underline{a}_1 }_{\underline{b}_2} \bar M_{n+m}{}^{\underline{a}_2}{}_{\underline{b}_1} ,  
\eqno(4.1.12)$$
We note that $M_{n}{}^{\underline{a}_1 \underline{a}_2}$ and $\bar M_{n}{}^{\underline{a}_1 \underline{a}_2}$ generate the 
algebra $SO(D)^+\otimes SO(D)^+$ which equals $(I_c(SO(D,D))^+$
\par
The generators of $I_c(D_D^+)$ are 
$$
J_{n}{}^{\underline{a}_1 \underline{a}_2}+J_{-n}{}^{\underline{a}_1 \underline{a}_2} ,\ S_{n}{}^{\underline{a}_1 \underline{a}_2}+S_{-n}{}^{\underline{a}_1 \underline{a}_2} , \ T_{n}{}^{\underline{a}_1 \underline{a}_2}-T_{-n}{}^{\underline{a}_1 \underline{a}_2} ,\ 
U_{n}{}^{\underline{a}_1 \underline{a}_2} -U_{-n}{}^{\underline{a}_1 \underline{a}_2} 
\eqno(4.1.13)$$
or equivalently 
$$
M_{n}{}^{\underline{a}_1 \underline{a}_2}+M_{-n}{}^{\underline{a}_1 \underline{a}_2} ,\ \bar M _{n}{}^{\underline{a}_1 \underline{a}_2}+\bar M_{-n}{}^{\underline{a}_1 \underline{a}_2} , \ V_{n}{}^{\underline{a}_1 \underline{a}_2}-V_{-n}{}^{\underline{a}_1 \underline{a}_2} ,\ 
\bar V_{n}{}^{\underline{a}_1 \underline{a}_2} -\bar V_{-n}{}^{\underline{a}_1 \underline{a}_2} 
\eqno(4.1.14)$$
Their commutators can be readily deduced from those above. The first two generators in the above equation belong to  the algebra  $I(SO(D)^+\otimes SO(D)^+)$ in terms of the above notation. The last two generators belong to a representation of this algebra. 
\par
We will now consider the vector representation of $D_D^+$ which contains the generators $P_{n \underline{a}}$ and $Q_n{}_{\underline{a}}$. Their commutators with the generators of  $D_D^+$ follow from those of equation (B.8) and (B.9) and are as follows 
$$
[J_{n}{}_{\underline{a}\underline{b}}, P_{m}{}_{\underline{c}}] =-2 \eta _{\underline{c}[\underline{a}} P_{n+m}{}_{\underline{b}]} ,\ [J_{\underline{a} \underline{b}}, Q_{m}{}_{\underline{c}}] = -2\eta _{\underline{c}[\underline{a}} Q_{n+m}{}_{\underline{b}]} ,
$$
$$
[S_n{}_{\underline{a} \underline{b}}, P_m{}_{\underline{c}}] =-2 \eta _{\underline{c}[\underline{a}} Q_{n+m}{}_{\underline{b}]} ,\ 
,\ [S_{n}{}_{\underline{a} \underline{b}}, Q_{m}{}_{\underline{c}}] =-2 \eta _{\underline{c}[\underline{a}} P_{n+m}{}_{\underline{b}]}
\eqno(4.1.15)$$
and 
$$
[T_{n}{}_{\underline{a} \underline{b}}, P_{m}{}_{\underline{c}}] =-2 \eta _{\underline{c}(\underline{a}} P_{n+m}{}_{\underline{b})} ,\ [T_{n}{}_{\underline{a} \underline{b}}, Q_{m}{}_{\underline{c}}] = +2\eta _{\underline{c}(\underline{a}} Q_{n+m}{}_{\underline{b})} ,\ 
$$
$$
[U_{n}{}_{\underline{a} \underline{b}}, P_{m}{}_{\underline{c}}] =-2 \eta _{\underline{c}[\underline{a}} Q_{n+m}{}_{\underline{b}]} ,\ 
,\ [U_{n}{}_{\underline{a} \underline{b}}, Q_{m}{}_{\underline{c}}] =2 \eta _{\underline{c}[\underline{a}} P_{n+m}{}_{\underline{b}]}
\eqno(4.1.16)$$
\par
It will be advantageous to formulate the commutators in terms of the generators 
$$
{\cal {\wp }}_n{}_{\underline{a}}= P_n{}_{\underline{a}}+Q_n{}_{\underline{a}} ,\  \bar {\cal {\wp }}_n{}_{\underline{a}} = P_n{}_{\underline{a}}-Q_n{}_{\underline{a}} ,\ 
\eqno(4.1.17)$$
Then the commutators with the generators of  $D_D^+$ follow from those of equation (B.11) and (B.12) and  are 
$$
[M_{n}{}_{\underline{a}_1 \underline{a}_2} , {\cal {\wp }}_{m}{}_{\underline{b}} ]= -2 \eta _{\underline{b} [\underline{a}_1} {\cal {\wp }}_{n+m}{}_{\underline{a}_2]} ,\ [\bar M_{n}{}^{\underline{a}_1 \underline{a}_2} , {\cal {\wp }}_{m}{}_{\underline{b}} ]=0 ,\ 
$$
$$
[\bar M_{n}{}_{\underline{a}_1 \underline{a}_2} , \bar {\cal {\wp }}_{m}{}_{\underline{b}} ]= -2 \eta _{\underline{b} [\underline{a}_1}\bar  {\cal {\wp }}_{n+m}{}_{\underline{a}_2]} ,\ [M_{n}{}^{\underline{a}_1 \underline{a}_2} , \bar {\cal {\wp }}_{\underline{b}} ]=0 ,\ 
\eqno(4.1.18)$$
and 
$$
[V_{n}{}_{\underline{a}_1 \underline{a}_2} ,  {\cal {\wp }}_{m}{}_{\underline{b}} ]= -2 \eta _{\underline{b} \underline{a}_1}  \bar  {\cal {\wp }}_{n+m}{}_{\underline{a}_2} ,\ [ V_{n}{}_{\underline{a}_1 \underline{a}_2} , \bar {\cal {\wp }}_{m}{}_{\underline{b}} ]=-2\eta_{\underline{b} \underline{a}_2} {\cal {\wp }}_{n+m}{}_{\underline{a}_1} ,\ 
$$
$$
[\bar V_{n}{}_{\underline{a}_1 \underline{a}_2} ,  \bar {\cal {\wp }}_{m}{}_{\underline{b}} ]= -2 \eta _{\underline{b} \underline{a}_1} {\cal {\wp }}_{n+m}{}_{\underline{a}_2} ,\ [\bar V_{n}{}_{\underline{a}_1 \underline{a}_2} ,  {\cal {\wp }}_{m}{}_{\underline{b}} ]=-2\eta_{\underline{b} \underline{a}_2} \bar {\cal {\wp }}_{n+m}{}_{\underline{a}_1} 
\eqno(4.1.19)$$
\par
The $D_D^+$  variations of the vector representation are  given by $\delta P_{n\underline{a}} = [\Lambda , P_{n\underline{a}}  ]$ and $\delta Q_{n\underline{a}} = [\Lambda , Q_{n\underline{a}}  ]$ where 
$$
\Lambda = \Lambda _n{}^{\underline{a} \underline{b}}J_n{}_{\underline{a} \underline{b}}+ \tilde \Lambda_n{} ^{\underline{a} \underline{b}}S_n{}_{\underline{a} \underline{b}}+ \Omega_n{}^{\underline{a} \underline{b}} T_n{}_{\underline{a} \underline{b}} + \hat \Omega_n{}^{\underline{a} \underline{b}} U_n{}_{\underline{a} \underline{b}}
$$
$$
=  \lambda_n{}^{\underline{a} \underline{b}}M_n{}_{\underline{a} \underline{b}}+ \tilde \lambda_n{} ^{\underline{a} \underline{b}}\bar M_n{} _{\underline{a} \underline{b}}+ \mu _n{}^{\underline{a} \underline{b}} V_n{}_{\underline{a} \underline{b}} + \tilde  \mu_n{}^{\underline{a} \underline{b}} \bar V_n{}_{\underline{a} \underline{b}}
\eqno(4.1.20)$$
We find,  generalising  equations (B.15) and (B.16),  that  the transformations on the vector representation are given by 
$$
 \delta {\cal {\wp }}_{n}{}_{\underline{a}} = -2(\Lambda_{m}{}+\tilde \Lambda_{m})_{\underline{a}}{}^{\underline{b}}  {\cal {\wp }}_{n+m}{}_{\underline{b}} -2(\Omega_m -\hat \Omega_m)_{\underline{a}}{}^{\underline{b}}  \bar {\cal {\wp }}_{n+m}{}_{\underline{b}}
$$
$$
= -2 \lambda_{m}{}_{\underline{a}}{}^{\underline{b}}  {\cal {\wp }}_{n+m}{}_{\underline{b}} -2( \mu_{m}{}_{\underline{a}}{}^{\underline{b}}+\bar \mu_{m}{}^{\underline{b}}{}_{\underline{a}} ) \bar {\cal {\wp }}_{n+m}{}_{\underline{b}} , 
$$
$$
\delta  \bar {\cal {\wp }}_{n}{}_{\underline{a}} = -2(\Lambda_{m}-\tilde \Lambda_{m}){}_{\underline{a}}{}^{\underline{b}}  \bar {\cal {\wp }}_{n+m}{}_{\underline{b}} -2(\Omega_{m}{} +\hat \Omega_{m}{})_{\underline{a}}{}^{\underline{b}}  {\cal {\wp }}_{n+m}{}_{\underline{b}} 
$$
$$
=  -2\tilde  \lambda_{m}{}_{\underline{a}}{}^{\underline{b}} \bar {\cal {\wp }}_{n+m}{}_{\underline{b}} - 2(\tilde  \mu_{m}{}_{\underline{a}}{}^{\underline{b}} +\mu_{m}{}^{\underline{b}}{}_{\underline{a}}) { \cal {\wp }}_{n+m}{}_{\underline{b}}, 
\eqno(4.1.21)$$
\par
Their   transformations under $I_c(D_D^+)$ are then given by 
$$
 \delta {\cal {\wp }}_{n}{}_{\underline{a}} = -2(\Lambda_{m}{}+\tilde \Lambda_{m})_{\underline{a}}{}^{\underline{b}}  ({\cal {\wp }}_{n+m}{}_{\underline{b}}+{\cal {\wp }}_{n-m}{}_{\underline{b}}) -2(\Omega_m -\hat \Omega_m)_{\underline{a}}{}^{\underline{b}} ( \bar {\cal {\wp }}_{n+m}{}_{\underline{b}} - \bar {\cal {\wp }}_{n-m}{}_{\underline{b}})
$$
$$
= -2 \lambda_{m}{}_{\underline{a}}{}^{\underline{b}}  ({\cal {\wp }}_{n+m}{}_{\underline{b}}+{\cal {\wp }}_{n-m}{}_{\underline{b}})  -2( \mu_{m}{}_{\underline{a}}{}^{\underline{b}}+\bar \mu_{m}{}^{\underline{b}}{}_{\underline{a}} )( \bar {\cal {\wp }}_{n+m}{}_{\underline{b}} - \bar {\cal {\wp }}_{n-m}{}_{\underline{b}}) , 
$$
$$
\delta  \bar {\cal {\wp }}_{n}{}_{\underline{a}} = -2(\Lambda_{m}-\tilde \Lambda_{m}){}_{\underline{a}}{}^{\underline{b}}  (\bar {\cal {\wp }}_{n+m}{}_{\underline{b}}+\bar {\cal {\wp }}_{n-m}{}_{\underline{b}}) -2(\Omega_{m}{} +\hat \Omega_{m}{})_{\underline{a}}{}^{\underline{b}} ( {\cal {\wp }}_{n+m}{}_{\underline{b}} -{\cal {\wp }}_{n-m}{}_{\underline{b}})
$$
$$
=  -2\tilde  \lambda_{m}{}_{\underline{a}}{}^{\underline{b}} (\bar {\cal {\wp }}_{n+m}{}_{\underline{b}}+\bar {\cal {\wp }}_{n-m}{}_{\underline{b}})-2(\tilde  \mu_{m}{}_{\underline{a}}{}^{\underline{b}} +\mu_{m}{}^{\underline{b}}{}_{\underline{a}})( { \cal {\wp }}_{n+m}{}_{\underline{b}} - { \cal {\wp }}_{n-m}{}_{\underline{b}}) , 
\eqno(4.1.22)$$
\par
The quantity  
$$
\sum_n (P_{n}{}_{\underline{a}}^2 + Q_{n}{}_{\underline{a}}^2) +  \sum_n (P_{n}{}_{\underline{a}}^2 -Q_{n}{}_{\underline{a}}^2) , \ \ \ {\rm or  
\ equivalently } \ \ \  \sum_n {\cal {\wp }}_{n}{}_{\underline{a}}^2 + \sum_n \bar  { \cal {\wp }}_{n}{}_{\underline{a}}^2
\eqno(4.1.23)$$
is invariant under $I_c(D_{D}^+)$ transformations. 
\medskip
{\bf 4.2 The toy string little algebra}
\medskip
Since we are seeking to mimic the behaviour of a string we expect that a component of its momentum $P_a$ and its string charge $Q_a$ are non-zero while all the other charges vanish. It may be tempting to take $P_0 = m$ and $ Q^1 = m$ non-zero however, this does not take into account the fact that the string has a SO(1,1) symmetry in its world sheet. As such we take the condition 
$$
P_a+\epsilon_a{}^b Q_b=0 
\eqno(4.2.1)$$
with all other charges zero, that is, 
$$
P_{na}=0=Q_{na} , \ \ \ {\rm and }\ \ \ P_{ni}=0=Q_{ni} , \ n=1,2,\ldots 
\eqno(4.2.2)$$ 
We have split the original indices $\underline a, \underline b,\ldots$ into $a,b,\ldots =0,1$ and $i,j,\ldots = 2,\ldots, D-1$. 
\par
For this choice of charges  the invariant of equation (4.1.23) vanishes. We can think of this as the equivalent of a BPS condition. We note that  $P_a$ and $Q_a$ are non zero and are not specified other than that  they  are related by equation (4.2.1).   Since the original Wigner method of irreducible representations  applied to point particle this subtlety would not have been encountered before. 
 \par
The little  algebra is the subalgebra of $D_{D-1}^+$ that preserves the choice of equations (4.2.1) and (4.2.2). To find it we consider the variations of  the combinations    $(P\pm Q)_{in}=0, n\ge 1$ and take $(P\pm Q)_{a }=(1\mp \epsilon)_a{}^b P_b \not= 0$ as required by equation (4.2.1). One finds the conditions 
$$
\Lambda_{n}{} _{i a}= \tilde \Lambda_{n}{} _{i b}\epsilon^b {}_a, \ \Omega_{n}{} _{i a}=  \hat \Omega _{n}{} _{i b}\epsilon^b {}_a
\eqno(4.2.3)$$
Taking the same equation for $(P\pm Q)_{an}=0, \ n\ge 1$ we find the conditions 
$$
\Omega_n{}_a{}^b-\hat \Omega_n{} _a {}^c \epsilon_c{}^b=0 , \  \ 
\Omega_{n}{} _{ab}=  \hat \Omega _{n}{} _{a c}\epsilon^c{}_b
\eqno(4.2.4)$$
\par
From these results,  and using the expression of the algebra element in  equation (4.1.20), we find the transformations  which  preserve equations (4.2.1) and (4.2.2) are  given by 
$$
{\cal H}= \{E^{(1)}_{nia}\equiv J_{nia}+ \epsilon_a{}^b S_{nib} +J_{-nia}+ \epsilon_a{}^b S_{-nib},\ 
E^{(2)}_{nia} \equiv T_{nia}+ \epsilon_a{}^b U_{nib} -U_{-nia}- \epsilon_a{}^b U_{-nib},\ 
$$
$$
E_n{}_{ab}\equiv T_{nab} -\eta_{ab} U_n - T_{-nab}  +\eta_{ab} U_{-n},\  J_{ab},\ S_{ab} ,\ \ I_c({SO(D-2, D-2)}^+)\}
\eqno(4.2.5)$$
where $ U_{n}{}  _{a b}= \epsilon_{ab}U_{n}{}$ and  $ I_c({SO(D-2, D-2)}^+)$ consists of the generators 
$$
 I_c (D_{D-2)}= \{M_{nij}+M_{-nij} ,\ \bar M_{nij}+\bar M_{-nij} ,\ T_{nij}-T_{-nij} , \ U_{nij}-U_{-nij} ,  \ n\ge 0\}
 \eqno(4.2.6)$$
\par
The generators $E^{(1)}_{nia}$, $E^{(2)}_{njb}$  and $E$ obey the commutators 
 $$
 [E^{(1)}_{nia} , E^{(1)}_{mjb}]=0,\ [E^{(2)}_{nia} , E^{(2)}_{mjb} ]=0 ,\ [E_{nia}^{(1)} , E_m{}_{b_1b_2}  ]= 0 ,\ [ E_{nia}^{(2)} , E_m{}_{b_1b_2}  ]= 0
 $$
 $$
  [ E^{(1)}_{nai} ,  E^{(2)}_{mbj}] = - \delta_{ij}   (E_{n+m}{}_{ab} -E _{n-m} {}_{ab}+E _{-n+m}{} _{ab}-E _{-n-m} {}_{ab})
 $$
 $$
[E_n{}^{a_1a_2} , E_m{}_{b_1b_2}  ]=-4\delta ^{[a_1}_{[b_1} (J_{n+m}{}^{a_2]}{}_{b_2]} -J_{n-m}{}^{a_2]}{}_{b_2]} -J_{-n+m}{}^{a_2]}{}_{b_2]} +J_{-n-m}{}^{a_2]}{}_{b_2]} )
  \eqno(4.2.7)$$
 Hence these three generators form a closed sub algebra. 
 \medskip
 {\bf 4.3 The toy string representation of the little algebra }
 \medskip
 We now wish to find an irreducible representation of $I_c(D_D)\otimes_s l_1$  following a similar  path to that taken by Wigner to find the irreducible representations of the Poincare group. The reader can look at section four  of reference [14] to see how this works for the case of $I_c(E_{11})\otimes_s l_1$ and the massless representation. In the last section we  found the subalgebra ${\cal H}$ of  $I_c(D_D^+)$ that preserves equation (2.1) which corresponds to a static string.  The next step is to choose  an irreducible representation of ${\cal H}\otimes_s l_1 $. Given this  we can then boost this irreducible representation  to find an irreducible representation of $I_c(D_{D}^+)\otimes_s l_1$
As the generators $E^{(1)}_{nia}$, $E^{(2)}_{njb}$  and $E_n{}_{ab}$ belong to a closed subalgebra    we can  take them  to be trivially realised. This step is similar to the case of the massless representations of Poincare algebra where one finds subalgebras  which we can  take to be trivially realised so as to ensure unitarity. We can also take the generators $J_{ab} $ and $S_{ab}$ to be trivially realised. This leaves  us to find  an irreducible representation of $I_c (SO(D-2, D-2)^+) $. The algebra  $SO(D-2, D-2)^+$ contains the subalgebra $SO(D-2)^+\otimes SO(D-2)^+$ and so,  as discussed above,  the algebra $I_c (SO(D-2, D-2)^+) $ contains the algebra $I(SO(D-2)^+)\otimes I(SO(D-2)^+)$ 
\par
In section five we will find the little algebra for the string and in section six we will examine if the string states carry a representation of this algebra. Here we will see if string like states can carry a representation of the  the little algebra of the toy string.  Our discussion will prove  useful for the case of the real superstring. We will first consider the open superstring  and in particular the NS sector which has the oscillators $b_r^i$.  
There is a well known relation between the oscillators that appear in string theory and Lie algebras.  These works involve strings whose momenta belong to the corresponding root,  or weight,  lattices. Indeed the suitably integrated string vertex operators have commutators that are those of the Lie algebra corresponding to  the root lattice. For a review see reference [21]. In our situation in this paper the momenta do not belong to a root  lattice, however,  it is also known that given  the $b_r^i$ oscillators we can form the objects 
$$
{\cal M}_n{}^{ij}= \sum_r : b_{n-r}^{[i}b_{r}^{j]} : , \ \ i,j,\ldots =2,\ldots , D-1
\eqno(4.3.1)$$
which generate the affine algebra $SO({D-2})^+$  [22,23] This follows  simply from the fact that these oscillators obey their usual anti-commutation relations
$$
\{ b_r^i , b_s^j\}= \delta_{r+s,0}\delta^{ij}
\eqno(4.3.2)$$
\par
As such  if we consider the $ b_r^i $ with $r<0$ acting on the vacuum $|0>$, which obeys $b_r^i |0>=0$ for $r>0$, then these states will form a representation of  $SO({D-2})^+$. Clearly they will also carry a representation of the subalgebra  $I(SO({D-2})^+)$.  
 These states do not, however, satisfy the physical state conditions of string theory even if we assigned the vacuum to carry a momentum. Acting on such a state with a generator, say with ${\cal M}_n^{i_1i_2}$, will change the level of the oscillators  which   act on the vacuum state, and so the level of the state,  without changing the momentum of the state. However, the momentum  of a string state  is related  to the level of the state and so even  if it was a string state to begin with it will not be after the action of the generator ${\cal M}_n{}^{ij}$. This problem can be rectified if we consider instead of the usual oscillators,  the corresponding DDF operators which in the N-S sector are denoted by $B_r^i$ [24]. These obey the commutator of equation (4.3.2) with $b_r^i $ replaced by $B_r^i $ and so we can construct the generator ${\cal M}_n{}^{ij}= \sum_r  : B_{n-r}^{[i}B_{r}^{j]} :$ and these will obey the algebra of $SO({D-2})^+$. Thus the string states 
$$
B_{-r_1}^{i_1}\ldots B_{-r_n}^{i_n}|p^{(0)}, 0> ,\ \  {\rm with } \ \ B_r^i  |p^{(0)}    > =0, \ r > 0 
\eqno(4.3.3)$$
will belong to a representation of $SO({D-2})^+$.  In this equation $(p^{(0)})^2={1\over 2\alpha'}$ and  $B_{-r}^i$ injects a momentum $-rk^i$ where  $p^{(0)}\cdot k= {1\over 2\alpha'}$ and $k^2=0$. These states  clearly also  carry a representation of  $I(SO({D-2})^+)$ which is part of  the little algebra  $I_c(SO(D-2, D-2)^+)$  for the toy string found in the previous section. 
\par
 Let us now consider the closed superstring, and particular  the NS-NS sector,  which has the oscillators 
$b^I_r\equiv ( b_r^i, \bar b_r^i )$ where the indices $I,J,\ldots$ each take $D-2$ values. From these we can construct the generators 
$$
 {\cal M}^c{}_n{}^{IJ}= \sum_r : b_{n-r}^{[I}b_{r}^{J]} : 
\eqno(4.3.4)$$
which obey the algebra $SO(D-2, D-2)^+$, Hence  the states formed  by $b^I_r$ with $r<0$ acting on the vacuum $|0>$, which obeys $b^I_r |0>=0$ for $r>0$, carry a representation of $SO(D-2, D-2)^+$. Clearly they also carry a representation of 
$I_c(SO(D-2, D-2)^+)$ which is our little algebra. 
\par
However, as for the open string these states do not obey the physical state conditions. As in this case we can  use the DDF operators $B^I_r$  instead of the $b_r^I$. They have the same anti-commutation relations and so the objects $\hat {\cal M}^c_n{}^{IJ}= \sum_r : B_{n-r}^{[I}B_{r}^{J]} : $ generate the algebra $SO(D-2, D-2)^+$. 
We can act with the closed string oscillators $B_{r}^{I}$ , for $r>0$,  on a vacuum $|0,p^{(0)}>$ which carries a suitable momentum $p^{(0)}$ and these states will carry a representation of $SO(D-2, D-2)^+$. 
These states will obey all the physical states conditions with the exception for  the GSO projection and the requirement of level matching. That this condition must be imposed by hand is a feature of the DDF construction of the physical states of all closed strings. One approach is to consider the enveloping algebra of $SO(D-2, D-2)^+$ and then only consider the action of  those elements which preserved level matching and the GSO condition. This  would be a symmetry which  preserve the physical state conditions.
\par
We can think of the open toy string as a restriction of the closed toy string and in particular by imposing the restriction $\sigma\to -\sigma$. This requires the two oscillators, say  $b_r^i$ and $\bar b_r^i$ to transform in the same way. This means that the algebra should be  is restricted to $I(SO(D-2)^+)$, as indeed we found. 
\par
 We will now present an alternative way of realising the symmetry on the physical states of the open superstring. Let is choose 
$p^{(0)}= (0,1, 0, \ldots, 0){1\over \sqrt {2\alpha'}}$  and $k= (-1,1, 0, \ldots, 0){1\over  \sqrt {2\alpha'}}$ for the momenta of our vacuum physical state, that is, the tachyon. Then the generators $(J_{01}-S_{01})$ acts on the momenta $p^{(0)}$ to give 
$p^{(0)}-k$. Instead of the oscillators $b_r^i$ we consider the oscillators 
$$
\hat b_r^i\equiv e^{r(J_{01}-S_{01})}b_r^i 
\eqno(4.3.5)$$
to act on the vacuum. These will carry a momentum injection  that is the one required by the physical state conditions. These oscillators also obey the same anti-commutators as $b_r^i$ and the bilinear of equation (4.3.1) with $b_r^i$ replaced by $\hat b_r^i$ also obey the algebra of $SO(D-2)^+$. As such  these states do indeed carry a representation of $I(SO(D-2)^+)$, that is, the part of the little algebra appropriate for the open superstring. In the above we have choosen  the string charge of the vacuum state  to be given by $q^{(0)}= (1,0, 0, \ldots, 0){1\over \sqrt {2\alpha'}}$. Then  under the above boost it changes by $q^{(0)}\to q^{(0)}+k$ and so preserves the condition of equation (4.2.1). 
\par
 Thus, neglecting level matching and the GSO projection,  we have seen that the fermionic oscillators NS-NS states of the closed superstring do indeed form a representation of the little algebra of the toy string and the NS states of the open string the corresponding subalgebra. As we will see in section 6, the little algebra for the actual superstring has some very significant differences but the discussions  in this section will prove useful. 
 \par
 We will close this section with some remarks concerning the oscillators   $\alpha_n^i$ , $\bar  \alpha_n^i$, $i=2,\ldots , D-1$ of the closed string which obey the usual commutation relations 
$$
[\alpha^i_n , \alpha^j_m ]= n\delta^{ij}\delta_{n+m,0} ,\ [\bar \alpha^i_n ,\bar  \alpha^j_m ]= n\delta^{ij}\delta_{n+m,0} , \ 
[ \alpha^i_n ,\bar  \alpha^j_m ]=0
\eqno(4.3.6)$$
We could  take them to have the same commutators with the generators of $SO(D-2, D-2)^+$ as do   ${\cal \wp}^i_n$ and $\bar {\cal \wp}^i_n$, that is, take them to belong to the same representation.  Making such a  replacement in equations (4.1.18) and  (4.1.19) we find that 
$$
[M_{n}{}_{i_1i_2 } , {\alpha }_{m}{}_{j} ]= -2 \eta _{j [i_1} {\cal {\alpha }}_{n+m}{}_{i_2]} ,\ [\bar M_{n}{}^{i_1i_2 } , {\cal {\alpha }}_{m}{}_{j} ]=0 ,\ 
$$
$$
[\bar M_{n}{}_{i_1i_2 } , \bar {\alpha }_{m}{}_{j} ]= -2 \eta _{j [i_1}\bar  {\alpha }_{n+m}{}_{i_2]} ,\ [M_{n}{}^{i_1i_2 } , \bar {\alpha}_{j} ]=0 ,\ 
\eqno(4.3.7)$$
and 
$$
[V_{n}{}_{i_1i_2 } ,  {\alpha}_{m}{}_{j} ]= -2 \eta _{j i_1}  \bar  {\alpha}_{n+m}{}_{i_2} ,\ [ V_{n}{}_{i_1i_2 } , \bar {\alpha}_{m}{}_{j} ]=-2\eta_{ji_2 } {\alpha}_{n+m}{}_{i_1} ,\ 
$$
$$
[\bar V_{n}{}_{i_1i_2 } ,  \bar {\alpha}_{m}{}_{j} ]= -2 \eta _{j i_1} {\alpha}_{n+m}{}_{i_2} ,\ [\bar V_{n}{}_{i_1i_2 } ,  {\alpha}_{m}{}_{j} ]=-2\eta_{ji_2 } \bar {\alpha}_{n+m}{}_{i_1} 
\eqno(4.3.8)$$
\par
At first sight we can apply these transformations to the physical states involving the bosonic oscillators and then conclude  that these states which have the above bosonic oscillators carry a representation of the string little algebra $SO(D-2,D-2)^+$. However, the transformations of equation (4.3.7) and (4.3.8) do not respect the commutators of equation (4.3.6)  and so it is inconsistent to apply the above transformations of the oscillators. Nonetheless this discussion will be useful in the case for the actual superstring in section six which has a significantly different little algebra.

%%%%%%%%%%%%%%%%%%%%%%%%%%%%%%%%%%%%%%%%%%%%%%%%%

\eject

\medskip
{{\bf 5. The string little  algebra at all levels}}
\medskip

In this section we will find the generators in the string  little algebra and calculate their algebra. 
\medskip
{{\bf 5.1 The generators in the  string little  algebra}}
\medskip
In section three we found all of the generators of the IIA string little  algebra ${\cal H}$ at level zero and level one, and three of the generators at level two. These are listed in equations (3.11-3.14). While these generators appear to be  rather different they can, with two exceptions mentioned below,  all be written in the form 
$$
L^{\alpha} = e^{-R} R^{\alpha} e^{R} + I_c(e^{-R} R^{\alpha} e^{-R}) \  , \ \ \ R = {1 \over 2} \varepsilon_{ab} R^{ab}\ \ . \eqno(5.1.1)$$
Here $\alpha$ is a positive root of $E_9$  and so the generator $R^\alpha$   has SL(10)  indices which only take the values  $i, j,... = 2,...,9$. Equation (5.1.1) can be re-written with the  action of $I_c$ in  the second term carried out to find 
$$
L_{\alpha} = e^{-R} R^{\alpha} e^{R} + e^{+\overline{R}} \overline{R}_{\alpha} e^{-\overline{R}} \ \ , 
\eqno(5.1.2)$$
where $\overline{R} = - I_c(R)  = {1 \over 2} \varepsilon^{ab} R_{ab}$ and $\overline{R}_{\alpha} = I_c(R^{\alpha}) = \pm R_{\alpha}$. Which  sign it is  in this  last equation can be read off  from the action of $I_c$ given in equations (2.7) to (2.9). 
 \par 
The reader  can, for example,  check that the generators  $L_i^{(1)}$ and $L_{i_1 i_2 i_3}^{(1)}$ in equations (3.12) can be written in the form of equation (5.1.1); 
$$
L_i^{(1)} = e^{-R} R^i e^{R} + I_c(e^{-R} R^i e^{R}) = S_i + {1 \over 2} \varepsilon^{a_1 a_2} S_{a_1 a_2 i} \ ,  
\eqno(5.1.3)$$
$$
L_{i_1 i_2 i_3}^{(1)} = e^{-R} R^{i_1 i_2 i_3} e^{R} + I_c(e^{-R} R^{i_1 i_2 i_3} e^{R} )= S_{i_1 i_2 i_3} + \varepsilon^{a_1 a_2} S_{a_1 a_2 i_1 i_2 i_3} \ .  
\eqno(5.1.4)$$
 The remaining generators in (3.12-3.15)  at levels one and two can also be shown to be of the form of equation (5.1.1). 
 \par
The IIA level zero generators in ${\cal H}$ which were given in equation (3.15) and so they are well understood but we now comment on how they fit into the framework of equation (5.1.1).  The generators $J_{ij}$ and $S_{ij}$ commute with $R$ and  $\overline{R} $ and,   by taking $R^\alpha$  to be the  positive root  $E_9$ generators $K^i{}_j, \ j > i $ and $R^{ij11} $ respectively, we see that they are of the form of equation (5.1.1). The   $J_{ij}$ and $S_{ij}$  generate $SO(8)\otimes SO(8)$. The generators $L_{\alpha}$ at  a given IIA level belong to a representation of this algebra . The generators $J_{ij}$ transform the indices  $i,j,\ldots =2,\ldots ,9$ indices on the generators  in the expected way while the generators $S_{ij}$ transform the different generators at the same IIA level into each other. Thus the generators of equation (5.1.1) at a given IIA level belong to a representation of $SO(8)\otimes SO(8)$.
 \par
  The generators $J_{ab}$ and $S_{ab}$  belong to ${\cal H}$ and have the algebra  $SO(2)\otimes SO(2)$. The generator $J_{ab}$ commutes with $R$ and  $\overline{R} $ and so they can also be written in the form of  equation (5.1.1). 
  It just transforms any $a,b,\ldots $ indices on the generators but such generators only occur at IIA level zero and their commutators are given in section three. The generator $S_{ab}$  is contained in $S={1\over 2} \epsilon_{ab}S^{ab}=R-\bar R$. Using the equations later in this section it can be shown that $[S, L_\alpha]=- lL_\alpha$ if $R^\alpha$ has level $l$. 
  As such it is consistent to take the generators $J_{ab}$ and $S_{ab}$ to be trivially realised.  
 \par
 The IIA level zero generators $L^{(0)}_{ci}$  given in equation (3.11) are in ${\cal H}$,  are  of the form of equation (5.1.1) as they can be written as  
 $$
 L^{(0)}_{ci}= J_{ci}+\epsilon _c{}^e S_{ei}= e^{-R} (-\hat K^i{}_c)  e^{R} + I_c(e^{-R} (-\hat K^i{}_c) e^{R})
 \eqno(5.1.5)$$
 However, the generator $-\hat K^i{}_c$  does not belong to $E_9$ and so it is an exception to the formulation below equation (5.1.1).  We will show later in this paper that these  generators commute with themselves and also all higher level generators of the string little algebra. As such we can take them to be trivially realised and so  they will also play no role in our further discussions  on the string little algebra.  Hence all the generators  in the string little algebra of interest are indeed of the form of equation (5.1.1). 
\par
 The generator $L_{\alpha}$ of equation (5.1.1) consists of two parts, namely the first and second terms.  If the generator $ R^{\alpha} $ of $E_9$ has IIA  level $l\ge 1$ then  all the terms in the first part  will have level $l$. This follows from the fact that  $R$ has level zero and so its commutator  with $R^{\alpha}$ also has  level $l$. The second part in equation (5.1.1)  will then have level $-l$ by a similar argument.   Thus a generator $L^{\alpha} $,  as given in equation (5.1.1), has one term of level $l $ and another term  of level $-l$. We will refer to such a  generator as being of levels $\pm l$. Since the generator $R$ is inert under commutators with the SL(8) generators of $E_{9}$ all the contributions that appear in the first part  of $L^{\alpha} $ have the same SL(8) character and similarly for the second piece.  
 \par
 At first sight the expression of equation (5.1.1) involves an infinite number of terms as it involves  more and more commutators with $R$, or $\overline{R}$,  coming form the exponentials. However, these  series always terminate. We  previously found the  IIA levels $\pm 1$  generators in ${\cal H}$  that we  found    in equations (4.1.3) and (4.1.4) had only  two terms and so these  series terminate after only one commutator.  While the IIA level $\pm 2$  generators in ${\cal H}$ that we previously found had three terms. We will prove below in section (5.4)  that the IIA levels $\pm l$ generators in ${\cal H}$ have only $l+1$ terms. 
\par 
We now show that generators of the form  of equation (5.1.1),  with the $R^\alpha$ as specified,  belong to  the string little algebra ${\cal H}$. For this to be true  they must correspond to transformations  of  the brane charges which vanish when only the brane charges for the string are non-zero.   The non-zero string brane charges  are the level one brane charges $P_a$ and $Q^a$ which obey the relation of equation (3.1).  To be more precise a generator $L_{\alpha}$ belongs to  ${\cal H}$ if the commutator $[L_{\alpha},Z^{\beta}]$, where  $Z^{\beta}$  is any  any brane charge,  vanishes when we take the brane charges to which it equals to be those for the string.  
\par
The commutator $[L_{\alpha},Z^{\beta}]$ is of  the form 
$$
[L_{\alpha},Z^{\beta}]= [e^{-R} R^{\alpha} e^R,Z^{\beta}]  + [e^{\overline{R}} \overline R_{\alpha} e^{-\overline{R}},Z^{\beta}] 
\eqno(5.1.6)$$
If $Z^{\beta}$ is an element  in the $l_1$ representation  with  level $\tilde{l}=1,2,\dots $and  $L_{\alpha}$ is a generator with  levels $\pm l$  then the first term in equation (5.1.6) has level $\tilde l+l$. This has a level which greater than one and as such  it can not be equal to the  string brane charges,  which  are the only non-zero charges,  Hence it must vanish. We are therefore left with the second term of equation (5.1.5). This has level  $\tilde l-l $ and it will only  be non-zero if $\tilde l-l =0.$ in which case it must  be of the form 
$$
[e^{\overline{R}} \overline{R}_{\alpha} e^{-\overline{R}},Z^{\beta}] = \lambda^a P_a + \mu_b Q^b = ( \lambda^b- \mu_a  \varepsilon^{ab})  P_b
\eqno(5.1.7)$$
where $\lambda^a$ and $\mu_b$ are numbers and we have used equation (3.1) in the last step. 
\par 
 The brane charge  $Z^{\beta}$ has it's $SL(8) \otimes SL(2)$ indices which are up and  consist of the SL(8) indices   $i,j,\dots =2,\ldots 9$ and  SL(2) indices $a,b, \ldots=0,1$. However  the $E_{9}$ negative level generator $ \overline{R}_{\alpha} $  has only  SL(8) indices $i,j,\dots =2,\ldots 9$. which are down.    The  $E_{11}$ commutators respect the up and down positions  of the $SL(8) \otimes SL(2)$  indices carried by  the generators. Indeed  the  result of the commutator  $ [\overline{R}_{\alpha} , Z^{\beta}] $ is equal to a delta symbol relating the up and down indices times a brane charge which has its indices up. It follows that this commutator can only result in the brane charge $Q^a$ and not the brane charge $P_a$ as the latter world  index is a down index and this is not present on the left-hand side of this commutator.  In this case  the $\beta $ indices on $Z^{\beta}$ must consist of $i, j,\ldots =2,\ldots , 9$ indices, denoted $\beta'$, and one index $a=0,1$.  The $i, j, \ldots =2,\ldots , 9$ indices on $Z^{\beta}$ form a delta symbol with the $i, j, \ldots =2,\ldots , 9$ indices on $\overline{R}_{\alpha}$, denoted $\delta^{\beta'}_\alpha$, and the $a$ index is  inherited onto $Q^a$. Using these arguments we find  that the commutator between $\overline{R}_{\alpha}$ and $Z^{\beta}$ takes the form 
$$
[\overline{R}_{\alpha},Z^{\beta}] = f \delta_{\alpha}^{\beta'} Q^a
 \eqno(5.1.8)$$ 
where $f$ is a number.  
\par
 The generator $R=-R^{0111}$ where $R^{0111}$ is the generator in the eleven dimensional theory, while 
$\overline{R}= -R_{0111}$. It follows, using the previous type of arguments,  that $ [\overline{R},Z^{\beta}] = 0$ for  the $Z^{\beta}$  discussed above which gave a non-zero commutator with  $\overline{R}_{\alpha}$ when only the string brane charges were non-zero. Using this result we can re-write (5.1.7) as
$$
e^{\overline{R}} [\overline{R}_{\alpha},Z^{\beta}]e^{-\overline{R}} = e^{\overline{R}} f \delta_{\alpha}^{\beta'} Q^a e^{-\overline{R}} =  f \delta^{\alpha}_{\beta'} (Q^a + \varepsilon^{ab} P_b) =0
 \eqno(5.1.9)$$
where we used the commutators in equation (A.7)  in the appendix and,  in the last step,  we used the relation between the string charges $P_a$ and $Q^a$ given in equation (3.1). 
\par
Thus we find that a generator $L_{\alpha}$ of equation (5.1.1) obeys the relation 
$$
[L_{\alpha},Z^{\beta}] = 0. 
\eqno(5.1.10)$$  
for all brane charges $Z^{\beta}$ when only the string charges are non-zero after the commutator has been evaluated. It follows that $L_{\alpha}$ corresponds to transformations of the brane charges that preserve the string charges and and so $L_{\alpha}$ belongs to the string little algebra ${\cal H}$. We observe that this is an all orders result in the context of $E_{11}$. This does not show that all generators in ${\cal H}$ are of the form of (5.1.1). However, we believe this is  the case as it is true at IIA levels one and two.  
\medskip
{{\bf 5.2 The commutators  of the  string little  algebra}}
\medskip
In this section we will derive  the commutators of generators of the string little algebra, that is, generators of the form of equation (5.1.1), or equivalently equation (5.1.2), for the $R^\alpha$  with $\alpha$ a positive root of $E_9$. Using equation (5.1.2) the commutator of $L_{\alpha}$  with $L_{\beta}$ can be written, as
$$
[L_{\alpha},L_{\beta}] = e^{-R} [R^{\alpha} , R^{\beta}] e^R  + e^{\overline{R}} [\overline{R}_{\alpha} , \overline{R}_{\beta}] e^{-\overline{R}} 
$$
$$
+ [e^{-R} R^{\alpha} e^R , e^{\overline{R}} \overline{R}_{\beta} e^{-\overline{R}}] + [e^{\overline{R}} \overline{R}_{\alpha} e^{-\overline{R}},e^{-R} R^{\beta} e^R ] .
 \eqno(5.2.1)$$
\par
If $R^\alpha$ and $R^\beta$ have IIA  levels  $l$ and $l'$ then the first, second , third and fourth terms in equation (5.2.1) have IIA levels $l+l^\prime$, $-l-l^\prime$, $l-l^\prime$ and $-l+l^\prime$ respectively. 
We will show below that the sum of the  third and fourth terms of  levels $l-l^\prime$ and $-l+l^\prime$ respectively vanish. We will now refer to these terms as {\bf cross terms}. This will be proved in the next section but we will assume it to be the case for the rest of this section. As such   the commutator of equation (5.2.1) is equal to 
$$
[L_{\alpha},L_{\beta}] = e^{-R} [R^{\alpha} , R^{\beta}] e^R  + e^{\overline{R}} [\overline{R}_{\alpha} , \overline{R}_{\beta}] e^{-\overline{R}} =  e^{-R} [R^{\alpha} , R^{\beta}] e^R  +I_c( e^{-R} [R^{\alpha} , R^{\beta}] e^R  )
 \eqno(5.2.2)$$
\par
The commutator of  two positive level generators $R^{\alpha}$ and $ R^{\beta}$ of $E_9$ leads to another such positive level generator and so we may write it as 
$$
[R^{\alpha} , R^{\beta}] = f^{\alpha \beta}{}_{\gamma} R^{\gamma}
 \eqno(5.2.3)$$
It follows, using the properties of the $I_c$ involution,  that the commutator of  $L^{\alpha}$ and $L^{\beta}$ is given by 
$$
[L^{\alpha} , L^{\beta}] = f^{\alpha \beta}{}_{\gamma} L^{\gamma}
 \eqno(5.2.4)$$
Thus the commutator of two elements of the string algebra is  of  the form given in  equation (5.1.1) and so it also belongs to  $I_c(E_{11})$ and  to  the string little algebra ${\cal H}$.  This shows that the string little algebra of equation (5.1.1) does indeed form a closed algebra. 
\par
 Since the generators are of the form of equation (5.1.1), that is, they are convoluted with $R$,  or $\overline R$, it is easy to compute their commutators. For example, for the generators of equations (5.1.3) and (5.1.4) we readily find that 
$$
[L_{i}^{(1)},L_{j}^{(1)}] = e^{-R}[R^i,R^j]e^{R} + I_c(e^{-R}[R^i,R^j]e^{R}) = 0  \ ,
\eqno(5.2.5)$$
$$
[L_{i}^{(1)},L_{j_1 j_2 j_3}^{(1)}] = e^{-R}[R^i,R^{j_1 j_2 j_3}]e^{R} + I_c(e^{-R}[R^i,R^{j_1 j_2 j_3}]e^{R}) = 0 
 \eqno(5.2.6)$$
and
$$
[L_{i_1 i_2 i_3}^{(1)},L_{j_1 j_2 j_3}^{(1)}] =  e^{-R}[R^{i_1 i_2 i_3},R^{j_1 j_2 j_3}]e^{R} + I_c(e^{-R}[R^{i_1 i_2 i_3},R^{j_1 j_2 j_3}]e^{R}) $$
$$
= 2 e^{-R} R^{i_1 i_2 i_3 j_1 j_2 j_3} e^{R} + 2 I_c(e^{-R}R^{i_1 i_2 i_3 j_1 j_2 j_3}e^{R}) 
= 2 L_{i_1 i_2 i_3 j_1 j_2 j_3}^{(2)} 
 \eqno(5.2.7)$$
A more complicated example is given by 
$$
[L_{i_1 i_2 i_3}^{(1)},L_{j_1 \ldots j_5}^{(1)}] =  e^{-R}[R^{i_1 i_2 i_3},R^{j_1 \ldots j_5}]e^{R} + I_c(e^{-R}[R^{i_1 i_2 i_3},R^{j_1 \ldots j_5}]e^{R}) $$
$$
= e^{-R} R^{i_1 i_2 i_3j_1 \ldots j_5}e^{R} + I_c(e^{-R} R^{i_1 i_2 i_3j_1 \ldots j_5}e^{R} ) 
$$
$$
- 5 e^{-R} R^{i_1 i_2 i_3[j_1 \ldots  j_4,j_5]}e^{R}  - 5 I_c(e^{-R} R^{i_1 i_2 i_3[j_1 \ldots  j_4,j_5]}e^{R}) 
= L_{i_1 i_2 i_3 j_1 \ldots j_5}^{(2)} - 5 L_{i_1 i_2 i_3 [j_1 \ldots j_4,j_5]}^{(2)} 
 \eqno(5.2.8)$$

%%%%%%%%%%%%%%%%
\medskip
{{\bf 5.3 The vanishing of the cross terms}}
\medskip
In this section we will show that the cross terms, that is, the level $l-l^\prime$ and level $-l+l^\prime$ terms in the commutator 
of equation (5.2.1) vanish.   In section three we computed the algebra of the level $\pm 1$ generators of the string little algebra and we did indeed find that the  cross terms do vanish, see equations (3.16).  We recall that in these sections  we are no longer considering  the level zero generators in the string little algebra unless stated to the contrary. To show the cross terms vanish in general we will need a stream lined notation. We use the notation 
$$
1 \wedge B = B, \ \ A \wedge B = [A,B], \ \ A^{(2)} \wedge B = [A,[A,B]], \ \ {\rm etc...}
\eqno(5.3.1)$$
 so  that 
 $$
e^{-R} R^{\alpha} e^R = e^{-R} \wedge  R^{\alpha}= \sum_p R^p\wedge R^{\alpha} 
\eqno(5.3.2)$$
In fact this sum terminates when $p=l$ as we will show in section 5.4 that 
$$
 R^{l+1} \wedge R^{\alpha} = 0 \ \ \ , \ \ \  \overline{R}^{l'+1} \wedge \overline{R}_{\beta} = 0. 
 \eqno(5.3.3)$$
if $R^{\alpha} $ and $ \overline{R}_{\beta}$ have levels $l$ and $l^\prime$ respectively. The second equation  is just the Cartan involution of the first equation. As such the sum in equation (5.3.2) terminates when $p=l$. 
\par 
Using the above notation and the above result the first cross term in equation (5.2.1) can be written as 
$$
[e^{-R} R^{\alpha} e^R , e^{\overline{R}} \overline{R}_{\beta} e^{-\overline{R}}]  = \sum_{p=0}^{\ell} \sum_{q=0}^{\ell'} {(-1)^p \over p!} {1 \over q!} [X^{(p)},Y^{(q)}] 
\eqno(5.3.4)$$
where 
$$
X^{(p)} \equiv  R^{p} \wedge R^{\alpha} \ \ , \ \ Y^{(p)} \equiv \overline{R}^{p} \wedge R_{\beta} \ \ .  
\eqno(5.3.5)$$
\par
To evaluate the expression of equation (5.3.4) we will need some identities. We define 
$$
\tilde{D} = K^0{}_0+    K^1{}_1 = \hat   K^0{}_0+   \hat   K^1{}_1+{1\over 3} \tilde R= {2\over 3} ( \hat   K^0{}_0+   \hat   K^1{}_1+\hat K^{11}{}_{11} ) - {1\over 3} \sum _{a=2}^9 \hat K^a{}_a
\eqno(5.3.6)$$
where we have used the correspondence with the generators in the eleven dimensional theory given in equations (2.2). We recall that a hat refers to generators as they appear in the eleven dimensional theory. This generator obeys the following commutators 
$$
[R,\overline{R}] = - \tilde{D} \ \ , \ \ [\tilde{D},R] = + 2 R \ \ , \ \ [\tilde{D},\overline{R}] = - 2 \overline{R} \ \ .
 \eqno(5.3.7)$$
It can be shown that 
$$
[\tilde{R},R^{\alpha}] = - 3 l R^{\alpha} \ \ , \ \ [\tilde{R},\overline{R}_{\beta}] = + 3 l' \overline{R}_{\beta} \ \ . 
\eqno(5.3.8)$$
where $R^{\alpha} $ is an $E_{11}$ generator with IIA positive  level $l\ge 0$ and $\overline{R}_{\beta} $ is a an $E_{11}$ generator with negative level $-l^\prime \le 0$. 
It follows from equation (5.3.6) that $$
[\tilde{D},R^{\alpha}] = - l R^{\alpha} \ \ , \ \ [\tilde{D},\overline{R}_{\beta}] = + l' \overline{R}_{\beta} \ \ .
 \eqno(5.3.9)$$
as there generators have no $0$ or $1$ indices as they belong to $E_9$. Using the Jacobi identity and equation (5.3.9) we find that 
$$
[\tilde D , Y^{(p)} ]= [\overline {R} , [\tilde D , Y^{(p-1)}]] -2  [\overline {R} ,  Y^{(p-1)}] 
= (l^\prime -2p ) Y^{(p)}
 \eqno(5.3.10)$$
Using the Cartan involution,  in particular $I_c(\tilde D)= -\tilde D$ and  $I_c(Y^{(p)} )= (-1)^p X^{(p)}$, on this last equation we find that 
$$
[\tilde D , X^{(p)} ]= (l -2p ) X^{(p)}
 \eqno(5.3.11)$$
One can show that 
$$
[R,Y^{(p)}] = - p (l' - (p-1)) Y^{(p-1)} \ \ {\rm and }\ \ [\overline{R},X^{(p)}] = - p (l - (p-1))X^{(p-1)} 
\eqno(5.3.12)$$
To derive the first equation we commute $R$ through the $\overline {R}$'s in $Y^{(p)}$ using equation (5.3.11) and then use equation $[R, \overline {R}_\beta ]=0$. Using similar arguments one can show that 
$$
R^{(q)} \wedge Y^{(p)} = (-1)^q {p!\over (p-q)!}   {( l'-(p-q))!\over  (l' - p) !}Y^{(p-q)} 
\eqno(5.3.13)$$
and
$$
\overline{R}^{(q)} \wedge X^{(p)} =  (-1)^q {p!\over (p-q)!}   {( l-(p-q))!\over  (l - p) !}X^{(p-q)} 
\eqno(5.3.14)$$
\par
Finally, using the Jacobi identities and equation (5.3.12)   one can  show two  recursion relations for $[X^{(p)},Y^{(q)}]$ ; 
$$
[X^{(p)},Y^{(q)}] = [R,[X^{(p-1)},Y^{(q)}]] + q (l' - (q-1)) [X^{(p-1)},Y^{(q-1)}]
 \eqno(5.3.15)$$
and
$$
[X^{(p)},Y^{(q)}] = [\overline{R},[X^{(p)},Y^{(q-1)}]] + p (l - (p-1)) [X^{(p-1)},Y^{(q-1)}] 
\eqno(5.3.16)$$
\par
Finally we can evaluate the object of interest,  the commutator of equation (5.3.4). We will do this for  a level one generator with a level $l'$ generator. Using equation (5.3.3) this commutator becomes 
$$
[e^{-R} R^{\alpha} e^R , e^{\overline{R}} \overline{R}_{\beta} e^{-\overline{R}}]  = \sum_{p=0}^1 \sum_{q=0}^{\ell'} {(-1)^p \over p!} {1 \over q!} [X^{(p)},Y^{(q)}] 
\eqno(5.3.17)$$
Using the recursion relation (5.3.15) we find that the terms involving $X^{(1)}$ can be evaluated as 
$$
- {1 \over (q + 1)!}[X^{(1)},Y^{(q+1)}] = - {1 \over (q + 1)!}\{ [R,[R^{\alpha},Y^{(q+1)}]] + (q + 1)(l' - q)[R^{\alpha},Y^{(q)}] \} $$
$$
 = - {1 \over (q + 1)!} \{R \wedge \overline{R}^{(q+1)} \wedge [R^{\alpha},\overline{R}_{\beta}] + (q + 1)(l' - q)[R^{\alpha},Y^{(q)}] \}$$
$$
= - {1 \over (q + 1)!} \{ [ - (q + 1)(l' - 1 - q') + (q + 1)(l' - q )][R^{\alpha},Y^{(q)}] \} = - {1 \over q!} [R^{\alpha},Y^{(q)}] \} . 
\eqno(5.3.18)$$
This term  cancels the ${1 \over (q)!}[X^{(0)},Y^{(q)}] $ term in equation (5.3.17). Carrying out all such cancelations we find that 
 only the $p = 0$, $q = l'$ term remains, and so 
 $$
[e^{-R} R^{\alpha} e^R , e^{\overline{R}} \overline{R}_{\beta} e^{-\overline{R}}]  = 
{1 \over l'!}[R^{\alpha},Y^{(l')}] = {1 \over l'!}[R^{\alpha},\overline{R}^{l'} \wedge \overline{R}_{\beta}]  = {1 \over l'!} \overline{R}^{l'} \wedge [R^{\alpha},\overline{R}_{\beta}] = 0.
 \eqno(5.3.19)$$
In the above we have used the  Jacobi identity, as well as the relation $[R^{\alpha},\overline{R}] = 0$. The second to last step in the above equation contains the generators  
$R^{\alpha}$ of  level $ 1$ and $\overline{R}_{\beta}$ of  level $-l'$ and so the commutator $[R^{\alpha},\overline{R}_{\beta}]$ is at level $-(l' - 1)$. Using equation (5.3.3) we find that the final expression vanishes. The same result holds for the other cross term in equation (5.2.2) as it is the Cartan involution of the term we just showed vanished. 
\par
Thus we have shown that  the cross terms vanish for $[L_{\alpha},L_{\beta}]$ when $L_{\alpha}$ is at levels $\pm 1$  and $L_{\beta}$ is at levels $\pm l'$. However, from the commutators of these  two elements of the string little algebra we can find all elements of this algebra and as a a result show that their commutators  have  no cross terms. The commutator of any  two generators $L_\alpha$ and $L_\beta$ can be written as 
$$
[L_\alpha , L_\beta ]= [L_\gamma , L_{\alpha^{\prime} }], L_\beta ]= [L_\gamma , [ L_{\alpha^{\prime} }, L_\beta ]]
+ [ [L_\gamma ,  L_\beta ] , L_{\alpha^{\prime} }]
 \eqno(5.3.20)$$
 where $L_\alpha$ has been given by $L_\alpha= [L_\gamma , L_{\alpha^{\prime}} ]$ where $L_\gamma$ has level $\pm1$. The generator $ L_{\alpha^{\prime}} $ has a level which is one less than $L_\alpha$. Proceeding in this way we can write the commutator $[L_{\alpha},L_{\beta}]$  in terms of commutators of multiple  level one generators and one other element of the string little algebra. From the result just above it follows that the commutator has no cross terms. 
\par
The reader may like to show directly that  the commutator of two generators in the string little algebra that have any   levels have no cross terms. A useful identity is given by 
$$
[X^{(p)},Y^{(q)}] =\cases {{p! \ l!\over (l-p)!}  [R^{\alpha},\overline{R}_{\beta}] ,   \quad {\rm if \ l^\prime > l }\cr
{p!\  l^\prime!\over (l^\prime- p)!}  [R^{\alpha},\overline{R}_{\beta }] , \quad {\rm if \ l > l ^\prime} \cr}
\eqno(5.3.21)$$
While evaluating the cross terms of the commutator $ [X^{(p)},Y^{(q)}] $ for general $p$ and $q$  directly, rather than using the above use of level $\pm 1$ generators, one has to be careful when the commutator $[R^{\alpha},\overline{R}_{\beta}]$ not only has IIA level zero but also $E_{11}$ level zero, that is, when it equals a combination of the generators $\hat K^i{}_j$ and 
$\hat K^{11}{}_{11}$. The problem is that these latter generators do not commute with either $R$ or $\overline R$. However keeping such terms in the calculation one finds as already shown that the cross terms vanish. 
\par
There is one important exception to the fact that the  cross terms in the commutators of the generators of the string little algebra vanish, namely for those generators that are at level zero. These are the generators 
$$
J_{ij}=\eta_{ik} K^k{}_j -\eta_{jk} K^k{}_i ,\  S^{ij}= \hat R^{ij11}- \eta^{ik}\eta^{jl} \hat R_{kl11} ,\ 
L_{ai}^{(0)} = -e^{-R} \hat K^k{}_a  e^{R} \eta _{ki}+ e^{\overline R} \hat K^d{}_i e^{-\overline R}\eta_{ad}
\eqno(5.3.22)$$
The cross terms in the commutators of these generators are non-zero . The generators $J_{ij}$ and $S^{ij}$ obey the algebra of $SO(8)\otimes SO(8)$ while the commutators of the generators $L_{ai}^{(0)} $  with $J_{ij}$ and $S^{ij}$ give again $L_{ai}^{(0)} $, see equation (3.16). 
\par
As we will now show the level zero generator $L_{ai}^{(0)} $ commutes with all the level greater than zero generators in the string algebra. The general commutator is of the form 
$$
[L_{ai}^{(0)} , L_\beta ]=
[  X_{ia} , e^{-R} R^\beta  e^{R}]+
[   X_{ia} , e^{\overline R} \overline R_\beta e^{-\overline R} ]
+ I_c ( [  X_{ia} , e^{-R} R^\beta  e^{R}]+[   X_{ia} , e^{\overline R} \overline R_\beta e^{-\overline R} ])
\eqno(5.3.23)$$
where $X_{ia}\equiv   -e^{-R} \hat K^k{}_a  e^{R} \eta _{ki}$
The first term has the simple form 
$$
 -e^{-R} [\hat K^k{}_a   ,  R^\beta ]e^{R}=0 
 \eqno(5.3.24)$$
It is zero as $R^\beta  $ belongs to $E_9$, with $\beta$ being a positive root,  which means that it possesses  no $a,b$ indices  and as a result  it commutes with $ \hat K^k{}_a $.
The second term can be treated much like the level one and level $l^\prime $ generators in equation (5.3.17) and evaluated as in equation (5.3.18) using  the equation $[\tilde D , \hat K^i{}_a ]= - \hat K^i{}_a$. As a result we find  very similar cancellations leaving the expression ${1\over l'!}{\overline R}^{l^\prime} \wedge [ \hat K^i{}_a , \overline {R}_\beta]\eta_{ik}=0$. One can show that this vanishes at low levels. We believe  it vanishes at all levels and we take this to be the case.  As such the generators $L_{ai}^{(0)} $ commute with all the generators in the string little group except for $J_{ij}$ and $S^{ij}$ under  which it is a representation. As such it is consistent to take the generators $L_{ai}^{(0)}$ to be trivially realised. 
\par
Thus the non-trivial part of the string little algebra ${\cal H}$ contains at level zero $SO(8)\otimes SO(8)$ and at other levels the generators are in one to one correspondence with the positive root generators of $E_9$. They  also obey the same algebra as these generators. Put another way the generators have the same algebra as the Borel subalgebra of $E_9$ except for those at IIA level zero which have the algebra of $SO(8)\otimes SO(8)$. 
This little algebra  is quite different to  the little algebra found for the massless point particle which was found to be $I_c(E_9)$ [14]. Although the generators $L_\alpha$  given in   equation (5.1.1) contain the expression $R^\alpha +I_c(R^\alpha)$,  which is in $I_c(E_9)$, they also contains commutators with $R$ and $\bar R$ and as a result  $L_\alpha$  belong to $I_c(E_{11})$ rather than $I_c(E_9)$. Indeed it is this latter point that changes the commutators from those of $I_c(E_9)$ to those of the Borel subalgebra of $E_9$.  Thus although the commutators are the same as those of the Borel subalgebra of $E_9$ the generators   are not the Borel generators of the $E_9$ that is evident from the Dynkin diagram of $E_{11}$. 

\medskip
{{\bf 5.4 Proof of  $R^{l+1} \wedge R^{\beta} = 0$}}
\medskip
In this subsection we will  prove the relation
$$
R^{l+1} \wedge R^{\beta} = 0\ \  
\eqno(5.4.1)$$
In this equation $\beta$  is  a positive root of  $E_9$ with IIA level $l$ and  $R = {1\over 2} \epsilon_{ab} \hat R^{ab\, 11}$.  This  equation was stated in equation (5.3.3) of subsection 4.3 and used to prove some of the crucial results in that section. Here  we will show that the root associated to the generator of equation (5.4.1)  is not present in the underlying $E_{11}$ algebra and so the commutators in equation  (5.4.1) must vanish. 

The generator $R = - \hat R^{12 \, 11}$ can be found by taking suitable commutators of $\hat K^a{}_b$ with the generators $R^{91011}$ and in particular 
$$
R^{12 \, 11} = \hat K^2{}_3 \wedge \hat K^3{}_4 \wedge \ldots \hat K^9{}_{10} \wedge \hat K^1{}_2 \wedge \hat K^2{}_3 \wedge \ldots \wedge \hat K^8{}_9 \wedge R^{9 10 \, 11} 
\eqno(5.4.2)$$
In agreement with much of the previous literature   we will be somewhat contrary-wise and take the indices to run from $1,2,\ldots , 11$ rather than $0,1,\ldots ,10$. The root $\alpha_{11}$ is associated to $\hat R^{9 10 \, 11}$, while the roots   $\alpha_1$  and $\alpha_2$  to $\hat K^1{}_2$ and $\hat K^2{}_3$ respectively etc It follows that the root associated to $R^{1 2 \, 11}$ is
$$
\delta = (\alpha_1 + 2 \alpha_2 + 2 \alpha_3 + \ldots + 2 \alpha_8 + \alpha_9) + \alpha_{11} 
\eqno(5.4.3)$$
It is straight forward to show that  $\delta^2 = 2$ which must be the case as it belongs to the subalgebra $D_{10}$ of $E_{11}$. We note that $\delta$ possess no root $ \alpha_{10} $ and  has level 2 with respect to node 2. 
\par
We will now consider the $E_{11}$ algebra when decomposed into $A_1\otimes D_8$ as indicated by the deleted nodes in the  $E_{11}$ Dynkin diagram given by 
$$
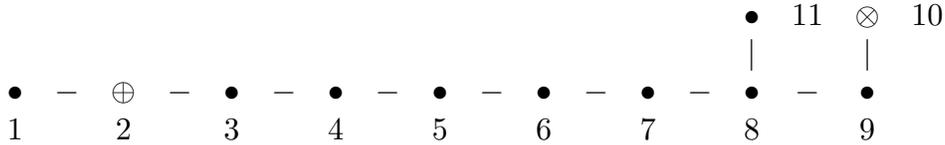
\matrix{
& & & & & & & & & & & & & & \bullet  & 11 & \otimes & 10 &  \cr
& & & & & & & & & & & & & & | & & | & & \cr
\bullet & - & \oplus & - & \bullet & - & \bullet & - & \bullet & - & \bullet & - & \bullet & - & \bullet & - & \bullet & & \cr
1 & & 2 & & 3 & & 4 & & 5 & & 6 & & 7 & & 8 & & 9 & &  \cr
}
$$
The $E_{11}$ root $\alpha$ associated to the generator $R^{s} \wedge R^{\beta}$  can  be written as
$$
\alpha = \epsilon + 2s \alpha_2 + \gamma+ l  \alpha_{10} 
\eqno(5.4.4)$$
Here $\epsilon $ is a root of the Lie algebra $A_1$ which is associated with node 1, while $\gamma$ is a root of $D_8$ which is associated with nodes 3,4,5,6,7,8, 9 and 11. The roots $\alpha_2$ and $\alpha_{10}$ are the simple roots of $E_{11}$ and their coefficients $2s$ and $l$ correspond to the fact that the root has levels $2s$ and $l$ corresponding to nodes 2 and 10 respectively. We recall that node ten is the node that we delete to find the ten dimensional IIA theory  and so level $l$ relates to the fact that $R^\beta $ has IIA level $l$. 
\par
By construction $R^\beta $ belongs to $E_9$ which is associated with nodes 3-10 and node 11 and it has, by assumption,  IIA level $l$., However, it  does not contain the root $\alpha_2$ and so it has level zero for this node. In contrast we see from equation (5.4.3) that $R$ has level two with respect to node 2 but  it has IIA level zero, Consequently $R^s$ has level $2s$ and 0 with respect to nodes 2 and 10. As such the root of equation (5.4.4), which corresponds to $R^{s} \wedge R^{\beta}$,  has levels $2s$ and $l$ with respect to nodes 2 and 10. 
\par
Following, for example,  the review of chapter 16 of [19] we can analyse the $A_1\otimes D_8$ representations that occur in the roots of the form of equation (5.4.4). The first step is to write the roots $\alpha_2$ and $\alpha_{10}$ in the form 
$$
\alpha_2 = y - \nu_1 - \mu \ \ \ \ \, \ \ \ \ \alpha_{10} = x - \nu_7
\eqno(5.4.5) $$
where $\mu$ is the  fundamental weight of $A_1$ and $ \nu_1$ and $ \nu_7$ are the fundamental weights  of $D_8$  corresponding to nodes 3 and 9 in the above Dynkin diagram respectively. The quantities $x$ and $y$ are orthogonal to all the roots, and so the weights,  of  $A_1$ and $D_8$. 
These roots must have length squared two and so 
$$
 \alpha_2^2 = y^2 + \nu_1^2 + \mu^2 = y^2 + 1 + {1 \over 2}=2 \ \ \ {\rm and \ so } \ \ \ y^2 = {1 \over 2} 
\eqno(5.4.6)$$
and 
$$
 \alpha_{10}^2 = x^2 + \nu_7^2= x^2+2=2 \ \ \  {\rm and \ so }  \ \ \ x^2 = 0 
\eqno(5.4.7)$$
In evaluating these equations we used the scalar products of the fundamental weights which can be found, for example, in 
 appendix D of [19].  It is straight forward to check that they have the correct scalar products with the other $E_{11}$ roots with one exception, namely $ \alpha_2\cdot \alpha_9$ which requires more care. We have that 
$$
\alpha_2\cdot \alpha_9= \nu_1\cdot \nu_7+x\cdot y={1\over 2}+x\cdot y=0 \ \ \  {\rm and \ so }  \ \ \ x\cdot y = -{1\over 2} 
\eqno(5.4.8)$$
This novel feature arises as we have deleted two nodes. 
\par
Substituting these expressions for the roots $\alpha_2$ and $\alpha_{10}$ in the root $\alpha$ of equation (5.4.4) is given by 
$$
\alpha =\epsilon - 2s\mu +2sy  + \gamma -2s\nu_1- l  \nu_7+lx
\eqno(5.4.9)$$
\par
We can analyse the $A_1\otimes D_8$ representations that occur in the roots of the form of equation (5.4.4). The part of the root in equation (5.4.4) in the space of the $A_1$ algebra is $\epsilon - s\mu$ and this must, for suitable $\epsilon$,  be a highest weight of $A_1$. We therefore find the equation 
$$
q\mu=2s\mu-\epsilon= 2s\mu -n\hat \alpha_1 ,\ \  {\rm and \ so } \ \  {q\over 2} =s -n ,\ \ {\rm for}\ \  n\ge 0
\eqno(5.4.10)$$
where $q$ is the Dynkin index of the highest representation and $\hat \alpha_1$ is the simple root of $A_1$. 
Proceeding in a similar way for the $D_8$ part of the algebra we find that 
$$
\sum_{j \in \{3,..,9,11\}} p_j \lambda_j =  l  \nu_7 + 2s \nu_1 -\gamma= l  \nu_7 + 2s \nu_1 -\sum_{j \in \{3,..,9,11\}} m_j\hat  \alpha_j ,\ \ {\rm for}\ \ m_j \ge 0
\eqno(5.4.11)$$
We have written $\gamma=  \sum_{j \in \{3,..,9,11\}} m_j\hat  \alpha_j $ where $\hat  \alpha_j $, $j=3,..,9,11$ are the simple roots of $D_8$ and $\lambda_j$ its fundamental weights. 
Taking the dot product of  equation (5.4.11) with $\lambda_i$ and using the definition of the inverse $D_8$ Cartan matrix $(A_{ij}^{D_8})^{-1} = \lambda_i \cdot \lambda_j$, we find
$$
\sum_{j \in \{3,..,9,11\}} p_j (A_{ji}^{D_8})^{-1}  = l \nu_7 \cdot \lambda_i + 2s \nu_1 \cdot \lambda_i - m_i 
\eqno(5.4.12)$$
By analysing  equations (4.4.11) and (4.4.12) one can find at low levels the possible representations of $A_1\otimes D_8$ that can occur.
\par
 In any  Kac-Moody algebra the lengths of the roots are bounded by $2$. Using equations (4.4.10) and (4.4.11) we find that 
$$
\alpha^2 = {q^2 \over 2} + \sum_{i,j} p_i (A_{ij}^{D_8})^{-1} p_j + {2s^2 } - 2s l \leq 2 . 
\eqno(5.4.13)$$
This will only be less than or equal to 2 if $s \le l$ holds.   This indeed proves the identity of equation (5.4.1) since the root corresponding to $R^{l+1}\wedge R^\beta$ does not exist in the $E_{11}$ algebra.

%%%%%%%%%%%%%%%%%%

\medskip
{{\bf 6 String states and an  irreducible representation of the string  little algebra }}
\medskip

We begin this section by clarifying how the string  little algebra sits in $E_{11}$ in relation to the well known subalgebras $E_9$, $E_8$ and its Cartan involution subalgebra SO(16). 

\medskip
{{\bf 6.1 Relationship of the string little algebra to $E_9$}}
\medskip

In section five  we found the subalgebra ${\cal H}$ of $I_c(E_{11})\otimes_s l_1$ that preserves the string charges. We referred to this   as the string little algebra.   It contained the generators $J_{ab}$ and  $S_{ab}$,  $a,b=0,1$, which generate  $SO(2)\otimes SO(2)$,  and the  generator $L_{ai}^{(1)}$.   It also contained the generators 
$$
{\cal H}_r= \{J_{ij},S_{ij} ; L_{\alpha} \ ; \ \alpha \in {\rm positive \ roots \ of \ } E_9 \}. 
\eqno(6.1.1) $$
where $L_{\alpha} $ is defined in equation (5.1.1). As explained in section 5.2 the generators in equation (6.1.1) obey an algebra that has the algebra $SO(8)\otimes SO(8)$ at IIA level zero but at higher levels it has commutators that  have the same algebra as the  Borel subalgebra of $E_9$.  
\par
In this paper we wish to construct the string analogue of the irreducible particle representations first constructed by Wigner [13]. To do this we must choose a representation of the string little algebra and then carry out the corresponding boost. As discussed in the previous section we may choose the generators  $J_{ab}$, $S_{ab}$ and $L_{ai}^{(1)}$ of ${\cal H}$ to be  trivially realised.  As such we just have to choose a representation of the algebra obeyed by the algebra ${\cal H}_r$ of equation (6.1.1). 
\par
To better understand the algebra ${\cal H}_r$ it will be useful to understand in detail its relation to $E_{9}$. 
The Dynkin diagram of $E_9$ is found  by deleting the first two  nodes, labelled one and two and indicated by a  $\oplus$ ,  from the Dynkin diagram of $E_{11}$: 
$$
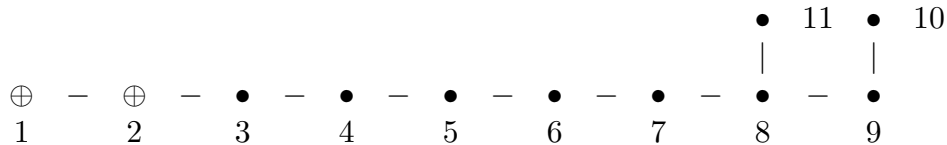
\matrix{
& & & & & & & & & & & & & & \bullet & 11 &\bullet &  10\cr
& & & & & & & & & & & & & & | & &| \cr
\oplus & - & \oplus & - & \bullet & - & \bullet & - & \bullet & - & \bullet & - & \bullet & - & \bullet & - & \bullet &  &  \cr
1 & & 2 & & 3 & & 4 & & 5 & & 6 & & 7 & & 8 & & 9 & &  \cr
}
$$
 Since we are considering the ten dimensional IIA theory we delete  node ten, called the  IIA node, in  the above Dynkin diagram. We will  decompose the  $E_9$ algebra with respect to the remaining $D_8 = {\rm SO}(8,8)$ subalgebra as shown by the diagram
$$
\matrix{
& & & & & & & & & & & & & & \bullet & 11 &\oplus &  10\cr
& & & & & & & & & & & & & & | & &| \cr
\oplus & - & \oplus & - & \bullet & - & \bullet & - & \bullet & - & \bullet & - & \bullet & - & \bullet & - & \bullet &  &  \cr
1 & & 2 & & 3 & & 4 & & 5 & & 6 & & 7 & & 8 & & 9 & &  \cr
}
$$
\par
 The low level generators in such a decomposition can be found by using Simplie [16]. Deleting node 10 we find that $E_9$  decomposes into the 120 dimensional adjoint representation of $D_8$ at level  zero and the 128 dimensional spinor representation of  $D_8$ at level one. This pattern repeats itself, at all  even levels and odd levels we find the adjoint  and spinor representations of $D_8$ respectively. In order to recognise this result in terms of generators we are familiar with we will not only delete the IIA node ten but also node eleven to result in a decomposition to  the algebra SL(8) which corresponds in the Dynkin diagram consisting of  nodes 3 to 9 inclusive.  At IIA level zero, the 120 generators which belong to the adjoint of $D_8= S0(8,8)$ is given by 
$$
K_{(0)}{}^i{}_j = K^i{}_j  \ , \ R^{i j}_{(0)} = R^{i j} \ , \ R_{(0)}{}_{i j} = R_{i j} ,\ \ \ i,j,\ldots = 3,4,\ldots , 10
\eqno(6.1.2) $$
 Here the $K^i{}_j$ generate GL(8)  and together with $R^{i j} $ and $R_{i j} $ they generate SO(8,8). 
 The subscripts here, and in the equations that follow,  denote the IIA level of the generators. At level zero only one finds an additional generator with no SL(8) indices which we have omitted to write. It is the analogue of the dilaton generator in the IIA theory. 
 \par
At  IIA level one we find the generators
$$
R^i_{(1)} = R^i \ , \ R^{i_1 i_2 i_3}_{(1)} = R^{i_1 i_2 i_3} \ , \ R^{i_1 \ldots i_5}_{(1)} = R^{i_1 \ldots i_5} \ , \ R^{i_1 \ldots i_7}_{(1)} = R^{i_1 \ldots i_7} \ , 
\eqno(6.1.3)  $$
These 128 generators belong to the Majorana-Weyl  spinor representation of $D_8$.   Similarly at level minus we find another 128 dimensional  spinor representation that is given by
$$
R_{(-1)}{}_i = R_i \ , \ R_{(-1)}{}_{i_1 i_2 i_3} = R_{i_1 i_2 i_3} \ , \ R_{(-1)}{}_{i_1 \ldots i_5} = R_{i_1 \ldots i_5} \ , \ R_{(-1)}{}_{i_1 \ldots i_7} = R_{i_1 \ldots i_7} \ , 
\eqno(6.1.4)  $$
\par
At level two we find the generators
$$
K_{(2)}{}^j{}_i = {1 \over 7!} \varepsilon_{i k_1 \ldots k_7} R^{k_1 \ldots k_7,j}  \ , \ \tilde{R}_{(2)} = {1 \over 8!} \varepsilon_{k_1 \ldots k_8} R^{k_1 \ldots k_8} \ , 
$$
$$
R^{i j}_{(2)} = {1 \over 6!} \varepsilon^{i j}{}_{k_1 \ldots k_6} R^{k_1 \ldots k_6} \ , \ R_{(2)}{}_{i j} = {1 \over 8!} \varepsilon_{k_1 \ldots k_8} R^{k_1 \ldots k_8, i j} \ , 
\eqno(6.1.5) $$
which belong to another copy of the 120 dimensional adjoint representation of $D_8$. The first and second generator 
generate GL(8). While at level minus we have the same representation 
$$
K_{(-2)}{}^i{}_j = {1 \over 7!} \varepsilon^{i k_1 \ldots k_7} R_{k_1 \ldots k_7,j}  \ , \ \tilde{R}_{(-2)} = {1 \over 8!} \varepsilon^{k_1 \ldots k_8} R_{k_1 \ldots k_8} \ ,
 $$
$$
R^{i j}_{(-2)} = {1 \over 6!} \varepsilon^{i j k_1 \ldots k_6} R_{k_1 \ldots k_6} \ , \ R_{(-2)}{}_{i j} = {1 \over 8!} \varepsilon^{k_1 \ldots k_8} R_{k_1 \ldots k_8, i j} \ , 
\eqno(6.1.6) $$
\par
At level three  we find that the generators belong  to the 128 dimensional  spinor representation of $D_8$. They are given by 
$$
R_{(3)}^i = {1 \over 8!} \varepsilon_{k_1 \ldots k_8} R^{k_1 \ldots k_8,i} \ , \ R^{i_1 i_2 i_3}_{(3)} = {1 \over 8!} \varepsilon_{k_1 \ldots k_8} R^{k_1 \ldots k_8, i_1 i_2 i_3} \ , 
$$
$$
R^{i_1 \ldots i_5}_{(3)} = {1 \over 8!} \varepsilon_{k_1 \ldots k_8} R^{k_1 \ldots k_8,i_1 \ldots i_5} \ , \ R^{i_1 \ldots i_7}_{(3)} = {1 \over 8!} \varepsilon_{k_1 \ldots k_8} R^{k_1 \ldots k_8,i_1 \ldots i_7} \ , 
\eqno(6.1.7)  $$
Similarly at  level minus three we find another spinor of $D_8$, 
$$
R_{(-3)}{}_i = {1 \over 8!} \varepsilon^{k_1 \ldots k_8} R_{k_1 \ldots k_8,i} \ , \ R_{(-3)}{}_{i_1 i_2 i_3} = {1 \over 8!} \varepsilon^{k_1 \ldots k_8} R_{k_1 \ldots k_8, i_1 i_2 i_3} \ , $$
$$
R_{(-3)}{}_{i_1 \ldots i_5} = {1 \over 8!} \varepsilon^{k_1 \ldots k_8} R_{k_1 \ldots k_8,i_1 \ldots i_5} \ , \ R_{(-3)}{}_{i_1 \ldots i_7} = {1 \over 8!} \varepsilon^{k_1 \ldots k_8} R_{k_1 \ldots k_8,i_1 \ldots i_7} \ , 
\eqno(6.1.8)  $$
\par
In section three we did not list the correspondence between the $E_{11}$ generators that appear in the eleven dimensional theory at IIA level three and the generators as they appear in the IIA  theory, that is, when the  node ten is deleted. We now give this correspondence
$$
R^{\underline{a}_1 \ldots \underline{a}_8,\underline{b}} = \hat{R}^{\underline{a}_1 \ldots \underline{a}_8,\underline{b}} \ , \ R^{\underline{a}_1 \ldots \underline{a}_8,\underline{b}_1 \underline{b}_2 \underline{b}_3} = \hat{R}^{\underline{a}_1 \ldots \underline{a}_8 11,\underline{b}_1 \underline{b}_2 \underline{b}_3} \ , 
 $$
$$
R^{\underline{a}_1 \ldots \underline{a}_8,\underline{b}_1 \ldots \underline{b}_5} = \hat{R}^{\underline{a}_1 \ldots \underline{a}_8 11,\underline{b}_1 \ldots \underline{b}_5 11} \ , \ R^{\underline{a}_1 \ldots \underline{a}_8,\underline{b}_1 \ldots \underline{b}_7} = \hat{R}^{\underline{a}_1 \ldots \underline{a}_8 11,\underline{b}_1 \ldots \underline{b}_7 11, 11} \ .\eqno(6.1.9) $$
The identifications at  level minus three are given by 
$$
R_{\underline{a}_1 \ldots \underline{a}_8,\underline{b}} = \hat{R}_{\underline{a}_1 \ldots \underline{a}_8,\underline{b}} \ , \ R_{\underline{a}_1 \ldots \underline{a}_8,\underline{b}_1 \underline{b}_2 \underline{b}_3} = \hat{R}_{\underline{a}_1 \ldots \underline{a}_8 11,\underline{b}_1 \underline{b}_2 \underline{b}_3} \ , \ $$
$$
R_{\underline{a}_1 \ldots \underline{a}_8,\underline{b}_1 \ldots \underline{b}_5} = \hat{R}_{\underline{a}_1 \ldots \underline{a}_8 11,\underline{b}_1 \ldots \underline{b}_5 11} \ , \ R^{\underline{a}_1 \ldots \underline{a}_8,\underline{b}_1 \ldots \underline{b}_5} = \hat{R}_{\underline{a}_1 \ldots \underline{a}_8 11,\underline{b}_1 \ldots \underline{b}_7 11, 11} \ .\eqno(6.1.10)  $$
\par
The pattern that emerges is obvious, at even IIA  levels $2n=0, \pm 2, \pm 4, \ldots $  the $E_9$ algebra contains  the adjoint representation of ${\rm SO}(8,8)$, which we denote by ${\rm SO}(8,8)_{2n}$, with generators
$$
K_{(2n)}{}^i{}_j \ \ , \ \ R_{(2n)}^{ij} \ \ , \ \ R_{(2n)}{}_{ij} \ \ , 
\eqno(6.1.11)$$
While at each  odd $E_9$ level, $2n+1= 1,3,\ldots $ ,  we find a Majorana-Weyl spinor representation of $D_8$, which we denote by ${\bf 128}_{2n+1}$,   with generators 
$$
R_{(2n+1)}^i \ , \ R_{(2n+1)}^{i_1 i_2 i_3} \ , \ R_{(2n+1)}^{i_1 \ldots i_5} \ , \ R_{(2n + 1)}^{i_1 \ldots i_7} \ \ , \ \ n=1, \pm 3, \pm 5, \ldots   \ .
\eqno(6.1.12)$$
We can thus write $E_9$ as 
$$
E_9 = \sum_{n \in  Z} {\rm SO}(8,8)_{2n} \oplus {\bf 128}_{2n+1} \ . 
\eqno(6.1.13)$$
\par
The algebra of $E_9$ reflects  its affine character and can be written as  
$$
[{\bf 120}_{2n},{\bf 120}_{2m}] = {\bf 120}_{2(n+m)} \ , \ [{\bf 120}_{2n},{\bf 128}_{2m+1}] = {\bf 128}_{2(n+m)+1} \ , $$
$$
[{\bf 128}_{2n+1},{\bf 128}_{2m+1}] = {\bf 120}_{2(n+m+1)}
\eqno(6.1.14)$$
\par 
The Cartan involution invariant algebra of $E_9$ has generators that are given by $R^\alpha +I_c(R^\alpha)$ where $R^\alpha$ is a positive root generator of $E_9$.  Using the action of the Cartan involution given in equation (2.7) we find at IIA level zero that  the Cartan involution invariant generators are given by 
$$
K^i{}_j- K^j{}_i ,\ \ R^{i_1i_2}-R_{i_1i_2}
\eqno(6.1.15)$$
 which generate $SO(8)\otimes SO(8)$. While examining equation (2.8)  the  Cartan involution invariant generators arising from IIA  levels $\pm 1$ are given by 
$$
 R^i -R_i \ , \  R^{i_1 i_2 i_3} - R_{i_1 i_2 i_3} \ ,  R^{i_1 \ldots i_5} + R_{i_1 \ldots i_5}\ , \  R^{i_1 \ldots i_7}-R_{i_1 \ldots i_7}   
\eqno(6.1.16)  $$
The pattern at higher levels is obvious. 
\par
We can now discuss the string little algebra in more detail and in particular the generators given in  equation (6.1.1). 
The generator $L_{\alpha}$ has  the form 
$$
L_{\alpha}= R^\alpha +I_c(R^\alpha)+ \ {\rm   terms\  involving\ commutators \  with \ R\  or \  \overline R}
\eqno(6.1.17)$$
We recognise the two  terms shown explicitly  in $L_{\alpha}$  in equation (6.1.17) as an element of $I_c(E_9)$ but , as indicated,  it also contains  terms involving the commutators with  $R$ or  $\overline R$.   Thus the  generators in the string little algebra ${\cal H}_r$ are in one to one correspondence with those of $I_c(E_{9})$. However,  $L_{\alpha}$  belongs to  $I_c(E_{11})$ rather than $I_c(E_{9})$ as the generators  $R$ and $\overline R$. are not in $E_9$. 
\par
At IIA level zero there are no $R$ and $\overline R$ commutators as these vanish  and the generators in  ${\cal H}_r$  are those of equation (6.1.15). Thus, at level zero, the algebra of ${\cal H}_r$ and $I_c(E_{9})$ are the same, namely  they  just contain  the generators of $SO(8)\otimes SO(8)$. However, for all higher  levels the  generators  are very different. The commutators of the generators in   ${\cal H}_r$ with  IIA level greater than zero have the same algebra as $R^\alpha$,  where $\alpha$ is a positive root of  $E_9$, that is,  their algebra is isomorphic to the  part of the  Borel subalgebra of $E_9$ that contains generators corresponding to positive roots. In particular the generators in ${\cal H}_r$  that contain the positive roots of $D_8^+$  obey the same algebra as the commutators of the Borel algebra of $D_8^+$. 
\par
As is well known $E_9$ is nothing but affine $E_8$, that is, $E_8^+$. However, so far we have hardly  mentioned $E_8$,  or it's Cartan involution invariant subalgebra $I_c(E_8)=SO(16)$ which both play such an important role in the eleven dimensional $E_{11}$ theory. In this theory one deletes  node 11 to leave a GL(11) algebra. This  contrasts with the IIA theory being studied in this paper in which one deletes the IIA node, that is, node ten which leaves the algebra $D_{10}$.  To be relevant to our current  context we will first delete the nodes 1 and 2 as shown in the Dynkin diagram in figure above  to leave $E_9$ rather than consider the full $E_{11}$ algebra.  Further deleting node eleven we find the algebra $A_8=SL(9)$ however, this is not the deletion that corresponds to the IIA theory which is of interest to us here. 
 \par
 The algebra  $E_9$ is affine $E_8$ but this $E_8$   arises in the $E_{11}$ Dynkin diagram consisting of  nodes 4-10 and node 11 and is found by deleting node three.  It is useful to  first analyse how the generators of $E_8$ appear from the eleven dimensional theory by deleting node eleven and the resulting SL(8) algebra which corresponds to nodes 4-10.  The resulting Dynkin diagram is given by 
$$
\matrix{
& & & & & & & & & & & & & & \oplus & 11 &\bullet &  10\cr
& & & & & & & & & & & & & & | & &| \cr
\oplus & - & \oplus & - & \oplus& - & \bullet & - & \bullet & - & \bullet & - & \bullet & - & \bullet & - & \bullet &  &  \cr
1 & & 2 & & 3 & & 4 & & 5 & & 6 & & 7 & & 8 & & 9 & &  \cr}
$$
\par
The generators of $E_8$ are given by 
$$
\hat{K}^{\underline {i}}{}_{\underline{j}}, \ \hat R^{\underline{i}_1 \underline{i}_2 \underline{i}_3}, 
\ \epsilon _{\underline k_1 \underline k_2\underline{i}_1 \ldots \underline{i}_6}, \hat R^{\underline{i}_1 \ldots \underline{i}_6}, 
\epsilon_{\underline{i}_1 \ldots \underline{i}_8,} \hat R^{\underline{i}_1 \ldots \underline{i}_8,\underline{j}}  ,\ \  \eqno(6.1.18)$$ 
where the indices take the values $\underline{i},   \underline{j}, \underline{i}_1,\underline{i}_2,\ldots = 4,\ldots,11$.
 These generators have levels 
zero, one, two and three respectively  with respect to the level corresponding to node eleven. Their  negative level analogues 
are given by 
$$
\hat R_{\underline{i}_1 \underline{i}_2 \underline{i}_3}, 
\ \epsilon^{\underline k_1\underline k_2 \underline{i}_1 \ldots \underline{i}_6} \hat R_{\underline{i}_1 \ldots \underline{i}_6},
 \ \epsilon ^{\underline{i}_1 \ldots \underline{i}_8,} \hat R_{\underline{i}_1 \ldots \underline{i}_8,\underline{j}}
\eqno(6.1.19)$$
The Cartan involution acts on these generators as specified in equation (2.7) and the invariant algebra is generated by 
$$
\hat{K}^{\underline{i}}{}_{\underline{j}}-\hat{K}^{\underline{j}}{}_{\underline{i}} , 
\ \hat R^{\underline{i}_1 \underline{i}_2 \underline{i}_3}-\hat R_{\underline{i}_1 \underline{i}_2 \underline{i}_3}, 
\ \hat R^{\underline{i}_1 \ldots \underline{i}_6}+\hat R_{\underline{i}_1 \ldots \underline{i}_6}, \ 
\hat R^{\underline{i}_1 \ldots \underline{i}_8,\underline{j}}-\hat R_{\underline{i}_1 \ldots \underline{i}_8,\underline{j}}
\eqno(6.1.20)$$ 
The second and third generators obey the algebra of $SO(8)\otimes SO(8)$ while the second and fourth generators 
belong to the $(8_s, 8_c)$ representations of this algebra where $8_s$ and $8_c$ are the spinor representations of $SO\otimes SO(8)$. Taken  together  these 120 generators obey the algebra of SO(16). 
A detailed discussion can be found in section four of reference [15]. The 128 Cartan involution odd generators in $E_8$ have the opposite signs to those in equation (6.1.18) and they belong to the Majorana-Weyl  spinor representation of this SO(16). 
\par
The $E_9$ is then given by the  affinisation of the $E_8$ algebra. It generators are 
$$
E_9= \sum_n 248_n= \sum_n  (SO(16) _n\oplus 128 _n)\ ,\ \  n=0,\pm 1, \ldots
\eqno(6.1.21)$$
where 248 is the adjoint representation of $E_8$ and the second line contains the decomposition of $E_9$ that we just discussed into its Cartan involution invariant sublagebra SO(16)  of $E_8$. 
\par
From the IIA perspective studied in this paper we have the same $E_9$ but we deleted the IIA node, node 10, to find $D_8=SO(8,8)$ and we decomposed the $E_9$ algebra in terms of this algebra. To aid our analysis we also deleted node eleven and decomposed to the remaining $A_8$ algebra associated with nodes 3-9. While the algebra  $ D_8$
 appears in both  the eleven dimensional and IIA viewpoints they are not the same algebra. 
This is  easily seen once we realise that in the IIA viewpoint node ten is deleted but  from the eleven dimensional viewpoint this node it is part of $E_8$ Dynkin diagram. Thus in the IIA and eleven dimensional viewpoints different nodes are deleted. Hence the  $D_8=SO(8,8) $  we discussed in the IIA theory is not  subalgebra  of $E_8$ as it involves node three which is  not a node of the $E_8$ that appears from the elven dimensional viewpoint. In contrast the $E_8$ of  the eleven dimensional theory requires node ten to be active but this is deleted in the IIA theory. It is intriguing that despite this $E_9$ has a very similar decomposition into the two different $D_8$'s. In the IIA case we can  write $E_9 $ as in equation (6.1.13) while from the eleven dimensional viewpoint we write it as in equation (6.1.19). One can think of it as a kind of duality. 

%%%%%%%%%%%%%%%%%%%
\medskip
{{\bf 6.2 The relation of the string little algebra to the string states. }}
\medskip
As we explained in the previous section the string little algebra involves an algebra that is the same as the Borel subalgebra of $E_9$ at higher IIA levels and $SO(8)\otimes SO(8)$ at IIA level zero. Such a little algebra is unexpected and is quite different to the little algebras for the point particles. At first sight  it is not clear why it has such a  simple form. 
\par
In principle we should now choose an irreducible  representation of the string little algebra and then carry out the boost to find the full irreducible representation of $I_c(E_{11})\otimes_s l_1$. We will  not be able to show that the well known string states  do carry a representation of the full string little algebra, rather we will make progress towards this goal. In particular the  algebra  $SO(8,8)^+$ is a subalgebra of $E_9$ and  we will focus  on considering a representation of the Borel subalgebra of  $SO(8,8)^+$ part of the string little algebra. 
\par
We first consider the open  superstring and in particular the NS sector. As explained in section 4.3, from  the $b_r^i$ oscillators one can construct a bilinear quantity whose generators are those of $SO(8)^+$, see equation (4.3.1). 
However, these generators  do not act on  physical states to give physical states as they do not boost the momentum of the states to that required by the physical state conditions. As explained at the end of section four, one can replace them by their DDF analogues, see equation (4.3.3),  or use the boost of equation (4.3.5), so that they act on  physical  states. to give physical states.  Thus the open string states in the NS sector  do carry a representation of  $SO(8)^+$ and so the Borel subalgebra of $SO(8)^+$.  Since the generators are bilinear in the oscillators they act on the physical states so as to take states with an even (odd) number of oscillators to states which also have an even (odd) number of oscillators. As such the physical states do not carry an irreducible representation of  $SO(8)^+$ but contain two irreducible representations which have an odd and even in the number of the oscillators. If we choose to have the irreducible representation that is odd in oscillators then we make the GSO [25]  projection in this sector. Thus the GSO projection can be understood as a restriction to have an irreducible representation of $SO(8)^+$. 
\par
Let us now consider the NS-NS sector of the closed superstring as discussed in section 4.3. Now we have the oscillators 
$b^I_r\equiv ( b_r^i, \bar b_r^i )$ where the indices $I,J,\ldots$ take $8+8=16$ values and the generators 
$ {\cal M}^c{}_n{}^{IJ}= \sum_r : b_{n-r}^{[I}b_{r}^{J]} : $ which obey the algebra $SO(8, 8)^+$.  As for the open string these do not take physical states to physical states, however, we can use their DDF analogues or do a boost similar to that in equation 
(4.3.5). At first sight  the string states in the NS-NS sector do carry a representation of $SO(8, 8)^+$ and so   the Borel subalgebra of $SO(8, 8)^+$  at  IIA levels greater than one and $SO(8)\otimes SO(8)$ at level zero which a large part of the string  little algebra. 
 \par
 However, this statement does not take account of the GSO projection which ensures that one has odd number of $b_r^i$ oscillators and also odd numbers of $\bar b_r^i$. This choice is  preserved by $SO(8)^+\otimes SO(8)^+$ but not the generators $ {\cal M}^c{}_0{}^{IJ}$ for $I=1,\ldots, 8$ and $j=9,\ldots, 16$. However, it is preserved by the action of bilinears of the latter generators. There is also the level matching condition to take account of. This would only be preserved if we took suitable combinations of generators. However, one could also allow windings,  or achieve matching by also acting with the bosonic oscillators. The string states come so close to carrying a representation of $SO(8, 8)^+$ that it is 
 tempting to think that in a more complete treatment would work better. 
 \par
 One could also consider generators that transform the NS-NS into the R-R sectors. Such generators could take the generic form 
 $$
 -2D_k |R > <N | b_{{1\over 2}}{}^{(i}\bar b_{{1\over 2}}^{k)} + D^{i}{}_{ j_1 j_2} |R > <N | b_{{1\over 2}}^{j_1}\bar b_{{1\over 2}}^{j_2}+\ldots 
 $$
 $$
 9D^{[i_1} |R > <N | b_{{1\over 2}}{}^{i_2}\bar b_{{1\over 2}}^{i_3]} -6 D^{k [i_1 i_2} |R > <N | b_{{1\over 2}}^{(i_3]}\bar b_{{1\over 2}}^{k)}+\ldots 
 $$
 where $| R>$ and $| N>$ are the vacuum states in the R-R and NS-NS sectors, $D^k= d_0^k+i\bar d_0^k$ and $+\ldots $ are higher level terms. In fact these generators are the Borel part of  the transformations of the massless states that appear when one considers the irreducible representation of $I_c(E_{11})\otimes l_1$ suitable to the IIA theory. This can be deduced from the eleven dimensional case given in references [14,15] by carrying out a dimensional reduction. 
 \par
 In the  R and R-R sectors of the open and closed  sectors of the superstring we have the $d_n^i$  and the $d_n^I= (d_n^i , \bar d_n^i)$ oscillators respectively. From these one can form bilinear expressions that  generate the algebras $SO(8)^+$ and  $SO(8,8)^+$ respectively. It would be interesting to find  their replacement by the analogue DDF operators, or  include momentum   boosts,  which take physical states to physical sates. 
 \par
 We now study the situation from the view point of the Green-Schwarz  formulation of the IIA superstring. In the  light-cone
 gauge one has a  residual SO(8) symmetry of the Lorentz algebra. In addition to the bosonic oscillators the theory contains    two fermionic spinor oscillators   $S_n^\alpha $ and $\bar S_n^\alpha$ with $\alpha, \beta, \ldots =1,\ldots, 8$. These  are Majorana-Weyl spinors of Spin(8) of opposite chirality. They obey the anti-commutators 
 $$
\{ S_n^\alpha , S_m^\beta\}=\delta^{\alpha , \beta}\delta_{n+m,0} , \   \{ \bar S_n^\alpha , \bar S_m^\beta\}=\delta^{\alpha , \beta}\delta_{n+m,0} , \  \{ S_n^\alpha , \bar S_m^\beta\}=0
 \eqno(6.2.1)$$
 \par
We will need the properties of such spinors.  Looking at section 5.6 of the book [19] on finds for this Euclidean case that we may choose $B=C=I$ and $\rho=1$. Then $\epsilon =1$ and $(\gamma^i)^*= \gamma^i$ and  
 $(\gamma^i)^T= \gamma^i$. It follows that the matrices 
 $$
 \gamma^i , \  \gamma^{i_1\ldots i_4} ,\  \gamma^{i_1\ldots i_5}  ,\  \gamma^{i_1\ldots i_8} ,\  
 \eqno(6.2.2)$$
 are symmetric and the others are antisymmetric. We also find that if $\gamma_9\chi=\chi$, where $\gamma_9=\gamma_1\ldots \gamma_8$, then $\bar \chi \gamma_9=\bar \chi$ for a Majorana-Weyl spinor and as such only the expressions $\bar \chi \gamma^{i_1 i_2}$ and $\bar \chi \gamma^{i_1\ldots i_6}\chi$ are non zero. However, these two expressions are related by a duality transformation. 
 \par
Taking into account the above properties we can form the operators 
$$
M_n^{i_1i_2}=  \sum_p   S_{n-p} \gamma^{i_1i_2} S_p ,\ \  \bar M_n^{i_1i_2}=\sum_p  \bar S_{n-p} \gamma^{i_1i_2} \bar S_p
\eqno(6.2.3)$$
In this equation   $S_{n-p} \gamma^{i_1i_2} S_p=S_{n-p}^\alpha( \gamma^{i_1i_2})_{\alpha\beta} S_p ^\beta$. The first expression is  the  only  bilinear object   in  $S_n^\alpha $ alone that we can construct. The same applies for $\bar S_n^\alpha$. Using the relations of equation (6.2.1) we find that the generators of equation (6.2.3) have the algebra
 $SO(8)^+\otimes SO(8)^+$ which we can think of acting on the left and right moving  parts of the string. 
\par
For the mixed bilinear expressions  we have spinors of opposite Weyl character and so we can only have 
$$
L^I_n\equiv  \sum_p   S_{n-p} \gamma^{I} \bar S_p 
\eqno(6.2.4)$$
where $\gamma^I= (\gamma^i, \gamma^{i_1i_2i_3})$. We have taken account of the fact that the  gamma matrices between the spinors obey duality relations. 
\par
The commutators of the $L^I_n$ with $M_n^{i_1i_2}+\bar M_n^{i_1i_2}$  is given by 
$$
[M_n^{i_1i_2}+\bar M_n^{i_1i_2} , L^I_n ]=2 \sum _p S_{n+m-p} [\gamma ^{i_1i_2}, \gamma^I ]\bar S_p
\eqno(6.2.5)$$
which is just the usual affine rotation. With $M_n^{i_1i_2}-\bar M_n^{i_1i_2}$ we find that 
$$
[M_n^{i_1i_2}-\bar M_n^{i_1i_2} , L^I_n ]=2 \sum _p S_{n+m-p} \{\gamma ^{i_1i_2} , \gamma^I \}\bar S_p
\eqno(6.2.6)$$
Thus the $L^I_n$ belong to a representation of  $SO(8)^+\otimes SO(8)^+$ and indeed they belong to the $(8_s, 8_c)_n$ representation. 
\par
The commutators of the  $L^I_n$ are given by 
$$
[L^I_n , L^J_m ]= \sum_p S_{n+m-p} \gamma^J\gamma^I S_p \eta_I-  \sum_p \bar S_{n+m-p} \gamma^I\gamma^J \bar S_p\eta_I
\eqno(6.2.7)$$
where $\eta_I=-1$ if $\gamma^I$ equals $ \gamma ^i$ and $\eta_I=1$ if $ \gamma ^{i_1\ldots i_5}$.  In view of the above comments these commutators can only belong to $SO(8)^+\otimes SO(8)^+$. As such the generators of equations (6.2.3) and (6.2.4 ) close to give the algebra $SO(8, 8)^+$. 
\par
In the open superstring,  the states are formed by the oscillators $S_n^\alpha$ and  the bosonic oscillators acting on the 16 dimensional  $8_v\oplus 8_c$ states which belong to a representation of $S_0^\alpha$ which itself belongs to the $8_s$ representation of spin(8). This is part of the spacetime supersymmetry and it  takes the 8 spin one  states into the 8 spin "one" "half" states and vice-versa. The $8_v\oplus 8_c$  are the massless states of the open string. For a review of states in the superstring in the Green-Schwarz formulation see section 5.3 of reference [26].
\par
The $SO(8)^+$ generators $M_n^{i_1i_2}$ clearly act on the states of the open superstring  to give other states which will be physical states provided  we also inject a corresponding momenta. The latter could either be achieved  by the analogue of a DDF construction  or the boosts we considered at the end of section four.  Assuming this to be possible  the open string states belong to a representation of $SO(8)^+$ in agreement with our result in the Neveu-Schwarz formulation. We note that this is a representation of the bosons and also the fermion as the generator $M_n^{i_1i_2}$ does not mix these two. 
\par
The closed string physical states result from the left moving oscillators $S_n^\alpha$ and the right moving oscillators $\bar S_n^\alpha$ and also  the bosonic oscillators acting on the $256=16\otimes 16$ dimensional formed from the tensor product of the left and right moving states $(8_v+8_c)\otimes(8_v+8_s)$. The states 
$8_v\otimes 8_v$ belong to the $1+28+35_v$ dimensional representations of spin(8) which are the massless states in the NS-NS sector. While the $8_c\otimes 8_s$  belong to the $8_v+56_v$ representation of spin(8) which are the massless states of the R-R sector. The generators of equation (6.2.3) and (6.3.4) act on the physical states.  The left moving generators $M_n^{i_1i_2}$ and $\bar M_n^{i_1i_2} $ give affine  Lorentz rotations of the left moving states and the right moving sectors. The generators $L^I$ act on the left and right movers so as to transform the NS-NS states into the  R-R states into each other. Assuming we can provide the required boost of the momentum when these generators act  on the physical states they will carry a representation of the algebra $SO(8,8)^+$ and also take account of the level matching. This latter constraint can perhaps be achieved by taking certain combinations of generators, or   involving the bosonic operators,  or allowing winding, 
\par
So far we did not discuss the role of the bosonic oscillators  $\alpha_n^i$ , $\bar  \alpha_n^i$, $i=2,\ldots , 9$. They should also play an important and one would expect the above symmetries to also act on these oscillators.  However, as we explained at the end of section four,  the natural transformations of $SO(8,8)^+$ of the oscillators, see equations (4.3.7) and (4.3.8), do not respect their commutation relations  of equation (4.3.6). However, the string little algebra for the superstring  is not $I_c(E_9)$ but is essentially the Borel subalgebra of $E_9$. A  subalgebra of this algebra  is the Borel  subalgebra of $SO(8,8)^+$ which we can take to have  the generators $M_n^{ij}$, $\bar M_n^{ij}$,  $V_n^{ij}$ and $\bar V_n^{ij}$ for $n\le 0$. These act on the negatively moded oscillators as in equations  (4.3.7) and (4.3.8) so as to  take the negatively modeed into themselves. The physical states contain negatively moded bosonic oscillators acting on the vacuum and so, with  a suitable momentum boost,  these generators take physical states to physical states. While somewhat unconventional the action of these generators on the positively modded oscillators can be chosen to preserve the commutation relations. Thus if would appear that one can have the action of a significant part of the string little algebra. One could also apply this strategy to the fermionic oscillators. 
\par
In this section we have made a fragmented and partial attempt to understand if a representation of the string little algebra is carried by the well known physical states of the superstring. While there are many encouraging features there are also some unresolved issues. We have only investigated the Borel part of  $SO(8,8)^+$  and not the 128 dimensional spinor representation that occurs at even levels. We have also not taken a unified approach in that we did not consider the bosonic and fermionic oscillators in a unified way. It would be good to persue these matters further.   

\medskip
{{\bf 7 Discussion }}
\medskip
E theory provides, for the first time,  a unified approach to branes within which one can hope to treat point particles,  string and branes in the same way. Indeed one can take the attitude that they are all irreducible representations of $ I_c(E_{11}) \otimes_s l_1$ in much the same way  as particles, in our familiar  quantum field theories, are irreducible representations of the Poincare algebra [14]. In this paper we have applied this strategy to strings and in particular the IIA string. Taking the corresponding charges in the vector representation to be non zero  we found the little algebra for the string. It has an unusual form in that it is essentially the Borel subalgebra of $E_9$. However, this $E_9$ is not the $E_9$ that is obviously apparent form the Dynkin diagram of $E_{11}$ by deleting node two,  but only emerged after a rather involved calculation involving generators in $E_{11}$ which are outside this obvious $E_9$. The Borel character of the little algebra has its origins in the fact that the string charges involve the momenta and the string charge which satisfy a relation rather than each have specific values. This is a novel feature of finding the irreducible representations for extended objects. This   contrasts with the massless case where it is $I_c(E_{9})$. 
\par
The next step in the procedure would be  to take a representation of the little algebra. However in section six we investigated if the known states of the superstring did indeed carry a representation of essentially the Borel subalgebra of $E_9$. While we were not able to show that this was the case we did find quite a bit of evidence that there are additional symmetries which are in this algebra. Starting from the fact that bilinears in the known oscillators can generate the algebra $SO(8,8)^+$, whose Borel subalgebra,  is a substantial subalgebra of the string little algebra we investigated if the physical states carried a representation of this algebra. If one allowed winding or some other way of relaxing the level matching requirements,   and one could carry out the required momentum boosts, then it would appear that this was true. Our treatment was also provisional in that it did not consider the symmetry to act in a unified way on all the oscillators. We note that it is only required to show that the physical states carry a representation of this algebra and this may not have a simple expression in terms of the known oscillators. 
\par
Assuming that we had a representation of the string little algebra the next step would be to boost it up to a representation of 
$I_c(E_{11})\otimes_s l_1$. Such a step was discussed in reference [14] and it would be very interesting to do this for the string. The net result should be a wavefunction that depends on the string coordinates and obeys the physical state conditions. 
These conditions also follow from the string dynamics which can also be constructed from and $E_{11}$ viewpoint as a non-linear realisation of $I_c(E_{11})\otimes_s l_1$ with a suitable local subalgebra [7,27]. This local subalgebra has been found at low levels and it essentially agrees with the string little algebra found in this paper even though the two derivations are very different. It would be interesting to join up these two different approaches and this may help illuminate the unknown parts of   each approach. 
\par
It would be very interesting to extend the approach of this paper to the branes and in particular the M2 and M5 branes. The first step would be to find their little algebras. We would expect the calculation to have many points in common with the derivation of the string little algebra found in this paper. 
\par
String theory on a torus has additional symmetries which  can, for a suitable torus, be any of  the Lie algebras in the Killing-Cartan list. It is believed that these are a quirk of taking the string on  the torus and not  a sign of some underlying symmetry of string theory. A similar attitude was taken to the Cremmer-Julia symmetries in the maximal supergravity theories; they were thought just to be a consequence of the dimensional reduction on a torus. However, we now know that they are symmetries of the underlying theory, namely E theory, although this theory was not discovered by thinking in this way [1]. Perhaps we should think that the torus probes  aspects of string theory not present in our usual  spacetime and this allows us to see some hidden symmetries. 
\par
In E theory  space time has an infinite number of coordinates [2]. Taking string theory on a torus introduces the winding coordinates that are just a few of these coordinates. It is useful to think of the infinite coordinates as a kind of effective spacetime reflecting some of the properties of the underlying structure. The idea  that taking particular situations may display hidden symmetries has a precedent in particle physics. For example symmetries that are hidden at low energy in particle physics, as they are spontaneously broken,  will show up at high energy. Another example is the discovery of  asymptotic charges in general relativity when we consider spacetime that are not flat but asymptotically flat.

%%%%%%%%%%%%%%%%%%%%%
\medskip
{\bf Acknowledgement}
\medskip
Peter West  has benefitted from grant numbers ST/P000258/1  and ST/T000759/1  from STFC in the UK. Keith Glennon would like to thank Kings College  for support during his PhD studies. 
%%%%%%%%%%%%%%%%%%

\medskip
{\bf Appendix A: The IIA Commutators}
\medskip

In this appendix we will construct at low levels the commutators of the $E_{11}$ algebra in its  decomposition that leads to   the IIA theory. As described in section two  of this paper  this is in the decomposition to SO(10,10) which is for ease of recognition further decomposed into representation of GL(10). Given the generators at a low  IIA  levesl this algebra can be determined directly by considering all possible expressions consistent with the ${\rm SL}(10)$ character of the generators involved in a given commutator, the fact that the commutators preserve the level and then requiring that these expressions satisfy the Jacobi identities. Equivalently the algebra can be determined from  the algebra of the  $E_{11}$  algebra in its eleven dimensional decomposition as given, for example,  in [19]. We give the IIA algebra to level one, but only  include  those generators  which involve the underlying $E_{11}$ generators to $E_{11}$ level three for simplicity. This means for example that we will ignore $R^{{\underline{a}}_1 \ldots {\underline{a}}_9}= R^{{\underline{a}}_1 \ldots {\underline{a}}_9 11, 1111}$   in the algebra in this approximation as it has level four with respect to the deletion of node eleven. The algebra of the  generators at  IIA level zero is
$$
[K^{\underline{a}}{}_{\underline{b}},K^{\underline{c}}{}_{\underline{d}}] = \delta^{\underline{c}}{}_{\underline{b}} K^{\underline{a}}{}_{\underline{d}} - \delta^{\underline{a}}{}_{\underline{d}} K^{\underline{c}}{}_{\underline{b}} \ \ , \ \ [K^{\underline{a}}{}_{\underline{b}},\tilde{R}] = 0 \ \ ,  \ \ [K^{\underline{a}}{}_{\underline{b}},R^{\underline{c} \underline{d}}] = 2 \delta^{[\underline{c}}{}_{\underline{b}} R^{|\underline{a}| \underline{d}]} \ \ ,
$$
$$
\ \ [K^{\underline{a}}{}_{\underline{b}},R_{\underline{c} \underline{d}}] = - 2 \delta^{\underline{a}}{}_{[\underline{c}} R_{|\underline{b}| \underline{d}]} \ \ , \ \  [\tilde{R},\tilde{R}] = 0 \ \ , \ \ [\tilde{R},R^{\underline{a} \underline{b}}] = 0 \ \ , \ \ [\tilde{R},R_{\underline{a} \underline{b}}] = 0 \ \ ,
$$
$$
[R^{\underline{a} \underline{b}}, R^{\underline{c} \underline{d}}] = 0 \ \ , \ \ [R^{\underline{a} \underline{b}}, R_{\underline{c} \underline{d}}] = 4 \delta^{[\underline{a}}{}_{[\underline{c}} K^{\underline{b}]}{}_{\underline{d}]} \ \ , \ \ [R_{\underline{a} \underline{b}}, R_{\underline{c} \underline{d}}] = 0 \ \ .  
\eqno(A.1) $$
These generators obey the algebra of $D_{10}\otimes GL(1)=SO(10,10)\otimes GL(1)$ as they must. 
\par
The commutators between the  IIA  level one generators are  given by
$$
[R^{\underline{a}},R^{\underline{b}}] = 0 \ , \ [R^{\underline{a}},R^{\underline{b}_1 \underline{b}_2 \underline{b}_3}] = 0 \ , \ [R^{\underline{a}},R^{\underline{b}_1 \ldots \underline{b}_5}] = - R^{\underline{a} \, \underline{b}_1 \ldots \underline{b}_5} \ , \ $$
$$
[R^{\underline{a}},R^{\underline{b}_1 \ldots \underline{b}_7}] = - 2 R^{\underline{a} \, \underline{b}_1 \ldots \underline{b}_7} - 7 R^{\underline{a} [\underline{b}_1 \ldots \underline{b}_6 , \underline{b}_7]} \ , \ [R^{\underline{a}_1 \underline{a}_2 \underline{a}_3},R^{\underline{b}_1 \underline{b}_2 \underline{b}_3}] = 2 R^{\underline{a}_1 \underline{a}_2 \underline{a}_3 \underline{b}_1 \underline{b}_2 \underline{b}_3} \ ,$$
$$
[R^{\underline{a}_1 \underline{a}_2 \underline{a}_3},R^{\underline{b}_1 \ldots \underline{b}_5}] = - 3 R^{\underline{b}_1 \ldots \underline{b}_5[\underline{a}_1 \underline{a}_2 , \underline{a}_3]} \ , \ [R^{\underline{a}_1 \underline{a}_2 \underline{a}_3},R^{\underline{b}_1 \ldots \underline{b}_7}] = 0 \ ,
$$
$$
[R^{\underline{a}_1 \ldots \underline{a}_5},R^{\underline{b}_1 \ldots \underline{b}_5}] = 0 \ , \  [R^{\underline{a}_1 \ldots \underline{a}_5},R^{\underline{b}_1 \ldots \underline{b}_7}] = 0 \ , \ [R^{\underline{a}_1 \ldots \underline{a}_7},R^{\underline{b}_1 \ldots \underline{b}_7}] = 0 \ , 
\eqno(A.2)$$
and for those with  IIA  level minus one  by
$$
[R_{\underline{a}},R_{\underline{b}}] = 0 \ , \ [R_{\underline{a}},R_{\underline{b}_1 \underline{b}_2 \underline{b}_3}] = 0 \ , \ [R_{\underline{a}},R_{\underline{b}_1 \ldots \underline{b}_5}] = + R_{\underline{a} \, \underline{b}_1 \ldots \underline{b}_5} \ , \ $$
$$
[R_{\underline{a}},R_{\underline{b}_1 \ldots \underline{b}_7}] = + 2 R_{\underline{a} \, \underline{b}_1 \ldots \underline{b}_7} + 7 R_{\underline{a} [\underline{b}_1 \ldots \underline{b}_6 , \underline{b}_7]} \ , \ [R_{\underline{a}_1 \underline{a}_2 \underline{a}_3},R_{\underline{b}_1 \underline{b}_2 \underline{b}_3}] = 2 R_{\underline{a}_1 \underline{a}_2 \underline{a}_3 \underline{b}_1 \underline{b}_2 \underline{b}_3} \ , $$
$$
[R_{\underline{a}_1 \underline{a}_2 \underline{a}_3},R_{\underline{b}_1 \ldots \underline{b}_5}] = - 3 R_{\underline{b}_1 \ldots \underline{b}_5[\underline{a}_1 \underline{a}_2 , \underline{a}_3]} \ , \ [R_{\underline{a}_1 \underline{a}_2 \underline{a}_3},R_{\underline{b}_1 \ldots \underline{b}_7}] = 0 \ , 
$$
$$
[R_{\underline{a}_1 \ldots \underline{a}_5},R_{\underline{b}_1 \ldots \underline{b}_5}] = 0 \ , \  [R_{\underline{a}_1 \ldots \underline{a}_5},R_{\underline{b}_1 \ldots \underline{b}_7}] = 0 \ , \ [R_{\underline{a}_1 \ldots \underline{a}_7},R_{\underline{b}_1 \ldots \underline{b}_7}] = 0 \ . 
\eqno(A.3)$$
The algebra of IIA  level zero generators with IIA  level one generators are given by  
$$
[K^{\underline{a}}{}_{\underline{b}},R^{\underline{c}}] = R^{\underline{a}} \delta^{\underline{c}}{}_{\underline{b}} - {1 \over 2} \delta^{\underline{a}}{}_{\underline{b}} R^{\underline{c}} \ , \   [K^{\underline{a}}{}_{\underline{b}},R^{\underline{c}_1 \underline{c}_2 \underline{c}_3}] = 3 \delta^{[\underline{c}_1}{}_{\underline{b}} R^{|\underline{a}| \underline{c}_2 \underline{c}_3]} - {1 \over 2} \delta^{\underline{a}}{}_{\underline{b}} R^{\underline{c}_1 \underline{c}_2 \underline{c}_3}  \ ,
$$
$$
[K^{\underline{a}}{}_{\underline{b}},R^{\underline{c}_1 \ldots \underline{c}_5}] = 5 \delta^{[\underline{c}_1}{}_{\underline{b}} R^{|\underline{a}| \underline{c}_2 \ldots \underline{c}_5]} - {1 \over 2} \delta^{\underline{a}}{}_{\underline{b}} R^{\underline{c}_1 \ldots \underline{c}_5} \ , 
$$
$$
[K^{\underline{a}}{}_{\underline{b}},R^{\underline{c}_1 \ldots \underline{c}_7}] = 7 \delta^{[\underline{c}_1}{}_{\underline{b}} R^{|\underline{a}| \underline{c}_2 \ldots \underline{c}_7]} - {1 \over 2} \delta^{\underline{a}}{}_{\underline{b}} R^{\underline{c}_1 \ldots \underline{c}_7} \ \ ,
 $$
$$
[\tilde{R},R^{\underline{a}}] = - 3 R^{\underline{a}} \ \ , \ \ [\tilde{R},R^{\underline{a}_1 \underline{a}_2 \underline{a}_3}] = - 3 R^{\underline{a}_1 \underline{a}_2 \underline{a}_3} \ \ , \ \ [\tilde{R},R^{\underline{a}_1 \ldots \underline{a}_5}] = - 3 R^{\underline{a}_1 \ldots \underline{a}_5} \ \ ,$$
$$
[\tilde{R},R^{\underline{a}_1 \ldots \underline{a}_7}] = - 3 R^{\underline{a}_1 \ldots \underline{a}_7} \ \ , \ \ [R^{\underline{a}_1 \underline{a}_2},R^{\underline{b}}] = - R^{\underline{a}_1 \underline{a}_2 \underline{b}} \ , $$
$$
[R^{\underline{a}_1 \underline{a}_2},R^{\underline{b}_1 \underline{b}_2 \underline{b}_3}] = - 2 R^{\underline{a}_1 \underline{a}_2 \underline{b}_1 \underline{b}_2 \underline{b}_3} \ , \ [R^{\underline{a}_1 \underline{a}_2},R^{\underline{b}_1 .. \underline{b}_5}] = - R^{\underline{a}_1 \underline{a}_2 \underline{b}_1 .. \underline{b}_5}
$$
$$
[R^{\underline{a}_1 \underline{a}_2},R^{\underline{b}_1 .. \underline{b}_7}] = 0 \ , \ [R_{\underline{a}_1 \underline{a}_2},R^{\underline{b}}] = 0 \ , \ [R_{\underline{a}_1 \underline{a}_2},R^{\underline{b}_1 \underline{b}_2 \underline{b}_3}] = - 6 \delta^{[\underline{b}_1 \underline{b}_2}_{\underline{a}_1  \underline{a}_2} R^{\underline{b}_3]} \ , 
$$
$$
[R_{\underline{a}_1 \underline{a}_2},R^{\underline{b}_1 .. \underline{b}_5}] = - 10 R^{[\underline{b}_1 \underline{b}_2 \underline{b}_3} \delta^{\underline{b}_4 \underline{b}_5]}_{\underline{a}_1  \underline{a}_2}  \ , \  [R_{\underline{a}_1 \underline{a}_2},R^{\underline{b}_1 .. \underline{b}_7}] = - 42 R^{[\underline{b}_1 .. \underline{b}_5} \delta^{\underline{b}_6 \underline{b}_7]}_{\underline{a}_1  \underline{a}_2} \ , 
\eqno(A.4)$$
The commutators of the IIA  level zero with the  IIA  level minus one are given by 
$$
[K^{\underline{a}}{}_{\underline{b}},R_{\underline{c}}] = - R_{\underline{b}} \delta^{\underline{a}}{}_{\underline{c}} + {1 \over 2} \delta^{\underline{a}}{}_{\underline{b}} R_{\underline{c}} \ , \ [K^{\underline{a}}{}_{\underline{b}},R_{\underline{c}_1 \underline{c}_2 \underline{c}_3}] = - 3 \delta^{[\underline{a}}{}_{\underline{c}_1} R_{|\underline{b}| \underline{c}_2 \underline{c}_3]} + {1 \over 2} \delta^{\underline{a}}{}_{\underline{b}} R^{\underline{c}_1 \underline{c}_2 \underline{c}_3}  \ ,
$$
$$
[K^{\underline{a}}{}_{\underline{b}},R_{\underline{c}_1 \ldots \underline{c}_5}] = - 5 \delta^{[\underline{a}}{}_{\underline{c}_1} R_{|\underline{b}| \underline{c}_2 \ldots \underline{c}_5]} + {1 \over 2} \delta^{\underline{a}}{}_{\underline{b}} R_{\underline{c}_1 \ldots \underline{c}_5} \ \ ,
$$
$$
[K^{\underline{a}}{}_{\underline{b}},R_{\underline{c}_1 .. \underline{c}_7}] = - 7 \delta^{\underline{b}}{}_{[\underline{c}_1} R_{|\underline{b}| \underline{c}_2 .. \underline{c}_7]} + {1 \over 2} \delta^{\underline{a}}{}_{\underline{b}} R_{\underline{c}_1 .. \underline{c}_7} , 
$$
$$
[\tilde{R},R_{\underline{a}}] = 3 R_{\underline{a}} \ \ , \ \ [\tilde{R},R_{\underline{a}_1 \underline{a}_2 \underline{a}_3}] =  3 R_{\underline{a}_1 \underline{a}_2 \underline{a}_3} \ \ , \ \ [\tilde{R},R_{\underline{a}_1 \ldots \underline{a}_5}] = 3 R_{\underline{a}_1 \ldots \underline{a}_5} \ \ ,
$$
$$
[\tilde{R},R_{\underline{a}_1 \ldots \underline{a}_7}] = 3 R_{\underline{a}_1 \ldots \underline{a}_7} \ \ , \ \ [ R^{\underline{a}_1 \underline{a}_2},R_{\underline{b}}] = 0 \ \ , \ \ [R^{\underline{a}_1 \underline{a}_2},R_{\underline{b}_1 \underline{b}_2 \underline{b}_3}] = 6 \delta^{\underline{a}_1 \underline{a}_2}_{[\underline{b}_1 \underline{b}_2} R_{\underline{b}_3]} \ \ , 
$$
$$
[R^{\underline{a}_1 \underline{a}_2},R_{\underline{b}_1 .. \underline{b}_5}] = - 10 R_{[\underline{b}_1 \underline{b}_2 \underline{b}_3} \delta^{\underline{a}_1 \underline{a}_2}_{\underline{b}_4 \underline{b}_5]} \ \ , \ \ [R^{\underline{a}_1 \underline{a}_2},R_{\underline{b}_1 .. \underline{b}_7}] = - 42 R_{[\underline{b}_1 .. \underline{b}_5} \delta^{\underline{a}_1 \underline{a}_2}_{\underline{b}_6 \underline{b}_7]} \ \ , 
$$
$$
[R_{\underline{a}_1 \underline{a}_2},R_{\underline{b}}] = R_{\underline{a}_1 \underline{a}_2 \underline{b}} \ \ , \ \ [R_{\underline{a}_1 \underline{a}_2},R_{\underline{b}_1 \underline{b}_2 \underline{b}_3}] = - 2 R_{\underline{a}_1 \underline{a}_2 \underline{b}_1 \underline{b}_2 \underline{b}_3} \ \ , 
$$
$$
[R_{\underline{a}_1 \underline{a}_2},R_{\underline{b}_1 .. \underline{b}_5}] = - 2 R_{\underline{a}_1 \underline{a}_2 \underline{b}_1 .. \underline{b}_5} \ \ , \ \ [R_{\underline{a}_1 \underline{a}_2},R_{\underline{b}_1 .. \underline{b}_7}] = 0 \ \ . \eqno(A.5) $$
The IIA  level one generators with the IIA  level minus generators are giv en by 
$$
[R^{\underline{a}},R_{\underline{b}}] = K^{\underline{a}}{}_{\underline{b}} - {1 \over 6} \delta^{\underline{a}}{}_{\underline{b}} (3 D - \tilde{R}) \ \ , \ \ [R^{\underline{a}},R_{\underline{b}_! \underline{b}_2 \underline{b}_3}] = - 3 \delta^{\underline{a}}{}_{[\underline{b}_1} R_{\underline{b}_2 \underline{b}_3]} \ , 
$$
$$
[R^{\underline{a}},R_{\underline{b}_1 \ldots \underline{b}_5}] = 0 \ \ , \ \  [R^{\underline{a}},R_{\underline{b}_1 \ldots \underline{b}_7}] = 0 \ , \ [R^{\underline{a}_1\underline{a}_2\underline{a}_3},R_{\underline{b}}] = - 3 \delta^{[\underline{a}_1}{}_{\underline{b}} R^{\underline{a}_2 \underline{a}_3]} \ \ ,
$$
$$
[R^{\underline{a}_1 \underline{a}_2 \underline{a}_3},R_{\underline{b}_1 \underline{b}_2 \underline{b}_3}] = 18 \delta^{[\underline{a}_1 \underline{a}_2}_{[\underline{b}_1 \underline{b}_2} K^{\underline{a}_3]}{}_{\underline{b}_3]} - \delta^{\underline{a}_1 \underline{a}_2 \underline{a}_3}_{\underline{b}_1 \underline{b}_2 \underline{b}_3} (3D - \tilde{R}) \ \ , 
$$
$$
[R^{\underline{a}_1 \underline{a}_2 \underline{a}_3},R_{\underline{b}_1 .. \underline{b}_5}] = 30 \delta^{\underline{a}_1 \underline{a}_2 \underline{a}_3}_{[\underline{b}_1 \underline{b}_2 \underline{b}_3} R_{\underline{b}_4 \underline{b}_5]} \ , \ [R^{\underline{a}_1 \underline{a}_2 \underline{a}_3},R_{\underline{b}_1 .. \underline{b}_7}] = 0 \ , 
$$
$$
[R^{\underline{a}_1 .. \underline{a}_5},R_{\underline{b}}] = 0 \ , \ [R^{\underline{a}_1 .. \underline{a}_5},R_{\underline{b}_1 \underline{b}_2 \underline{b}_3}] = - 30 \delta^{[\underline{a}_1 \underline{a}_2 \underline{a}_3}_{[\underline{b}_1 \underline{b}_2 \underline{b}_3} R^{\underline{a}_4 \underline{a}_5]} \ ,
$$
$$
[R^{\underline{a}_1 .. \underline{a}_5},R_{\underline{b}_1 \ldots \underline{b}_5}] = - 5 \cdot 5 \cdot 6 \delta^{[\underline{a}_1 .. \underline{a}_4}_{[\underline{b}_1 \ldots \underline{b}_4} K^{\underline{a}_5]}{}_{\underline{b}_5]} + 5 \delta^{\underline{a}_1 .. \underline{a}_5}_{\underline{b}_1 \ldots \underline{b}_5} (3 D - \tilde{R})
$$
$$
[R^{\underline{a}_1 .. \underline{a}_5},R_{\underline{b}_1 \ldots \underline{b}_7}] = - 9 \cdot 70 \delta^{\underline{a}_1 .. \underline{a}_5}_{[\underline{b}_1 \ldots \underline{b}_5} R_{\underline{b}_6 \underline{b}_7]} \ , \ [R^{\underline{a}_1 .. \underline{a}_7},R_{\underline{b}}] = 0 \ ,$$
$$
[R^{\underline{a}_1 .. \underline{a}_7},R_{\underline{b}_1 \underline{b}_2 \underline{b}_3}] = 0 \ , \ [R^{\underline{a}_1 .. \underline{a}_7},R_{\underline{b}_1 \ldots \underline{b}_5}] =  9 \cdot 70 R^{[\underline{a}_1 \underline{a}_2} \delta^{\underline{a}_3 \ldots \underline{a}_7]}_{\underline{b}_1 \ldots \underline{b}_5} \ ,
 $$
$$
[R^{\underline{a}_1 .. \underline{a}_7},R_{\underline{b}_1 .. \underline{b}_7}] = 7 \cdot 7 \cdot 180 \delta^{[\underline{a}_1 \ldots \underline{a}_6}_{[\underline{b}_1 \ldots \underline{b}_6} K^{\underline{a}_7]}{}_{\underline{b}_7]} - 7 \cdot 6 \cdot 5  \delta^{\underline{a}_1 .. \underline{a}_7}_{\underline{b}_1 .. \underline{b}_7} (3D - \tilde{R}) \ \ ,  
\eqno(A.6)$$
where $D = \sum_{\underline{a}=0}^{9} K^{\underline{a}}{}_{\underline{a}}$.
\par
We now give  the commutators of the IIA generators with the $l_1$ generators. Again these results can be derived directly or equivalently from the commutators of the  $E_{11}\otimes_s l_1$ algebra in its eleven dimensional formulation [19]. Here we consider the algebra to level one in the IIA and $l_1$ representations and again only to $E_{11}$ level three. The IIA  level zero generators  with the IIA level zero $l_1$ generators have the commutators 
$$
[K^{\underline{a}}{}_{\underline{b}},P_{\underline{c}}] = - \delta^{\underline{a}}{}_{\underline{c}} P_{\underline{b}} \ \ , \ \ [\tilde{R},P_{\underline{c}}] = - 3 P_{\underline{c}} \ \ , \ \ [R^{\underline{a}_1\underline{a}_2},P_{\underline{b}}] = - 2 \delta^{[\underline{a}_1}{}_{\underline{b}} Q^{\underline{a}_2]} \ \ ,  
$$
$$
[R_{\underline{a}_1 \underline{a}_2},P_{\underline{b}}] = 0 \ \ , \ \  [K^{\underline{a}}{}_{\underline{b}},Q^{\underline{c}}] = + \delta^{\underline{c}}{}_{\underline{b}} Q^{\underline{a}} \ \ , \ \ [\tilde{R},Q^{\underline{c}}] = - 3 Q^{\underline{c}} \ \ , 
$$
$$
[R^{\underline{a}_1\underline{a}_2},Q^{\underline{c}}] = 0 \ \ , \ \ [R_{\underline{a}_1\underline{a}_2},Q^{\underline{c}}] = + 2 \delta^{\underline{c}}{}_{[\underline{a}_1} P_{\underline{a}_2]} \  \ , 
\eqno(A.7)$$
\par
The  IIA  level zero $IIA$ generators with the IIA  level one $l_1$ generators have the commutators 
$$
[K^{\underline{a}}{}_{\underline{b}},Z] = - {1 \over 2} \delta^{\underline{a}}{}_{\underline{b}} Z \ , \ [K^{\underline{a}}{}_{\underline{b}},Z^{\underline{c}_1 \underline{c}_2}] = 2 \delta^{[\underline{c}_1}{}_{\underline{b}} Z^{|\underline{a}| \underline{c}_2]} - {1 \over 2} \delta^{\underline{a}}{}_{\underline{b}} Z^{\underline{c}_1 \underline{c}_2} \ , \ 
$$
$$
[K^{\underline{a}}{}_{\underline{b}},Z^{\underline{c}_1 \ldots \underline{c}_4}] = 4 \delta^{[\underline{c}_1}{}_{\underline{b}} Z^{|\underline{a}| \underline{c}_2 \underline{c}_3 \underline{c}_4]} - {1 \over 2} \delta^{\underline{a}}{}_{\underline{b}} Z^{\underline{c}_1 \ldots \underline{c}_4} \ ,
$$
$$
[K^{\underline{a}}{}_{\underline{b}},Z^{\underline{c}_1 \ldots \underline{c}_6}] = 6 \delta^{[\underline{c}_1}{}_{\underline{b}} Z^{|\underline{a}| \underline{c}_2 \ldots \underline{c}_6]} - {1 \over 2} \delta^{\underline{a}}{}_{\underline{b}} Z^{\underline{c}_1 \ldots \underline{c}_6} \ \ ,
$$
$$
[\tilde{R},Z] = - 6 Z \ \ , \ \ [\tilde{R},Z^{\underline{a}_1 \underline{a}_2}] = - 6 Z^{\underline{a}_1 \underline{a}_2} \ \ , \ \ [\tilde{R},Z^{\underline{a}_1 \ldots \underline{a}_4}] = - 6 Z^{\underline{a}_1 \ldots \underline{a}_4} \ \ , 
$$
$$
[\tilde{R},Z^{\underline{a}_1 \ldots \underline{a}_6}] = - 6 Z^{\underline{a}_1 \ldots \underline{a}_6} \ \ , \ \ [R^{\underline{a}_1 \underline{a}_2},Z] = Z^{\underline{a}_1 \underline{a}_2} \ \ , \ \ [R_{\underline{a}_1 \underline{a}_2},Z] = 0 \ \ ,  
$$
$$
[R^{\underline{a}_1 \underline{a}_2},Z^{\underline{b}_1 \underline{b}_2}] = Z^{\underline{a}_1 \underline{a}_2 \underline{b}_1 \underline{b}_2} \ , \ [R_{\underline{a}_1 \underline{a}_2},Z^{\underline{b}_1 \underline{b}_2}] = 2 \delta^{\underline{a}_1 \underline{a}_2}_{\underline{b}_1 \underline{b}_2} Z \ ,
$$
$$
[R^{\underline{a}_1 \underline{a}_2},Z^{\underline{b}_1 .. \underline{b}_4}] = {1 \over 3} Z^{\underline{a}_1 \underline{a}_2 \underline{b}_1 .. \underline{b}_4} \ , \ [R_{\underline{a}_1 \underline{a}_2},Z^{\underline{b}_1 .. \underline{b}_4}] = 12 \delta^{[\underline{b}_1 \underline{b}_1}_{\underline{a}_1 \underline{a}_2} Z^{\underline{b}_3 \underline{b}_4]} \ , 
 $$
$$
[R^{\underline{a}_1 \underline{a}_2},Z^{\underline{b}_1 .. \underline{b}_6}] = 0 \ , \ [R^{\underline{a}_1 \underline{a}_2},Z^{\underline{b}_1 .. \underline{b}_6}] = 2 \cdot 45 \delta^{[\underline{b}_1 \underline{b}_2}_{\underline{a}_1 \underline{a}_2} Z^{\underline{b}_3 .. \underline{b}_6]} \ ,
 \eqno(A.8)$$
\par
The commutators of the   IIA  level one $IIA$ generators with level zero $l_1$ generators  are given by
$$
[R^{\underline{a}},P_{\underline{b}}] = - \delta^{\underline{a}}{}_{\underline{b}} Z \ , \ [R^{\underline{a}_1 \underline{a}_2 \underline{a}_3},P_{\underline{b}}] = 3 \delta^{[\underline{a}_1}{}_{\underline{b}} Z^{\underline{a}_2 \underline{a}_3]} \ , 
$$
$$
[R^{\underline{a}_1 \ldots \underline{a}_5},P_{\underline{b}}] = - {5 \over 2} \delta^{[\underline{a}_1}{}_{\underline{b}} Z^{\underline{a}_2 \ldots \underline{a}_5]} \ ,  \ [R^{\underline{a}_1 \ldots \underline{a}_7},P_{\underline{b}}] = {7 \over 6} \delta^{[\underline{a}_1}{}_{\underline{b}} R^{\underline{a}_2 \ldots \underline{a}_7]} \ , 
$$
$$
[R^{\underline{a}},Q^{\underline{b}}] = Z^{\underline{a} \underline{b}} \ , \ [R^{\underline{a}_1 \underline{a}_2 \underline{a}_3},Q^{\underline{b}}] = - Z^{\underline{a}_1 \underline{a}_2 \underline{a}_3 \underline{b}} \ ,  
$$
$$
[R^{\underline{a}_1 \ldots \underline{a}_5},Q^{\underline{b}}] = {1 \over 6} Z^{\underline{a}_1 \ldots \underline{a}_5 \underline{b}} \ , \ [R^{\underline{a}_1 \ldots \underline{a}_7},Q^{\underline{b}}] = 0 \ , 
 \eqno(A.9) $$ 
and the level minus one $IIA$ with level zero $l_1$ commutators are given by
$$
[R_{\underline{a}},P_{\underline{b}}] = 0 \ , \ [R_{\underline{a}_1 \underline{a}_2 \underline{a}_3},P_{\underline{b}}] = 0 \ , \ [R_{\underline{a}_1 \ldots \underline{a}_5},P_{\underline{b}}] = 0 \ , [R_{\underline{a}_1 \ldots \underline{a}_7},P_{\underline{b}}] = 0 \ , $$
$$
[R_{\underline{a}},Q^{\underline{b}}] = 0 \ , \ [R_{\underline{a}_1 \underline{a}_2 \underline{a}_3},Q^{\underline{b}}] = 0 \ , \ [R_{\underline{a}_1 \ldots \underline{a}_5},Q^{\underline{b}}] = 0 \ , \ [R_{\underline{a}_1 \ldots \underline{a}_7},Q^{\underline{b}}] = 0 \ . 
\eqno(A.10) $$ 
\par
The commutators of the level one $IIA$ generators with the level one $l_1$ commutators are given by
$$
[R^{\underline{a}},Z] = 0 \ , \ [R^{\underline{a}},Z^{\underline{b}_1 \underline{b}_2}] = 0 \ ,\ [R^{\underline{a}},Z^{\underline{b}_1 .. \underline{b}_4}] = Z^{\underline{a} \underline{b}_1 .. \underline{b}_4} \ , \ [R^{\underline{a}_1 \underline{a}_2 \underline{a}_3},Z] = 0 \ ,   
$$
$$
[R^{\underline{a}},Z^{\underline{b}_1 .. \underline{b}_6}] = Z^{\underline{a} \underline{b}_1 .. \underline{b}_6} + Z^{\underline{b}_1 .. \underline{b}_6,\underline{a}} \ , \ [R^{\underline{a}_1 \underline{a}_2 \underline{a}_3},Z^{\underline{b}_1 \underline{b}_2}] = Z^{\underline{a}_1 \underline{a}_2 \underline{a}_3 \underline{b}_1 \underline{b}_2}  
$$
$$
[R^{\underline{a}_1 \underline{a}_2 \underline{a}_3},Z^{\underline{b}_1 .. \underline{b}_4}] = Z^{\underline{b}_1 .. \underline{b}_4 [\underline{a}_1 \underline{a}_2 , \underline{a}_3]} - \overline{Z}^{\underline{a}_1 \underline{a}_2 \underline{a}_3 \underline{b}_1 .. \underline{b}_4} \ \ , \ [R^{\underline{a}_1 \underline{a}_2 \underline{a}_3},Z^{\underline{b}_1 .. \underline{b}_6}] = 0 \ ,  
$$
$$
[R^{\underline{a}_1 .. \underline{a}_5},Z] = {1 \over 2} R^{\underline{a}_1 .. \underline{a}_5} \ , \ [R^{\underline{a}_1 .. \underline{a}_5},Z^{\underline{b}_1 \underline{b}_2}] = - \overline{Z}^{\underline{a}_1 .. \underline{a}_5 \underline{b}_1 \underline{b}_2} - {1 \over 3} Z^{\underline{a}_1 .. \underline{a}_5 [\underline{b}_1 , \underline{b}_2]} \ ,  
$$
$$
[R^{\underline{a}_1 .. \underline{a}_5},Z^{\underline{b}_1 .. \underline{b}_4}] = 0 \ , \ [R^{\underline{a}_1 .. \underline{a}_5},Z^{\underline{b}_1 .. \underline{b}_6}] = 0 \ ,  
$$
$$
[R^{\underline{a}_1 .. \underline{a}_7},Z] = - {3 \over 2} \overline{Z}^{\underline{a}_1 .. \underline{a}_7} - {1 \over 6} Z^{\underline{a}_1 .. \underline{a}_7} \ , \ [R^{\underline{a}_1 .. \underline{a}_7},Z^{\underline{b}_1 \underline{b}_2}] = 0 \ ,  
$$
$$
[R^{\underline{a}_1 .. \underline{a}_7},Z^{\underline{b}_1 .. \underline{b}_4}] = 0 \ , \ [R^{\underline{a}_1 .. \underline{a}_7},Z^{\underline{b}_1 .. \underline{b}_6}] = 0 \ , 
\eqno(A.12)$$
and level minus one $IIA$ with the level one $l_1$ commutators are given by
$$
[R_{\underline{a}},Z] = - P_{\underline{a}} \ , \ [R_{\underline{a}},Z^{\underline{b}_1 \underline{b}_2}] = 2\delta^{[\underline{b}_1}{}_{\underline{a}} Q^{\underline{b}_2]} \ , \ [R_{\underline{a}},Z^{\underline{b}_1 .. \underline{b}_4}] = 0 \ , \ [R_{\underline{a}},Z^{\underline{b}_1 .. \underline{b}_6}] = 0 \ ,  
$$
$$
\ [R_{\underline{a}_1 \underline{a}_2 \underline{a}_3},Z] = 0 \ , \ [R_{\underline{a}_1 \underline{a}_2 \underline{a}_3},Z^{\underline{b}_1 \underline{b}_2}] = 6 \delta^{\underline{b}_1 \underline{b}_2}_{[\underline{a}_1 \underline{a}_2} P_{\underline{a}_3]} \ ,
$$
$$
[R_{\underline{a}_1 \underline{a}_2 \underline{a}_3},Z^{\underline{b}_1 .. \underline{b}_4}] = - 24 \delta^{[\underline{b}_1 \underline{b}_2 \underline{b}_3}_{\underline{a}_1 \underline{a}_2 \underline{a}_3} Q^{\underline{b}_4]} \ , \ [R_{\underline{a}_1 \underline{a}_2 \underline{a}_3},Z^{\underline{b}_1 .. \underline{b}_6}] = 0 \ ,
$$
$$
[R_{\underline{a}_1 .. \underline{a}_5},Z] = 0 , \ [R_{\underline{a}_1 .. \underline{a}_5},Z^{\underline{b}_1  \underline{b}_2}] = 0 \ , \ [R_{\underline{a}_1 .. \underline{a}_5},Z^{\underline{b}_1 .. \underline{b}_4}] = 72 \delta^{\underline{b}_1 .. \underline{b}_4}_{[\underline{a}_1 .. \underline{a}_4} P_{\underline{a}_5]} \ , 
$$
$$
[R_{\underline{a}_1 .. \underline{a}_5},Z^{\underline{b}_1 .. \underline{b}_6}] = - 135 \cdot 8 \delta_{\underline{a}_1 .. \underline{a}_5}^{[\underline{b}_1 .. \underline{b}_5} Q^{\underline{b}_6]} \ , \ [R_{\underline{a}_1 .. \underline{a}_7},Z] = 0 \ , \ [R_{\underline{a}_1 .. \underline{a}_7},Z^{\underline{b}_1 \underline{b}_2}] = 0 \ , $$
$$
[R_{\underline{a}_1 .. \underline{a}_7},Z^{\underline{b}_1 .. \underline{b}_4}] = 0 \ , \ [R_{\underline{a}_1 .. \underline{a}_7},Z^{\underline{b}_1 .. \underline{b}_6}] = 8 \cdot 7 \cdot 135 P_{[\underline{a}_1} \delta_{\underline{a}_1 .. \underline{a}_7]}^{\underline{b}_1 .. \underline{b}_6} \ . 
\eqno(A.13) $$

%%%%%%%%%%%%%%%%%%%%%%%%%%

\medskip
{\bf Appendix B} 
\medskip
In this section we will give some of commutators of the algebra $I_c(E_{11})\otimes_s l_1$ at level zero  which will be useful in section four of this paper. In  appendix  A of this paper  the $E_{11}\otimes_s l_1$ in its decomposition corresponding to the IIA theory was found by deleting node ten (the IIA node)  from the $E_{11}$ Dynkin diagram, leaving the algebra $SO(10,10)\otimes GL(1)$,  followed by deleting node eleven to decompose the result into representations of GL(10). 
The algebra of the generators   at IIA  level zero  was given in equation (A.1). In what follows we will discard the GL(1) generator $\tilde R$.  
\par
 At IIA level zero the vector representation contains the generators $P_a= \hat P_a$ and $Q^a= -\hat Z^{a11}$ and their commutators with $E_{11}$ was given in equation (A.13).  The Cartan involution acts as in equation (2.7)
\par
Defining $J_{ab}= K^a{}_b- \eta ^{ac}\eta_{bd}K^d{}_c$ , $S^{ab}= R^{ab}- \eta^{ac}\eta^{bd}R_{cd}$ we find the Cartan involution invariant algebra, $I_c(SO(10, 10))= I_c (D_{10})$ is given by 
$$
[J^{a_1a_2} , J_{b_1b_2} ]= -4 \delta ^{[a_1 }_{[b_1}J^{a_2]}{}_{b_2]} , \ 
[J^{a_1a_2} , S_{b_1b_2} ]= -4 \delta ^{[a_1 }_{[b_1}S^{a_2]}{}_{b_2]} , \ 
$$
$$
[S^{a_1a_2} , S_{b_1b_2} ]= -4 \delta ^{[a_1 }_{[b_1}J^{a_2]}{}_{b_2]} , \ 
\eqno(B.1)$$
Defining $M^{a_1a_2} ={1\over 2} (J^{a_1a_2} +S^{a_1a_2} )$, $\bar M^{a_1a_2} ={1\over 2} (J^{a_1a_2} -S^{a_1a_2} )$ these commutators become 
$$
[M^{a_1a_2} , M_{b_1b_2} ]= -4 \delta ^{[a_1 }_{[b_1}M^{a_2]}{}_{b_2]} ,
[\bar M^{a_1a_2} , \bar M _{b_1b_2} ]= -4 \delta ^{[a_1 }_{[b_1}\bar M ^{a_2]}{}_{b_2]} ,
[ M^{a_1a_2} , \bar M _{b_1b_2} ]=0
\eqno(B.2)$$
which we recognise as  the algebra $I_c(SO(10, 10))=SO(10)\otimes SO(10)$. In this section all our equations apply equally well to $SO(D)\otimes SO(D)$ with suitable adjustment of the range of the indices. 
\par
The Cartan involution odd generators are  $T_{ab}= \eta _{ac}K^c{}_b+ \eta _{bc}K^c{}_a$ , $U^{ab}= R^{ab}+ \eta ^{ac} \eta^{bd}R_{cd}$ and they obey the algebra 
$$
[J^{a_1a_2} , T_{b_1b_2} ]= -4 \delta ^{[a_1 }_{(b_1}T^{a_2]}{}_{b_2)} , \ 
[J^{a_1a_2} , U_{b_1b_2} ]= -4 \delta ^{[a_1 }_{[b_1}U^{a_2]}{}_{b_2]} , \ 
$$
$$
[S^{a_1a_2} , T_{b_1b_2} ]= 4 \delta ^{[a_1 }_{(b_1}U^{a_2]}{}_{b_2)} , \ 
[S^{a_1a_2} , U_{b_1b_2} ]= 4 \delta ^{[a_1 }_{[b_1}T^{a_2]}{}_{b_2]} 
\eqno(B.3)$$
and 
$$
[T^{a_1a_2} , T_{b_1b_2} ]= 4 \delta ^{(a_1 }_{(b_1}J^{a_2)}{}_{b_2)} , \ 
[U^{a_1a_2} , U_{b_1b_2} ]= 4 \delta ^{[a_1 }_{[b_1}J^{a_2]}{}_{b_2]} , \ 
[T^{a_1a_2} , U_{b_1b_2} ]= 4 \delta ^{(a_1 }_{[b_1}S^{a_2)}{}_{b_2]} 
\eqno(B.4)$$
\par
Defining  the generators 
$$
 V_{a_1a_2}= (T_{a_1a_2} -U_{a_1a_2} ) ,\  \bar V_{a_1a_2}=  (T_{a_1a_2} +U_{a_1a_2} ) 
\eqno(B.5)$$
the above commutators become  
$$
[M^{a_1a_2} , V_{b_1b_2} ]= -4 \delta ^{[a_1 }_{b_1}V^{a_2]}{}_{b_2} , \ 
[M^{a_1a_2} , \bar  V_{b_1b_2} ]= -4 \delta ^{[a_1 }_{b_2} V^{a_2]}{}_{b_1}  , \ 
$$
$$
[\bar M^{a_1a_2} , \bar V_{b_1b_2} ]= -4 \delta ^{[a_1 }_{b_1}\bar V^{a_2]}{}_{b_2} , \ 
[\bar M^{a_1a_2} , V_{b_1b_2} ]= -4 \delta ^{[a_1 }_{b_2} \bar V^{a_2]}{}_{b_1}  
\eqno(B.6)$$
While the odd with odd generators obey 
$$
[ V^{a_1a_2} ,   V_{b_1b_2} ]= 4 \delta ^{a_1 }_{b_1}\bar M^{a_2}{}_{b_2} +4 \delta ^{a_2 }_{b_2}M^{a_1}{}_{b_1} , \ 
[ \bar V^{a_1a_2} ,  \bar  V_{b_1b_2} ]= 4 \delta ^{a_1 }_{b_1}M^{a_2}{}_{b_2} +4 \delta ^{a_2 }_{b_2} \bar M^{a_1}{}_{b_1} , \ 
$$
$$
[ V^{a_1a_2} ,  \bar V_{b_1b_2} ]= 4 \delta ^{a_2 }_{b_1}M^{a_1}{}_{b_2} +4 \delta ^{a_1 }_{b_2} \bar M^{a_2}{}_{b_1} , \ 
\eqno(B.7)$$
\par
We will now consider the vector representation and it commutators with the odd and even parts of $D_{10}$. The Cartan involution invariant algebra $I_c (D_{10})$  acts on the vector representation as 
$$
[J_{ab}, P_c] =-2 \eta _{c[a} P_{b]} ,\ [J_{ab}, Q_c] = -2\eta _{c[a} Q_{b]} ,\ [S_{ab}, P_c] =-2 \eta _{c[a} Q_{b]} ,\ 
,\ [S_{ab}, Q_c] =-2 \eta _{c[a} P_{b]}
\eqno(B.8)$$
The above equations correct a number of typographical errors in reference [17]. While with the Cartan involution odd generators we have 
$$
[T_{ab}, P_c] =-2 \eta _{c(a} P_{b)} ,\ [T_{ab}, Q_c] = +2\eta _{c(a} Q_{b)} ,\ [U_{ab}, P_c] =-2 \eta _{c[a} Q_{b]} ,\ 
,\ [U_{ab}, Q_c] =2 \eta _{c[a} P_{b]}
\eqno(B.9)$$
\par
The commutators of the vector representation simplify if we write  the commutators in terms of 
$$
{\cal {\wp }}_a= P_a+Q_a ,\  \bar {\cal {\wp }}_a= P_a-Q_a ,\
\eqno(B.10)$$
Then 
$$
[M_{a_1a_2} , {\cal {\wp }}_b ]= -2 \eta _{c [a_1} {\cal {\wp }}_{a_2]} ,\ [\bar M^{a_1a_2} , {\cal {\wp }}_b ]=0 ,\ 
[\bar M_{a_1a_2} , \bar {\cal {\wp }}_b ]= -2 \eta _{c [a_1}\bar  {\cal {\wp }}_{a_2]} ,\ [M^{a_1a_2} , \bar {\cal {\wp }}_b ]=0 ,\ 
\eqno(B.11)$$
and 
$$
[V_{a_1a_2} ,  {\cal {\wp }}_b ]= -2 \eta _{b a_1}  \bar  {\cal {\wp }}_{a_2} ,\ [ V_{a_1a_2} , \bar {\cal {\wp }}_b ]=-2\eta_{b a_2} {\cal {\wp }}_{a_1} ,\ 
$$
$$
[\bar V_{a_1a_2} ,  \bar {\cal {\wp }}_b ]= -2 \eta _{b a_1} {\cal {\wp }}_{a_2} ,\ [\bar V_{a_1a_2} ,  {\cal {\wp }}_b ]=-2\eta_{ba_2} \bar {\cal {\wp }}_{a_1} ,\
\eqno(B.12)$$
\par
Thus the generators of $I_c(D_{10})\otimes_s l_1$ split into two  irreducible algebras  which are 
$$
M_{a_1a_2} ,\  {\cal {\wp }}_a ,\ \ \  {\rm and }\ \ \ 
\bar M_{a_1a_2} ,\  \bar  {\cal {\wp }}_b 
\eqno(B.13)$$
\par
Finally we give the explicit transformations of the vector representation which were given by $\delta P_a= [\Lambda , P_a]$ and $\delta Q_a= [\Lambda , Q_a ]$ where we take the Lie algebra element of $D_{10}$ to be given by 
$$
\Lambda = \Lambda ^{ab}J_{ab}+ \tilde \Lambda ^{ab}S_{ab}+ \Omega ^{ab} T_{ab} + \hat \Omega^{ab} U_{ab}
=  \lambda ^{ab}M_{ab}+ \tilde \lambda ^{ab}\bar M _{ab}+ \mu ^{ab} V_{ab} + \tilde  \mu^{ab} \bar V_{ab}
\eqno(B.14)$$
We find that under  $D_{10}$ transformations 
$$
 \delta {\cal {\wp }}_a= -2(\Lambda+\tilde \Lambda)_a{}^b  {\cal {\wp }}_b -2(\Omega -\hat \Omega)_a{}^b  \bar {\cal {\wp }}_b
= -2 \lambda_a{}^b  {\cal {\wp }}_b -2( \mu_a{}^b+\bar \mu^b{}_a ) \bar {\cal {\wp }}_b , 
$$
$$
\delta  \bar {\cal {\wp }}_a= -2(\Lambda-\tilde \Lambda)_a{}^b  \bar {\cal {\wp }}_b -2(\Omega +\hat \Omega)_a{}^b  {\cal {\wp }}_b 
=  -2\tilde  \lambda_a{}^b \bar {\cal {\wp }}_b-2(\tilde  \mu_a{}^b +\mu^b{}_a) { \cal {\wp }}_b, 
\eqno(B.15)$$
\par
It is straight forward to verify that 
$$
P_a^2 + Q_a^2 , \ \ \  {\rm and } \ \ \ P^a Q_a \ \ \ {\rm or  \ equivalently } \ \ \ {\cal {\wp }}_a^2 , \ \ \ {\rm and }\ \ \ \bar  { \cal {\wp }}_a^2
\eqno(B.16)$$
are invariant under $I_c(D_{10})$ transformations.

%%%%%%%%%%%%%%%%%
\medskip
{\bf References}
\medskip

\item{[1]} P. West, {\it $E_{11}$ and M Theory}, Class. Quant.  
Grav.  {\bf 18}, (2001) 4443, hep-th/ 0104081. 
\item{[2]} P. West, {\it $E_{11}$, SL(32) and Central Charges},
Phys. Lett. {\bf B 575} (2003) 333-342,  hep-th/0307098. 
\item{[3]} A. Tumanov and P. West, {\it E11 must be a symmetry of strings and branes },  Phys. Lett. {\bf  B759 }Ê(2016),
arXiv:1512.01644. 
\item{[4]} A. Tumanov and P. West, {\it E11 in 11D}, Phys.Lett. B758 (2016) 278, arXiv:1601.03974. 
\item{[5]} P. West,{\it A brief review of E theory}, Proceedings of Abdus Salam's 90th  Birthday meeting, 25-28 January 2016, NTU, Singapore, Editors L. Brink, M. Duff and K. Phua, World Scientific Publishing and IJMPA, {\bf Vol 31}, No 26 (2016) 

1630043, arXiv:1609.06863.
\item{[6]}  A. Kleinschmidt and P. West, {\it  Representations of G+++ and the role of space-time},  JHEP 0402 (2004) 033,  hep-th/0312247.
\item{[7]} P. West,  {\it Brane dynamics, central charges and $E_{11}$ }, Phys.Lett. {\bf B575} (2003) 333, , arXiv:hep-th/0412336. 
\item{[8]} P. Cook and P. West, {\it Charge multiplets and masses for E(11)},  JHEP {\bf 11} (2008) 091, arXiv:0805.4451.
\item{[9]} S. Elitzur, A. Giveon, D. Kutasov and E.  Rabinovici,  {\it
Algebraic aspects of matrix theory on $T^d$ }, {hep-th/9707217};  N. Obers,  B. Pioline and E.  Rabinovici, {\it M-theory and
U-duality on $T^d$ with gauge backgrounds}, { hep-th/9712084};  N. Obers and B. Pioline,~ {\it U-duality and  
M-theory, analgebraic approach}~, { hep-th/9812139};  N. Obers and B. Pioline,~ {\it U-duality and  M-theory}, {hep-th/9809039}. 
\item{[10]} P. West,  {\it $E_{11}$ origin of Brane charges and U-duality multiplets}, JHEP 0408 (2004) 052, hep-th/0406150. 
\item{[11]} P. Cook, {\it Exotic E11 branes as composite gravitational solutions}, Class.Quant.Grav.26 (2009) 235023, arXiv:0908.0485.
\item{[12]} P. Cook, {\it Bound States of String Theory and Beyond } ,  JHEP 1203 (2012) 028,   arXiv:1109.6595. 
\item{[13]} E. P. Wigner, ÒOn unitary representations of the inhomogeneous Lorentz group,Ó Annals Math.40(1939) 149.
\item{[14]} P. West,  {\it  Irreducible representations of E theory},  Int.J.Mod.Phys. A34 (2019) no.24, 1950133,  arXiv:1905.07324.
\item{[15]} K. Glennon and P. West, {\it The massless irreducible representation in E theory and how bosons can appear as spinors}, International Journal of Modern Physics A, p.2150096. arXiv:2102.02152
\item{[16]} T. Nutma, SimpLie, a simple program for Lie algebras, \break 
https://code.google.com/p/simplie/.
\item{[17]} P. West, {\it E11, generalised space-time and IIA string theory}, Phys.Lett.B696 (2011) 403-409,   arXiv:1009.2624.
\item{[18]} P. West, {\it Generalised BPS conditions}, Mod.Phys.Lett. A27 (2012) 1250202,  

arXiv:1208.3397.
\item{[19]} P. West, {\it Introduction to Strings and Branes}, Cambridge University Press, 2012.
\item{[20]} A. Kleinschmidt, R, Kohl, R. Lautenbacher and H. Nicolai, {\it Representations of involutory subalgebras of affine KAc-Moody algebras}, arXiv:2102.00870. 
\item{[21]} P. Goddard and D. Olive, {\it Algebras, Lattices and Strings}, Proceedings of Conference 10-17 November 1983 pages 51-96, PMRI, editors  J. Lepowsky, S. Mandelstam and I Singer, Springer 1985; {\it Kac-Moody and Virasoro algebras in relation to quantum physics},  IJMP, {\bf  A1}, (1986)  303. 
\item{[22]} l. Frenkel  and V.Kac, {\it Basic Representations of Affine Lie Algebras and Dual Resonance Models},  Inventionesmath. 62 (1980) 23
\item{[23]} D. Bernard and J Thierry-Meig, {\it Level one Representations of the simple Kac-Moody Algebras in their Homogeneous Gradations}, Commun. Math. Phys. {\bf 111} (1987)181. 
\item{[24]} E. Del Giudice, P. Di Vecchia and S. Fubini, {\it Generalised properties of the dual resonance model}, Ann Phys {\bf 70} (1972) 378;  R.C. Brower and P. Goddard, {\it Collinear algebra for the dual model}, Nucl. Phys. {\bf B40} (1972) 437; 
R.C. Brower, {\it Spectrum generating algebra and no-ghost theorem for the dual model} Phys. Rev. {\bf D6} (1972) 1655. 
\item{[25]} F. Gliozzi, D. Olive and J. Scherk,
{\it Supersymmetry, Supergravity Theories and the Dual Spinor Model}, Nucl. Phys. {\bf B122} (1977) 253.
\item{[26]} M. Green, J. Schwarz and E. Witten, {\it Superstring theory, volume 1}, Cambridge University Press,1987.
\item{[27]} P. West, {\it Generalised space-time and duality }, Phys.Lett.B693 (2010) 373, 

\item{} arXiv:1006.0893; {\it E11, Brane Dynamics and Duality Symmetries}, Int.J.Mod.Phys. A33 (2018) no.13, 1850080, arXiv:1801.00669; {\it A sketch of brane dynamics in seven and eight dimension using E theory }, Int.J.Mod.Phys. A33 (2018) no.32, 1850187, arXiv:1807.04176.

%%%%%%%%%%%%%%%%%%%%%%%

\end